\documentclass[%
 reprint,
 amsmath,amssymb,
 aps,
]{revtex4-1}

\usepackage[utf8x]{inputenc}
\usepackage[english]{babel}
\usepackage{natbib}
\setcitestyle{square, numbers}
\usepackage[T1]{fontenc}
\usepackage{comment}
\usepackage{xcolor}
\usepackage{color,soul}

\usepackage{bm}

\usepackage{amsmath}
\usepackage{amssymb}
\usepackage{mhchem}
\usepackage{graphicx}
\usepackage{subcaption}
\usepackage[colorinlistoftodos]{todonotes}
\usepackage[colorlinks=true, allcolors=blue]{hyperref}
\usepackage{booktabs}
\usepackage{dcolumn}
\usepackage{tabularx}
\usepackage[export]{adjustbox}
\newcolumntype{C}[1]{>{\centering\let\newline\\\arraybackslash\hspace{0pt}}m{#1}}

\newcommand{\density}{\ensuremath{\rho(\vec{x})}}

\newcommand{\fddens}[3]{\ensuremath{\rho^{#1#2#3}}}
\newcommand{\MCSHDescr}{\ensuremath{\bar{\lambda}_{R}^{(n)}}}

\DeclareRobustCommand{\hlcyan}[1]{{\sethlcolor{cyan}\hl{#1}}}

\begin{document}

\title{Design and Analysis of Machine Learning Exchange-Correlation Functionals via Rotationally Invariant Convolutional Descriptors}

\author{Xiangyun Lei}
\author{Andrew J. Medford}%
 \email{ajm@gatech.edu}
\affiliation{ 
School of Chemical and Biomolecular Engineering, Georgia Institute of Technology, Atlanta, GA, 30318 USA}%

\date{\today}


\begin{abstract}
In this work we explore the potential of a new data-driven approach to the design of exchange-correlation (XC) functionals. The approach, inspired by convolutional filters in computer vision and surrogate functions from optimization, utilizes convolutions of the electron density to form a feature space to represent local electronic environments and neural networks to map the features to the exchange-correlation energy density. These features are orbital free, and provide a systematic route to including information at various length scales. This work shows that convolutional descriptors are theoretically capable of an exact representation of the electron density, and proposes Maxwell-Cartesian spherical harmonic kernels as a class of rotationally invariant descriptors for the construction of machine-learned functionals. The approach is demonstrated using data from the B3LYP functional on a number of small-molecules containing C, H, O, and N along with a neural network regression model. The machine-learned functionals are compared to standard physical approximations and the accuracy is assessed for the absolute energy of each molecular system as well as formation energies. The results indicate that it is possible to reproduce B3LYP formation energies to within chemical accuracy using orbital-free descriptors with a spatial extent of 0.2 \AA{}. The findings provide empirical insight into the spatial range of electron exchange, and suggest that the combination of convolutional descriptors and machine-learning regression models is a promising new framework for XC functional design, although challenges remain in obtaining training data and generating models consistent with pseudopotentials.
\end{abstract}

\maketitle

\section{Introduction}

Since its introduction in the mid-1960s \cite{PhysRev.136.B864,PhysRev.140.A1133}, density functional theory (DFT) has become a much-used tool in the fields of chemistry, material science, biology and others. The popularity of DFT arises mainly from its simple formalism and low computational cost compared to wavefunction theories (WFTs). 
The basic formalism of DFT establishes that the exact ground-state energy can be written as functional of the electron density:

\begin{equation}
\label{eq:DFT}
E_{GS}[\density] = T[\density] + J[\density] + E_{ext}[\density] + E_{xc}[\density]
\end{equation}
where \density{} is the ground-state electron density, $E_{GS}$ is the ground-state energy, $T$ is the kinetic energy functional for the non-interacting system, $J$ is the classical Coulomb self-energy (or Hartree energy) functional, $E_{ext}$ is the energy due to external potential (e.g. atomic nuclei), and $E_{xc}$ is the exchange-correlation functional that accounts for the difference between classical and quantum-mechanical electronic repulsion as well as the difference in kinetic energy between the non-interacting and interacting systems \cite{PhysRev.140.A1133}. Of these, $T$, $J$, and $E_{xc}$ are independent of the external potential and are hence considered ``universal'' functionals, while $E_{ext}$ depends on the atomic coordinates of a chemical system. Furthermore, $T$, $J$ and $E_{ext}$ are known exactly, so the challenge is to determine $E_{xc}$. Although a universal and exact ground-state $E_{xc}$ functional does exist as proved by Hohenberg and Kohn \cite{PhysRev.136.B864}, the form of this functional remains unknown. Numerous strategies have been employed to construct density functional approximations (DFAs) for $E_{xc}$; however, despite over five decades of research and hundreds of trials, no existing functionals are universally comparable to the accuracy of wavefunction theories. 

Construction of DFA's relies on two main components: a model space that describes the electronic environment and a functional that connects the model space to the energy density. The concept of improving approximations by increasing the complexity of the model space is captured by Perdew's popular analogy to ``Jacob's ladder'' \cite{Perdew_2001}, which reveals an important trend: the more non-local information that is used to describe the electronic environment, the better the quality of the approximation. In the early days, Kohn and Sham approximated $E_{xc}$ by assuming a uniform electron gas with an electronic density equal to the (spin) density at a local point, referred to as the local density approximation (LDA)\cite{PhysRev.140.A1133}. This approximation is surprisingly accurate for delocalized systems such as metals, but its accuracy is considerably lower for molecules. The next major improvement came decades later when several authors proposed using the gradient of electron density as an additional input to the exchange-correlation functional \cite{BeckeA.D.1988Ceoa, PerdewJohnP.1986Aasd, Perdew1989EAas}. This led to the development of a family of DFA's known as generalized gradient approximation (GGA) functionals that provided an order of magnitude improvement in accuracy and started a rapid increase in the application of DFT. 
The next logical step would be inclusion of the second derivative in the spirit of a Taylor expansion; however, the kinetic energy density is more commonly used \cite{Perdew_2001}. This ``meta-GGA'' (mGGA) functional family includes a diverse range of physical and empirical approximations that have generally improved accuracy, although the improvements are not always systematic \cite{MedvedevMichaelG2017RtCo}. The next class of functionals deviates from the strategy of adding more semi-local information by including a component of fully non-local exact exchange. These ``hybrid functionals'', introduced by Becke and coworkers in the B3PW91 functional \cite{BeckeA.D.1988Deaw, BeckeA.D.1993DtIT}, exhibit another general (though not necessarily systematic) improvement in accuracy. Hybrid functionals such as the ubiquitous B3LYP functional \cite{BeckeA.D.1988Deaw,LeeC.1988DotC} are very popular, particularly in the chemistry community, due to their high accuracy and relatively low cost for molecular systems. However, hybrid functionals have considerably higher computational cost and are difficult to implement for extended systems (e.g. solids and surfaces), leading to screened hybrid functionals such as HSE06 \cite{HeydJochen2003Hfbo,KrukauAliaksandrV.2006Iote}. Numerous other strategies have also been employed to capture long-range interactions, including fully non-local approaches such as 100\% exchange functionals \cite{ZhaoY2006Dffs, BeckeAxelD.2003Armo}, approaches that combine multiple approximations \cite{Armiento_2005,de_Silva_2015}, double-hybrid functionals including wavefunction based correlation \cite{ZhaoYan2005Medf,AngyanJanosG.2005vdWf,GrimmeStefan2006Shdf}, and functionals including dispersion \cite{WuQin2002Ectd, GrimmeStefan2004Adov, GrimmeStefan2006SGdf,AnderssonY.1996VdWi,DobsonJ.F.1996Csil,DionM2004VdWd,ThonhauserT.2007VdWd}. These diverse options for model spaces indicate that inclusion of additional and increasingly non-local information increase the accuracy of DFA's; however, the improvement of model spaces has been based primarily on chemical intuition. This is advantageous in the derivation of functionals from chemical and physical principles, but also leads to difficulties in systematically improving models or deconvoluting multiple physical effects to avoid double counting.

An alternative approach to model space development is to construct model spaces that can be systematically expanded into a theoretically complete description of the system. This is similar to a common approach in the fields of image processing and computer vision where convolutions are used to extract features/information of varying length scales \cite{Witkin_1987,Lindeberg_1993}. A noteworthy recent triumph in the application of convolutions in image processing are convolutional neural networks (convNets) \cite{NIPS1989_293,Lecun98gradient-basedlearning} where convolutional kernels are determined through deep learning. ConvNets have achieved unprecedented accuracy in handwriting, object, and facial recognition, and have revolutionized the field of image analysis \cite{Lecun98gradient-basedlearning,KrizhevskySutskeverHinton17cacm,SimardP.Y.2003Bpfc,Vaillant94anoriginal,Nowlan95aconvolutional,LawrenceS.1997Frac}. This approach can be translated to the field of functional development since electron density data can be projected onto a finite-difference grid and treated as a 3D image. With 3D convolutions, any local and semi-local feature of the electronic environment can be extracted, analogous to 2D images. In this work we explore this approach to model space construction, and show that the convolutional descriptor model space is theoretically complete in the limit that the kernel has the same size as the system.

In addition to the model space, a functional requires a mathematical connection between the model space and the exchange-correlation energy. This challenge is at the core of most functional development, and there are two distinct philosophies. The reductionist approach applies physical principles and theoretical constraints to derive ``parameter-free'' functionals. The PBE functional is a well-known example of this philosophy \cite{PerdewJ.P.1996GGAM,PerdewJohnP.1997GGAM}. These derived functionals tend to have less systematic bias toward specific molecular systems, and have a predictable accuracy across all systems. The alternative approach is empiricism, with a more practical focus on maximizing the accuracy of DFT in specific applications. Most empirical functionals are based on derived functional forms where some parameters are optimized based on molecular data of the systems of interest. The data is usually obtained from experiments or higher-level calculations. These functionals are usually accurate for systems similar to those used in training, but the accuracy is typically lower for other systems or properties \cite{PhysRevB.85.235149}. The B3LYP functional and the Minnesota functionals are well-known examples of empirical functionals \cite{BeckeA.D.1988Deaw,LeeC.1988DotC,ZhaoYan2008TMso, PeveratiRoberto2011ItAo,PeveratiRoberto2012EFwG,YuHaoyuS.2016MAKg}. Recently, approaches from machine learning (ML) have taken the empirical approach to functional development to its logical extreme. In a seminal paper by Snyder et al. the idea of using ML to connect density and kinetic energy density of a 1D model system is introduced \cite{SnyderJohnC2012Fdfw}. The success of this approach inspired substantial interest and subsequent development of employing ML in many different ways related to DFT. An extensive review is beyond the scope of this work, but examples include the use of ML to develop molecular dynamics force-fields \cite{HuanTranDoan2017Ausf,YaoKun2018TTmc,RowePatrick2018Doam}, application of ML models to reproduce DFT results without the use of expensive QM calculations \cite{BrockherdeFelix2017BtKe,HegdeGanesh2017MatD,HimmetogluBurak2016Tbml,RuppMatthias2015MLfQ, HansenKatja2015MLPo,YaoKun2016KEoH,MillsKyle2017Dlat,PereiraFlorbela2017MLMt}, application of ML to improve the accuracy or speed of DFT \cite{BartokAlbertP2013Mafo,SeinoJunji2018Smke,GaoTing2016Amlc,LiuQin2017ItPo} and direct inclusion of ML models in the construction of density functionals \cite{Tozer_1996,PhysRevB.85.235149,WellendorffJess2014mAas,PhysRevB.93.235162,AldegundeManuel2016Doae}. These numerous strategies have illustrated the substantial promise of ML techniques in the field of functional development. 

The key concept of ML-based density functionals is that highly-flexible ``universal'' regression models with thousands or millions of parameters are applied to connect a model space to the exchange-correlation energy. The parameters are optimized using a large amount of known data from experiment or calculations. This strategy does not require any knowledge of the complex underlying physics, but instead the challenge arises from obtaining a sufficient amount of high-quality data and utilizing verification and validation approaches to avoid over-parameterization. Machine learning models also have the advantage of being systematically improvable by addition of training data and/or increase of the regression model complexity (i.e. increase of the number of fitting parameters). Some common choices of regression models are support vector regressors (SVR) \cite{VapnikVladimirNaumovich1995Tnos}, kernel ridge regression (KRR) \cite{MurphyKevinP.2012Mlap}, Gaussian process regressors  (GPR) \cite{RasmussenCarlEdward2006Gpfm}, and artificial neural networks (NN) \cite{McCullochWarren1990Alco,HornikKurt1989Mfna}. Neural networks are a particularly interesting and popular class of non-linear regression models due to their property of being theoretically capable of approximating any function to arbitrary accuracy, as proved by the universal approximation theorem \cite{HornikEtAl89,citeulike:3561150}. The complexity of a NN can be easily tuned by adding/removing neurons and layers, and prediction is very fast once training is complete. The quality of the approximation will ultimately depend on the amount of training data available and the heuristics applied during the training process, but in principle NN's provide a route to a systematically-improvable regression model to connect a given model space to the exchange-correlation energy.

In this work, we combine the ideas of convolutional fingerprinting of electronic environments and neural networks to propose a functional design framework with systematically improvable model spaces and regression-based functionals. We show that 3D convolutions can be used to re-formulate finite-difference DFT and are theoretically complete in the trivial limit that the convolutions are equivalent to the input density. We also developed a specific class of convolutional kernels to extract features (or ``descriptors'') and form model spaces that are complete and rotationally invariant, inspired by the work of Worral et al. \cite{WorrallDanielE.2016HNDT} and Applequist \cite{ApplequistJon2002Mshi}. This is combined with NN regression models and exchange-correlation (XC) data from the B3LYP hybrid functional for a range of small molecule systems to construct ``surrogate'' functionals. These surrogate functionals are inspected based on their accuracy as compared to the grid-projected B3LYP XC functional, which is chosen to be the ground truth in this study since it is widely used and there is no semi-local closed form for the exact exchange energy. The resulting convolutional surrogate functionals are orbital-free and have systematically increasing non-locality, providing a route to test the systematic improvement of the resulting functionals and gain insight into the locality of electron exchange. The results indicate that accuracy increases significantly with the size of the model space, and that it is possible to reproduce B3LYP energies to within chemical accuracy using a semi-local orbital-free functional with a range of $>$ 0.1 \AA{}. However, practical challenges remain in the finite-difference representation of all-electron systems, and access to spatially-resolved exchange-correlation data is currently limited. These obstacles are non-trivial, but addressing them represents an alternative strategy for XC functional development.

\section{Methods}
\label{sec:methods}

The DFT data are generated with the Psi4 package \cite{doi:10.1021/acs.jctc.7b00174}. Single-point spin-paired calculations are performed at the B3LYP/aug-cc-pvtz level with both density and energy convergence set to $10^{-12}$ Ha. The geometries of molecules are taken from computational chemistry comparison and benchmark database (CCCBDB) maintained by NIST \cite{CCCBDB} and are static for all calculations. The training set consists of 15 small-molecule systems: \ce{C_2H_2}, \ce{C_2H_4}, \ce{C_2H_6}, \ce{CH_3OH}, \ce{CH_4}, \ce{CO}, \ce{CO_2}, \ce{H_2}, \ce{H_2O}, \ce{HCN}, \ce{HNC}, \ce{N_2}, \ce{N_2O}, \ce{NH_3}, \ce{O_3}. These molecules contain 4 common atom types (C, H, O, N) and a diverse range of single, double, and triple bonds between them. An additional 7 molecular systems that have similar chemistry are used as an independent test set: \ce{CH_3CN}, \ce{CH_3CHO}, \ce{H_2CCO}, \ce{H_2CO}, \ce{H_2O_2}, \ce{HCOOH}, \ce{N_2H_4}. In addition, 3 extra molecular systems with different chemistry, \ce{CH_3NO_2}, \ce{NH_2CH_2COOH} (glycine), \ce{NCCN}, are used to test the models' ability to extrapolate. This ``extrapolation set'' is not included in the accuracy analysis, but is used to probe the generality of the model in Sec. \ref{sec:outliersAnalysis}. The converged electronic density ($\rho$) and exchange-correlation energy density ($\epsilon_{xc}$) of the systems are projected onto a uniform 3D finite-difference grid and stored as 3D arrays. The overall size of each grid is 10 \AA{} $\times$ 10 \AA{} $\times$ 10 \AA{}  with the molecule centered in the cell, and the grid-point spacing is 0.02 \AA{}. This results in a total of $500^3= 125,000,000$ data points per system. Due to memory limitations, domain decomposition with sub-grids of 2\AA{} $\times$ 2\AA{} $\times$ 2\AA{} are used for data manipulations including descriptor extraction, sub-sampling, and prediction.
The \texttt{scipy} \cite{Scipy} implementation of fast Fourier transform convolution (FFT convolution) is used to extract electronic environment descriptors. 
To ensure correct padding of the convolutions, each sub-grid is combined with the 26 neighboring sub-grids prior to FFT convolutions. 

The resulting data for each system is sub-sampled to reduce the computational burden of training. A ``near-uniform'' sub-sampling algorithm is developed to produce roughly uniform sampling density across the high-dimensional space. This improves sampling efficiency by ensuring that rare data points from the tails of the distribution are included in the training sample. An illustration of the procedure is shown in Figure \ref{fig:subsamplingExample} and detailed explanations of the procedure can be found in the Supplementary Information.

\begin{figure}
	\centering
	\begin{subfigure}[t]{0.24\textwidth}
		\centering
		\includegraphics[width=\linewidth]{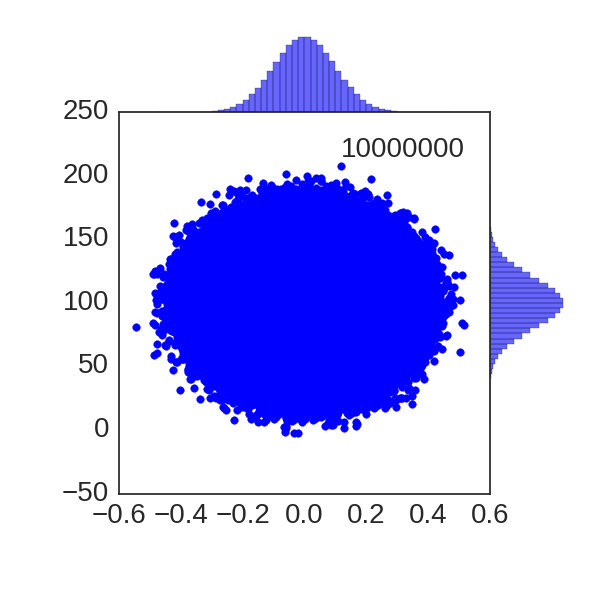}
        \caption{Before sub-sampling}
        \label{fig:subsamplingExampleBefore}
	\end{subfigure}%
	\begin{subfigure}[t]{0.24\textwidth}
		\centering
		\includegraphics[width=\linewidth]{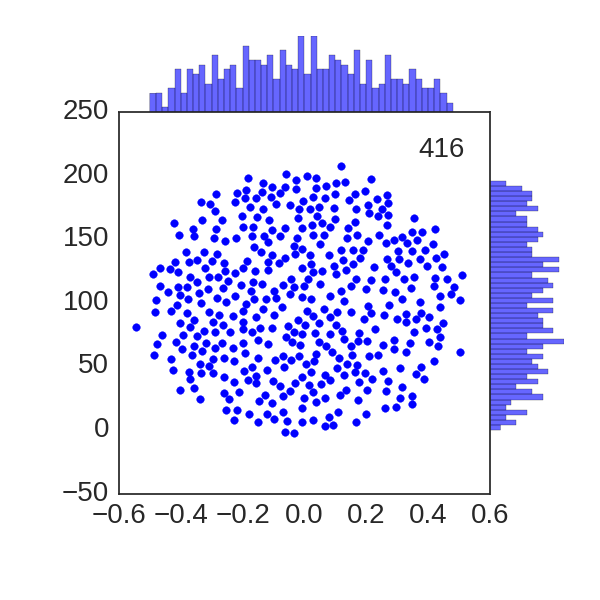}
        \caption{After near-uniform sub-sampling}
        \label{fig:subsamplingExampleAfter}
	\end{subfigure}
    \caption{Example of the near-uniform sub-sampling algorithm applied to a 2D random Gaussian distribution data set with hyper-parameters of $cutoff = 0.2, rate = 0.1$.}
	\label{fig:subsamplingExample}
\end{figure}


The near-uniform sub-sample is supplemented with a random sub-sample to provide information about the relative frequency of points in the original distribution. A fixed random sub-sample size of 10,000,000 in total for all training systems is selected, resulting in total sub-sample size of roughly 680,000 training points per molecule. The remaining 124,320,000 points in each molecular system are used to test the resulting models, corresponding to roughly 0.55\% training data and 99.45\% testing data, where $<$0.05\% of the training data is selected non-randomly to ensure that the tails of the distribution are represented. This is supplemented with fully independent validation sets from 7 additional molecular systems that are not used in model training; these systems are considered ``test'' systems in the remainder of the paper, while molecular systems used in model development are referred to as ``training'' systems although only 0.55\% of their data is actually used to train the NN models. The 3 additional molecular systems in the ``extrapolation set'' are also used to test the model and no points from these systems are used in training.

The machine-learning models are constructed using a framework similar to $\Delta$ machine learning \cite{RamakrishnanRaghunathan2015BDMQ} based on residuals from a re-fitted Vosko-Wilk-Nusair (VWN) \cite{VoskoS.H1980Asel} LDA model. The model is formulated as:
\begin{equation}
\begin{split}
\tilde{\epsilon}_{xc}(\vec{x}) = E_{r-VWN}(\density, C_1, \gamma, \alpha_1, \beta_1, \beta_2, \beta_3, \beta_4)\\
+ E_{NN}(\lambda_i(\density), W_{jk})
\end{split}
\label{eq:deltaML}
\end{equation}
where $\tilde{\epsilon}_{xc}(\vec{x})$ is the predicted XC energy density, $E_{r-VWN}$ is the energy of a VWN LDA model with the numerical values of its parameters ($C_1$, $\gamma$, $\alpha_1$, $\beta_1$, $\beta_2$, $\beta_3$, $\beta_4$) re-fitted to 1,000,000 randomly-selected data points from the B3LYP training systems. This re-parameterization is achieved with the Nelder-Mead algorithm \cite{GaoFuchang2012ItNs} as implemented in \texttt{scipy} \cite{Scipy} and the same r-VWN model is used for all surrogate functionals. The $E_{NN}$ term corresponds to a NN with a set of input descriptors, $\lambda_i$ and weights $W_{jk}$. The weights are optimized using the Adam algorithm for stochastic gradient descent \cite{KingmaDiederikP.2014AAMf} as implemented in the \texttt{Keras} package \cite{chollet2015keras}. A standard NN architecture of 2 hidden layers with 100 neurons and the ReLU activation function \cite{Nair:2010:RLU:3104322.3104425} is used for consistency. The training data is divided into separate steps with different learning rates and loss functions; details are available in the Supplementary Information.

The accuracy of the resulting models is assessed with three different error metrics at the chemical system level: sum of local absolute error ($\varepsilon_{absolute}$), energy prediction error ($\varepsilon_{predict}$) and formation energy prediction error ($\varepsilon_{formation}$), each probing different aspects of the models. The sum of local absolute error corresponds to the integral of absolute difference between the predicted and actual XC energy density:
\begin{equation}
\varepsilon_{absolute} = \sum_i | \epsilon_{xc}(\vec{x}_i) - \tilde{\epsilon}_{xc}(\vec{x}_i) | \times h^3
\end{equation}
This is a straightforward definition of the error from the perspective of model training, and is proportional to the mean absolute error. It directly probes the absolute accuracy of the model for a system and gives an upper bound for energy prediction error. The energy prediction error corresponds to the error of the integral of the XC energy density over the system, as approximated by the sum of the predicted energy at all grid points:
\begin{equation}
\begin{split}
    \varepsilon_{predict} &= E_{xc} - \sum_i \tilde{\epsilon}_{xc}(\vec{x}_i) \times h^3 \\
    &= \sum_i ( \epsilon_{xc}(\vec{x}_i) - \tilde{\epsilon}_{xc}(\vec{x}_i) )\times h^3
\end{split}
\end{equation}

This is proportional to the mean signed error, and cancellation of error will result in errors lower than the sum of local absolute error. The final metric of formation energy prediction error is the most practical since these relative quantities are most relevant in chemistry, and cancellation of systematic errors is a common feature of DFT functionals. The predicted formation energy error is obtained by computing the formation energy of each species relative to the following atomic reference states that are commonly employed in DFT studies: C: \ce{CH_4}, N: \ce{NH_3}, O: \ce{H_2O}, and H: \ce{H_2}. Formation energy errors enable the most cancellation of error, but anti-cancellation is also possible in the case of over-fitted models. Hence, formation energy errors compared to local and system level errors provide a convenient and practical measure of model accuracy and over-fitting.

The code for constructing descriptors, training, and evaluating models is available via the supporting information. 

\section{Results and Discussion}

The results are presented in three parts. In Sec. \ref{sec:reformulate} the theoretical motivation for using convolutions to construct model spaces is presented by re-formulating the XC energy in terms of convolution kernels, and some advantages and limitations are discussed. In Sec. \ref{sec:descriptors} a specific class of ``convolutional descriptors'' based on spherical harmonics are applied to the dataset of small molecules. In Sec. \ref{sec:regression} the machine learning framework for XC energy is introduced through discussion of both the re-parameterized VWN model and the NN models that are used to fit the residuals. The accuracy of models based on various model spaces are presented and discussed.

\subsection{Convolutional reformulation of exchange-correlation functional}
\label{sec:reformulate}

The theoretical motivation for using convolutions to form a basis set for XC model spaces is based on two properties: systematic increase of spatial range, and theoretical completeness in the limit that the range is equal to the system size. The spatial range can be increased by increasing the maximum size of the convolution kernels. To show completeness we re-formulate the XC functional, $E_{xc}[\density]$, as a function of convolutions between a set of kernels and the electron density. We show that the functional is equivalent to the function with the proper choice of kernels in the case where (i) the density is discretized onto a finite-difference (FD) grid, (ii) the maximum kernel size is equal to the system size, and (iii) the number of convolutions is equal to the number of points in the finite difference grid. This equivalence between functionals and functions for numerical representations in DFT has been noted before \cite{Kolb_2017}; here we briefly examine the specific case of finite difference representations and convolutions. We restrict the discussion to the spin-paired case for simplicity, but the same arguments hold in the case of spin-polarized systems. Similar arguments are expected to hold for any energy functional including the kinetic energy or full system energy, though we focus only on XC energy here. First, we consider spatially-resolved XC energy densities:

\begin{equation}
\label{eq:xcdensity}
E_{xc}[\density] = \int_{\mathbb{R}^3} \epsilon_{xc}[\density](\vec{x}) \mathrm{d}^3 \vec{x}
\end{equation}
where $\epsilon_{xc}[\density](\vec{x})$ is the exchange-correlation energy density defined at each point $\vec{x}$ in space. The existence of a locally-resolved XC energy density and methods to extract it have been examined previously \cite{Burke_1998}, although extracting this quantity presents a practical challenge for wave-function theories. Next, consider a finite-difference representation of the electron density:
\begin{equation}
\density = \fddens{x}{y}{z}
\end{equation}
where the density is represented on a 3D grid and $x$, $y$, $z$ are indices of each voxel along the (x, y, z) Cartesian axes. We note that these indices can be ``unraveled'' such that $\fddens{x}{y}{z}$ can be considered as a 1-dimensional vector of $N^3$ points with a single index, although it is conceptually simpler to consider $\fddens{x}{y}{z}$ as a 3-dimensional array of voxels. Each voxel has dimensions of $h_x$, $h_y$, $h_z$ \AA{} and a corresponding volume of $v = h_x h_y h_z$ \AA{}$^3$; for simplicity we consider the isotropic case where $h_x = h_y = h_z = h$. The finite difference representation is chosen because it is intuitive, convenient for convolutions, commonly used in solid-state codes \cite{ISI:000226735900040,GhoshSwarnava2017SAae1,GhoshSwarnava2017SAae2}, and systematically converges to the exact density in the limit of $h_i \rightarrow 0$. The XC energy can also be written in terms of a finite difference basis:

\begin{equation}
\label{eq:fdeps}
\epsilon_{xc}[\density](\vec{x}) = \epsilon^{lmn}(\fddens{x}{y}{z})
\end{equation}
where we have exploited the fact that a functional becomes a function when its argument is represented in a numerical basis (i.e. the XC energy density at each grid point is a function of the value of the electron density at every grid point). Finally, we introduce the concept of density convolutions:
\begin{equation}
\lambda_q^{xyz} = (C_q \circledast \rho)^{xyz}
\end{equation}
where $\lambda_q^{xyz}$ represents a vector of ``convolutional descriptors'' (indexed by $q$) spatially-resolved at each grid point (indexed by $xyz$), and $C_q$ is an arbitrary convolution kernel of size $n_x \times n_y \times n_z$. For convenience we restrict discussion to the case where $n_x = n_y = n_z = n$, and $n$ is odd, such that we can define the range $r_q$ of a convolution kernel $C_q$ as $r_q = h(n-1)/2$. Furthermore, we consider only the case of periodic boundary conditions to avoid issues of padding. In this case $\lambda_q^{xyz}$ is always the same dimension as $\rho^{xyz}$, corresponding to a feature vector indexed by $q$ at each grid point $xyz$. The restriction to periodic boundary conditions is a minor limitation, considering that any finite system can be represented as a periodic system with sufficient vacuum padding; this is commonly exploited in plane-wave codes. Considering a set of convolution kernels produces a set of $N_d$ ($q \leq N_d$) local descriptors for a point $xyz$ in space. These descriptors capture information out to a distance of a total range $R = max(r_q)$. If the largest dimension of the unit cell of the system is given by $L_{max}$ then in the limit of $R \rightarrow L_{max}/2$ and $N_d \rightarrow n^3$ the full non-local density can be recovered by using $n^3$ delta-function kernels:

\begin{equation}
\label{eq:nonlocaldesc}
\tilde{\lambda}_{xyz}^{lmn} = (\delta_{xyz} \circledast \rho)^{lmn} = \rho^{xyz}
\end{equation}
where $\delta_{xyz} = 1$ if $xyz = lmn$, 0 otherwise and $\tilde{\lambda}_{xyz}^{lmn}$ is the fully non-local descriptor set, equivalent to ``unraveling'' the entire density grid as a vector (indexed by $xyz$) at each spatially-resolved grid point (indexed by $lmn$). Substitution into Eq. \ref{eq:fdeps} yields:

\begin{equation}
\label{eq:fdepsexact}
\epsilon^{lmn}(\fddens{x}{y}{z}) = \epsilon^{lmn}(\tilde{\lambda}_{xyz}^{lmn})
\end{equation}
This expression is a trivial re-statement of Eq. \ref{eq:fdeps}, but it has the advantage of being  a system-independent mapping between a locally-centered electronic environment (as characterized by its convolutional descriptors) and a corresponding local XC energy density. However, in practice Eq. \ref{eq:fdepsexact} is no more efficient or practical than Equation \ref{eq:fdeps} since both ultimately require a fully non-local 6-dimensional evaluation of the energy functional. However, Eq. \ref{eq:fdepsexact} provides a natural starting point for establishing controlled orbital-free approximations to the XC energy density based on sets of descriptors $\lambda_{q}^{lmn}$ where $R << L_{max}/2$ and $N_d << n^3$. 

\begin{equation}
\epsilon^{lmn}(\fddens{x}{y}{z}) =  \epsilon^{lmn}(\tilde{\lambda}_{xyz}^{lmn}) \approx \epsilon^{lmn}(\lambda_q^{lmn})
\end{equation}

The two most common classes of orbital-free XC functionals, LDA and GGA, are easily reformulated in terms of convolutional descriptors. For example, the LDA functional approximates the exchange-correlation energy at a point $\vec{x}$ with the XC energy of the homogeneous electron gas with density equivalent to that point:

\begin{equation}
\epsilon_{LDA}[\density](\vec{x}) = \epsilon_{HEG}(\density)
\end{equation}
or, in convolutional notation:
\begin{equation}
\epsilon_{LDA}^{lmn}(\rho^{xyz}) = \epsilon_{HEG}^{lmn}(\rho^{lmn}) = \epsilon_{HEG}^{lmn}(\lambda_0^{lmn})
\end{equation}
where $\lambda_0^{lmn} = (\delta_{lmn} \circledast \rho)^{lmn} = \rho^{lmn}$. In the case of GGA functionals the XC energy density depends on the density and its gradient:
\begin{equation}
\epsilon_{GGA}[\density](\vec{x}) = \epsilon_{GGA}(\density, \nabla \density)
\end{equation}
or, in convolutional notation:
\begin{equation}
\epsilon_{GGA}^{lmn}(\rho^{xyz}) = \epsilon_{GGA}^{lmn}(\rho^{lmn}, \nabla \rho^{lmn}) = \epsilon_{GGA}^{lmn}(\lambda_0^{lmn}, \lambda_1^{lmn})
\end{equation}
where $\lambda_0^{lmn} = \rho^{lmn}$ as before, and $\lambda_1^{lmn} = (\nabla_1 \circledast \rho)^{lmn}$ where $\nabla_1$ is a finite difference stencil corresponding to the gradient. This pattern can be generalized to higher-order derivatives to produce a class of convolutional XC functionals based on differential stencils:

\begin{equation}
\begin{split}
\epsilon_{\nabla N}^{lmn}(\fddens{x}{y}{z}) \approx \epsilon^{lmn}(\nabla_0^{lmn}, \nabla_1^{lmn}, \nabla_2^{lmn}, ... \nabla_N^{lmn}) \\
=  \epsilon^{lmn}(\nabla_q^{lmn})
\end{split}
\label{eq:diff_class}
\end{equation}
where $\nabla_q^{xyz} = (\nabla_q \circledast \rho)^{xyz}$ and $\nabla_q$ is the $q^{\mathrm{th}}$ unmixed partial derivative stencil, with $\nabla_0 \equiv \delta_{xyz}$. Thus,  $\epsilon_{\nabla 0}^{lmn}$ corresponds to any fully local functional (e.g. LDA) and $\epsilon_{\nabla 1}^{lmn}$ corresponds to GGA functionals, while $\epsilon_{\nabla 2}^{lmn}$ corresponds to functionals that include the Laplacian \cite{Perdew_2007}, etc. The idea is analogous to that of Taylor series expansion in that it uses linear combinations of different orders of derivatives to approximate a function. Hence the functional should become more accurate as higher order derivatives are included and longer-range information is taken into account. However, the approach suffers from two issues, in theory and in practice: it's hard, if not impossible, to construct isotropic or rotation-invariant stencils for higher order derivatives, and higher order derivatives tend to become numerically unstable with practical grid spacing. It is clear that the descriptors must stay constant as the system rotates and translates. In other words, the stencils need to be rotation- and translation invariant. The magnitude of gradient operator and the Laplacian operator (trace of the Hessian) that are effectively adapted in GGA and mGGA, respectively, are known to be isotropic. However, the invariant norms of higher-order derivative tensors are not well known. Furthermore, issues with numerical stability are encountered when computing the Laplacian and higher order derivatives, even with analytical basis sets. For these reasons most functionals based on $\nabla_2^{xyz}$ have been abandoned for the ``meta-GGA’’ approach, which substitutes the kinetic energy density, $\tau^{xyz}$, for $\nabla_2^{xyz}$ \cite{CramerChristopherJ.2004Eocc}. The kinetic energy density is not orbital-free, and this transition deviates from the formalism of the Taylor expansion, making it unclear how to systematically improve model spaces beyond mGGA.

\subsection{Maxwell-Cartesian spherical harmonic descriptors}
\label{sec:descriptors}

\begin{figure*}
	\centering
    \begin{subfigure}{0.1\textwidth}
		\centering
        \includegraphics[width=\linewidth,frame]{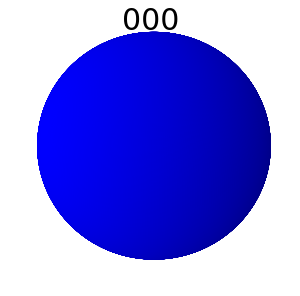}
        \caption{$S^{(0)}_{P(000)}$}
        \label{fig:LocalDensityStencil}
	\end{subfigure}\hspace{0.005\textwidth}%
    \begin{subfigure}{0.285\textwidth}
		\centering
        \includegraphics[width=\linewidth,frame]{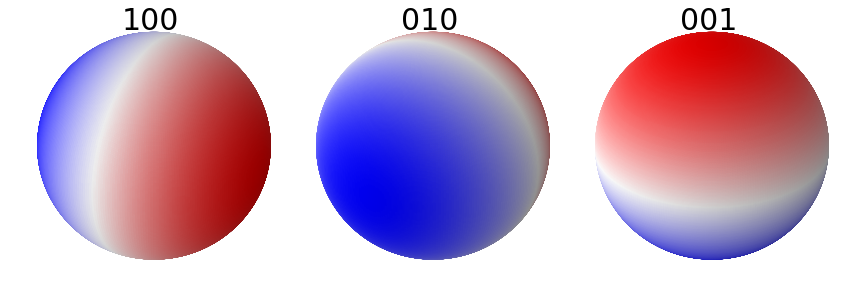}
        \caption{$S^{(1)}_{P(100)}$}
        \label{fig:AveDens006Stencil}
	\end{subfigure}\hspace{0.005\textwidth}%
    \begin{subfigure}{0.285\textwidth}
		\centering
        \includegraphics[width=\linewidth,frame]{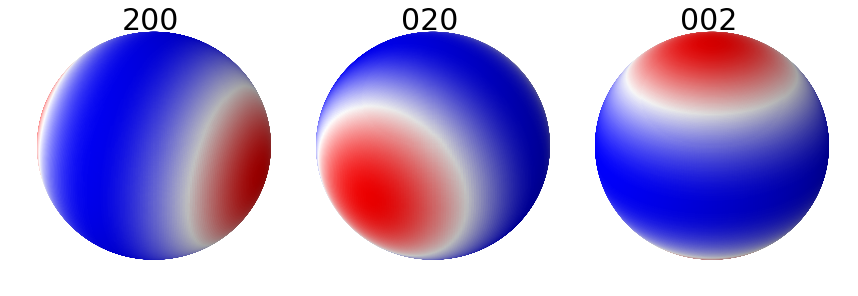}
        \caption{$S^{(2)}_{P(200)}$}
        \label{fig:LocalDensityStencil}
	\end{subfigure}\hspace{0.005\textwidth}%
    \begin{subfigure}{0.285\textwidth}
		\centering
        \includegraphics[width=\linewidth,frame]{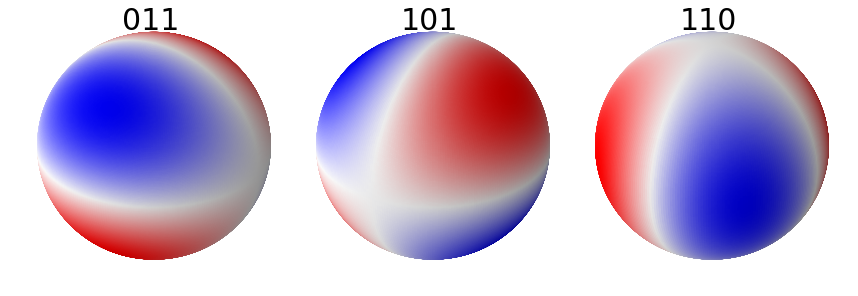}
        \caption{$S^{(2)}_{P(110)}$}
        \label{fig:AveDens006Stencil}
	\end{subfigure}
    \begin{subfigure}{0.285\textwidth}
		\centering
        \includegraphics[width=\linewidth,frame]{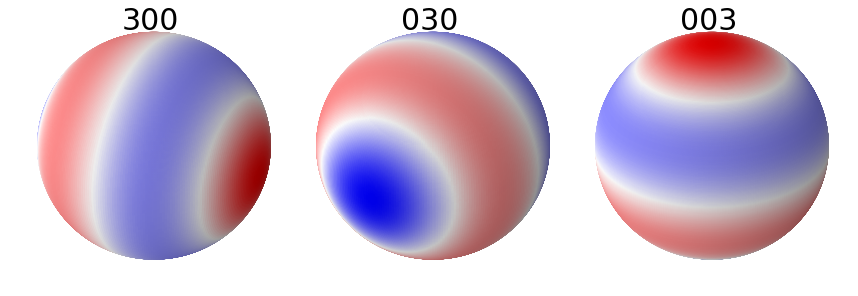}
        \caption{$S^{(3)}_{P(300)}$}
        \label{fig:AveDens006Stencil}
	\end{subfigure}\hspace{0.005\textwidth}%
    \begin{subfigure}{0.570\textwidth}
		\centering
        \includegraphics[width=\linewidth,frame]{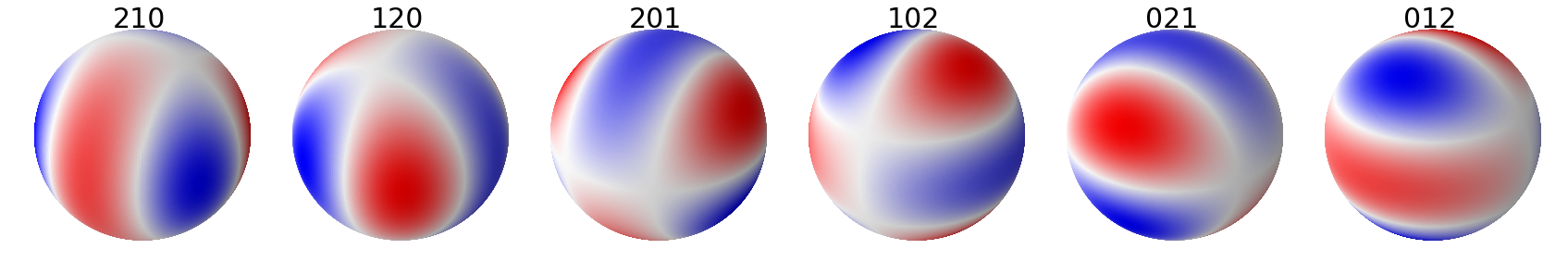}
        \caption{$S^{(3)}_{P(210)}$}
        \label{fig:LocalDensityStencil}
	\end{subfigure}\hspace{0.005\textwidth}%
    \begin{subfigure}{0.10\textwidth}
		\centering
        \includegraphics[width=\linewidth,frame]{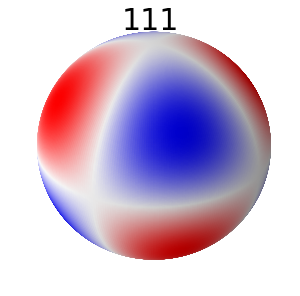}
        \caption{$S^{(3)}_{P(111)}$}
        \label{fig:AveDens006Stencil}
	\end{subfigure}
    \begin{subfigure}{0.29\textwidth}
		\centering
        \includegraphics[width=\linewidth,frame]{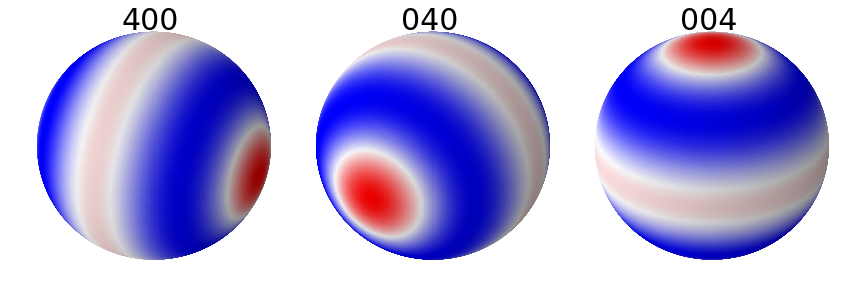}
        \caption{$S^{(4)}_{P(400)}$}
        \label{fig:AveDens006Stencil}
	\end{subfigure}\hspace{0.005\textwidth}%
    \begin{subfigure}{0.570\textwidth}
		\centering
        \includegraphics[width=\linewidth,frame]{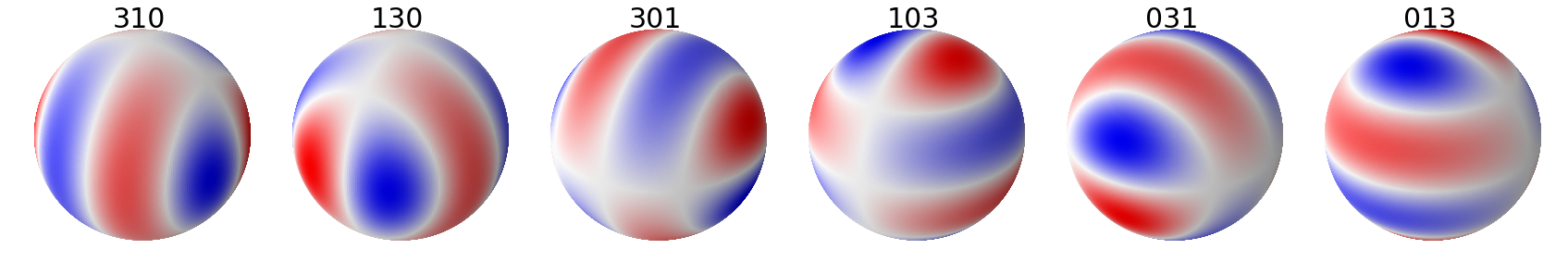}
        \caption{$S^{(4)}_{P(310)}$}
        \label{fig:AveDens006Stencil}
	\end{subfigure}
    \begin{subfigure}{0.285\textwidth}
		\centering
        \includegraphics[width=\linewidth,frame]{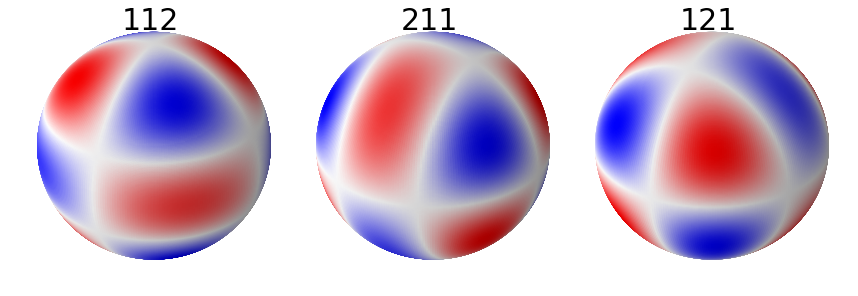}
        \caption{$S^{(4)}_{P(211)}$}
        \label{fig:LocalDensityStencil}
	\end{subfigure}\hspace{0.005\textwidth}%
    \begin{subfigure}{0.285\textwidth}
		\centering
        \includegraphics[width=\linewidth,frame]{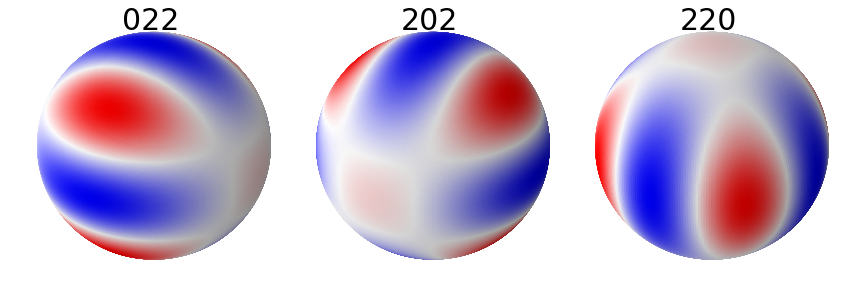}
        \caption{$S^{(4)}_{P(220)}$}
        \label{fig:AveDens006Stencil}
	\end{subfigure}
    \caption{Graphical illustrations of the first 4 orders of Maxwell-Cartesian spherical harmonics (MCSH) descriptor kernels denoted by $S^{(n)}_{P(ijk)}$. $n$ is the order and $P(ijk)$ denotes the permutation group of the index $ijk$. The Euclidean norm of MCSHs in each group as stencils gives 3D rotation-invariant descriptors}
    \label{fig:MCSH1}
\end{figure*}

Prior to selecting a convolutional descriptor set to fingerprint electronic environments, it is important to first define the necessary properties of the descriptor set. First the descriptor set needs to be complete. This means, at the limit of taking all the descriptors from the set, they should form a complete basis that can describe all possible variations in the electronic environment, or that of any 3D function in general. This complete set would be infinite, therefore the set should have a clear entry point and a systematic route toward convergence. Furthermore, the descriptors need to be invariant under translation and rotation, consistent with the symmetries of the Hamiltonian.

To find such descriptor set, the variation of 3D functions is decomposed into two parts: variations in angular coordinate (rotational variation) and variations in the radial coordinate (radial variation). The Maxwell-Cartesian spherical harmonics (MCSH) capture the rotational variations, and the radial variation is captured by varying a cutoff distance. This approach is inspired by the circular-harmonic-based 2D rotationally equivariant features developed by Worral et al. \cite{WorrallDanielE.2016HNDT} that has been generalized to 3D by the work of Thomas et al. \cite{ThomasNathaniel2018TfnR}, as well as Applequist's work \cite{ApplequistJon2002Mshi} on MCSHs. Specifically, the descriptor set is defined as follows:

\begin{equation}
\left\lbrace M^{(n)}_{r,ijk} = \sqrt{\sideset{}{_{P(i,j,k)}}\sum \mu_{r, ijk}^2} \mid i,j,k \in \mathbb{N}, r \in \mathbb{R}^{+}\right\rbrace
\label{eq:MCSHDescriptorSetDefinition}
\end{equation}
where $M$ denotes the descriptors, $P(i,j,k)$ denotes the permutation group of $i,j,k$, $n=i+j+k$ is the order of the descriptor, and $\mu_{r,ijk}$ is the convolution result using spherical harmonic $S_{ijk}$ with cutoff distance $r$ as the stencil:

\begin{equation}
\mu_{r,ijk} = input \circledast \overbrace{\left[ f_r(x,y,z) \times S_{ijk}^{(n)} (x,y,z)\right]}^\text{stencil}
\label{eq:MCSHDescriptorDefinition}
\end{equation}

$f_r$ is a step function that controls the cutoff distance, and $S^{(n)}_{ijk}$ is the Maxwell-Cartesian spherical harmonic:

\begin{equation}
f_r(x,y,z) = 
  \begin{cases} 
   1 & \text{if } \sqrt{x^2+y^2+z^2} \leq r \\
   0       & \text{if } x < 0
  \end{cases}
\label{eq:StepFunctionDefinition}
\end{equation}

\begin{equation}
\begin{split}
S_{ijk}^{(n)} (x,y,z) = \sum_{m_1=0}^{i/2}\sum_{m_2=0}^{j/2}\sum_{m_3=0}^{k/2} (-1)^m (2n-2m-1)!! \\
\times \left[\begin{array}{l}i\\m_1\end{array}\right] \left[\begin{array}{l}j\\m_2\end{array}\right] \left[\begin{array}{l}k\\m_3\end{array}\right] r^{2m} x^{i-2m_1} y^{j-2m_2} z^{k-2m_3}
\end{split}
\label{eq:MCSHDefinition}
\end{equation}

where $m=m_1+m_2+m_3$, and 

\begin{equation}
\left[\begin{array}{l}a\\b\end{array}\right] = \frac{a!}{2^bb!(a-2b)!}
\label{eq:MCSHDefinitionHelper3}
\end{equation}

Thus, each descriptor is the Euclidean norm of $\mu_{r,ijk}$ with all possible combination of $i,j,k$, and the whole set can be written as:

\begin{widetext}
\begin{equation}
\left\lbrace\begin{array}{l}
    \sqrt{\mu_{r_1,000}^2}, \sqrt{\mu_{r_1,100}^2 + \mu_{r_1,010}^2 + \mu_{r_1,001}^2}, \sqrt{\mu_{r_1,200}^2 + \mu_{r_1,020}^2 + \mu_{r_1,002}^2}, \sqrt{\mu_{r_1,110}^2 + \mu_{r_1,101}^2 + \mu_{r_1,011}^2}, \dots \\
    \sqrt{\mu_{r_2,000}^2}, \sqrt{\mu_{r_2,100}^2 + \mu_{r_2,010}^2 + \mu_{r_2,001}^2}, \sqrt{\mu_{r_2,200}^2 + \mu_{r_2,020}^2 + \mu_{r_2,002}^2}, \sqrt{\mu_{r_2,110}^2 + \mu_{r_2,101}^2 + \mu_{r_2,011}^2}, \dots \\
    \dots
  \end{array}\right\rbrace
\label{eq:MCSHDescriptorSet}
\end{equation}
\end{widetext}

\begin{table}[]
\centering
\begin{tabular}{C{0.2cm}C{0.7cm}C{2.0cm}|C{0.2cm}C{0.7cm}C{3.5cm}}
\toprule
n & \{ijk\} & $S^{(n)}_{ijk}$ & n & \{ijk\} & $S^{(n)}_{ijk}$         \\ \midrule
0 & 000     & 1                        & 4 & 400     & $105\hat{x}^4-90\hat{x}^2+9$\\
1 & 100     & $\hat{x}$ 			   &   & 040     & $105\hat{y}^4-90\hat{y}^2+9$\\
  & 010     & $\hat{y}$				   &   & 004     & $105\hat{z}^4-90\hat{z}^2+9$\\
  & 001     & $\hat{z}$ 			   &   & 310     & $105\hat{x}^3\hat{y}-45\hat{x}\hat{y}$\\
2 & 200     & $3\hat{x}^2-1$                  & & 301     & $105\hat{x}^3\hat{z}-45\hat{x}\hat{z}$\\
  & 020     & $3\hat{y}^2-1$                  & & 031     & $105\hat{y}^3\hat{z}-45\hat{y}\hat{z}$\\
  & 002     & $3\hat{z}^2-1$                  & & 130     & $105\hat{x}\hat{y}^3-45\hat{x}\hat{y}$\\
  & 110     & $3\hat{x}\hat{y}$               & & 103     & $105\hat{x}\hat{z}^3-45\hat{x}\hat{z}$\\
  & 101     & $3\hat{x}\hat{z}$               & & 013     & $105\hat{y}\hat{z}^3-45\hat{y}\hat{z}$\\
  & 011     & $3\hat{y}\hat{z}$               & & 220     & $105\hat{x}^2\hat{y}^2-15\hat{x}^2-15\hat{y}^2+3$\\
3 & 300     & $15\hat{x}^3-9\hat{x}$          & & 202     & $105\hat{x}^2\hat{z}^2-15\hat{x}^2-15\hat{z}^2+3$\\
  & 030     & $15\hat{y}^3-9\hat{y}$          & & 022     & $105\hat{y}^2\hat{z}^2-15\hat{y}^2-15\hat{z}^2+3$\\
  & 003     & $15\hat{z}^3-9\hat{z}$          & & 211     & $105\hat{x}^2\hat{y}\hat{z}-3\hat{y}\hat{z}$\\
  & 210     & $15\hat{x}^2\hat{y}-3\hat{y}$   & & 121     & $105\hat{x}\hat{y}^2\hat{z}-3\hat{x}\hat{z}$\\
  & 201     & $15\hat{x}^2\hat{z}-3\hat{z}$   & & 112     & $105\hat{x}\hat{y}\hat{z}^2-3\hat{x}\hat{y}$\\
  & 021     & $15\hat{y}^2\hat{z}-3\hat{z}$                         \\
  & 120     & $15\hat{x}\hat{y}^2-3\hat{x}$                         \\
  & 102     & $15\hat{x}\hat{z}^2-3\hat{x}$                         \\
  & 012     & $15\hat{y}\hat{z}^2-3\hat{y}$                         \\
  & 111     & $15\hat{x}\hat{y}\hat{z}$                         \\
 \bottomrule
\end{tabular}
\caption{The analytical expressions of first 4 orders of MCSH denoted by $S^{(n)}_{ijk}$.}
\label{tab:MCSH}
\end{table}

The first four orders of the MCSHs are listed in Table \ref{tab:MCSH} and illustrated in Figure \ref{fig:MCSH1}. The detailed properties of MCSH are introduced in the work of Applequist \cite{ApplequistJon2002Mshi}. The MCSHs are used to construct the descriptors because it is known that spherical harmonics form a complete basis for functions defined on the 3D unit sphere. This idea is analogous to that of multi-pole expansion, where the original 3D function is expressed as a linear combination of terms with progressively finer angular features \cite{maxwell1873treatise}. Examining the MCSHs in Figure \ref{fig:MCSH1} reveals that the order 0 MCSH corresponds to the monopole, and captures features that are constant and independent of angle (order 0 angular feature); the order 1 MCSH corresponds to the dipole, and captures features that vary once, from positive to negative, with angle (order 1 angular features); order 2 MCSH corresponds to the quadrupole, and captures features that varies more quickly with angle (order 2 angular features), and so on. Any rotational variations can be approximated by this linear combinations of angular features of different order, and will be exact in the limit of the entire series. The descriptors are empirically verified as rotation-invariant (see Supporting Information) and the mathematical proof is in progress but is beyond the scope of this work. In addition to the rotational variations it is necessary to capture radial variations. This is achieved by taking the rotation-invariant descriptors with different cutoff distances through the cutoff function $f_r$ in Eq. \ref{eq:StepFunctionDefinition}, where cutoff radii are discretized based on the underlying finite-difference grid. We conjecture that this descriptor set provides a complete basis on the rotation-invariant subspace of the 3D finite difference grid. Moreover, $\mu_{0,000}^2$, which is equivalent to fully local information (i.e. $\rho$), is the clear entry point of this descriptor set. There are 3 directions for systematic expansion: higher orders of MCSH, longer cutoff distances, and a finer grid for discretization. In this work we fix the grid spacing and explore the impact of higher orders of MCSH and longer cutoff distances.

The MCSH descriptors provide a route to include semi-local and non-local information about a local electronic environment. MCSH descriptor sets are defined by their maximum range ($R$) and the maximum order of spherical harmonics ($n$) that is the same as the maximum order of angular features captured, and are denoted as \MCSHDescr{}. Here, we consider 13 MCSH descriptor sets with ranges of 0 \AA{} , 0.02 \AA{}, 0.04 \AA{} , 0.08 \AA{} and 0.2 \AA{}, and orders of 0, 1 and 2. The descriptor sets are designed such that information of longer range and higher order of angular feature are gradually added:

\begin{equation}
\begin{split}
&\bar{\lambda}_{(0.0)}^{(0)} = \lbrace M^{(0)}_{(0.0,000)} \rbrace = \lbrace \rho \rbrace \\
&\bar{\lambda}_{(0.02)}^{(0)} = \lbrace \rho, M^{(0)}_{(0.02,000)} \rbrace \\
&\bar{\lambda}_{(0.02)}^{(1)} = \lbrace \rho, M^{(0)}_{(0.02,000)}, M^{(1)}_{(0.02,100)} \rbrace \\
&\bar{\lambda}_{(0.02)}^{(2)} = \lbrace \rho, M^{(0)}_{(0.02,000)}, M^{(1)}_{(0.02,100)}, M^{(2)}_{(0.02,200)}, M^{(2)}_{(0.02,110)} \rbrace \\
&\bar{\lambda}_{(0.04)}^{(0)} = \lbrace \rho, M^{(0)}_{(0.02,000)}, M^{(0)}_{(0.04,000)} \rbrace \\
&\bar{\lambda}_{(0.04)}^{(1)} = \lbrace \rho, M^{(0)}_{(0.02,000)}, M^{(1)}_{(0.02,100)}, M^{(0)}_{(0.04,000)}, M^{(1)}_{(0.04,100)} \rbrace 
\label{fig:DescriptorSetDefinition}
\end{split}
\end{equation}

\begin{table}[]
\centering
\begin{tabular}{C{2.0cm}C{2.0cm}|C{2.0cm}C{2.0cm}}
\toprule
Descriptor Set & Number of Descriptors & Descriptor Set & Number of Descriptors         \\ \midrule
$\bar{\lambda}_{(0.00)}^{(0)}$      & 1 & $\bar{\lambda}_{(0.08)}^{(1)}$      & 9\\
$\bar{\lambda}_{(0.02)}^{(0)}$      & 2 & $\bar{\lambda}_{(0.2)}^{(1)}$       & 21\\
$\bar{\lambda}_{(0.04)}^{(0)}$      & 3 & $\bar{\lambda}_{(0.02)}^{(2)}$      & 5\\
$\bar{\lambda}_{(0.08)}^{(0)}$      & 5 & $\bar{\lambda}_{(0.04)}^{(2)}$      & 9\\
$\bar{\lambda}_{(0.2)}^{(0)}$       & 11& $\bar{\lambda}_{(0.08)}^{(2)}$      & 17\\
$\bar{\lambda}_{(0.02)}^{(1)}$      & 3 & $\bar{\lambda}_{(0.2)}^{(2)}$       & 41\\
$\bar{\lambda}_{(0.04)}^{(1)}$      & 5 & &\\

 \bottomrule
\end{tabular}
\caption{List of number of features for each of the descriptor sets}
\label{tab:ModelDimensionality}
\end{table}

The total number of descriptors for each of the descriptor sets are listed in Table \ref{tab:ModelDimensionality}.

\subsection{Regression models for exchange-correlation energy}
\label{sec:regression}

The functional form linking the MCSH descriptors and the exchange correlation (XC) energy density is not known. While reductionist approaches may be feasible, the empirical approach is more pragmatic since the physical meaning of the descriptors is not obvious. In this work we employ a machine-learning strategy based on a function with two terms: a local-density term based on a re-parameterization of the VWN functional form of LDA (r-VWN), and a descriptor-based term using a NN with ReLU activation functions. The r-VWN term is static for all regression models, and the NN is trained using the residuals of the r-VWN model (Eq. \ref{eq:deltaML}); this is similar to the $\Delta$ machine learning strategy proposed previously \cite{RamakrishnanRaghunathan2015BDMQ}. This section first discusses the results of the r-VWN model and a NN based solely on the local density, and subsequently addresses the performance of NNs based on the convolutional descriptors.

\subsubsection{r-VWN and NN LDA model}
\label{sec:LDAmodels}

\begin{figure*}
\centering
	\begin{subfigure}{0.33\textwidth}
		\centering
		\includegraphics[width=\linewidth]{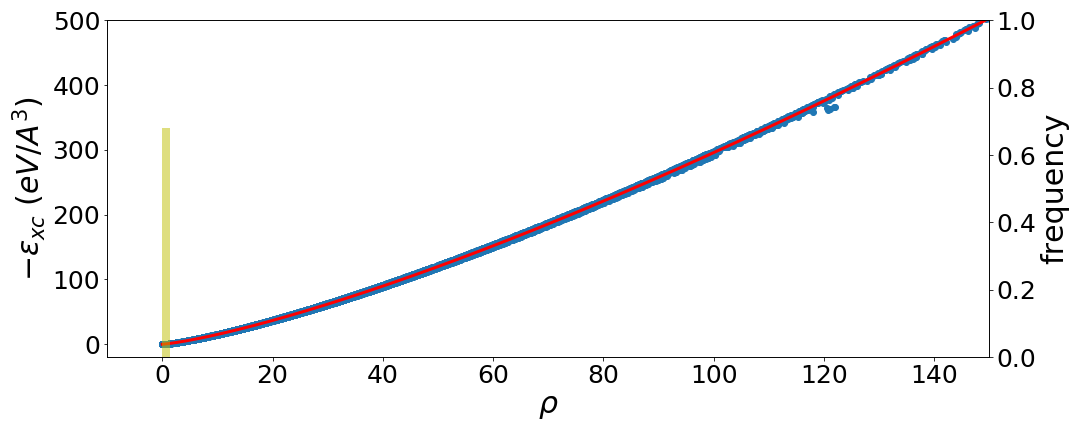}
        \caption{$\epsilon_{xc}$ vs. $\rho$}
        \label{fig:EngDensVsElecDensRealReal}
	\end{subfigure}%
	\begin{subfigure}{0.33\textwidth}
		\centering
		\includegraphics[width=\linewidth]{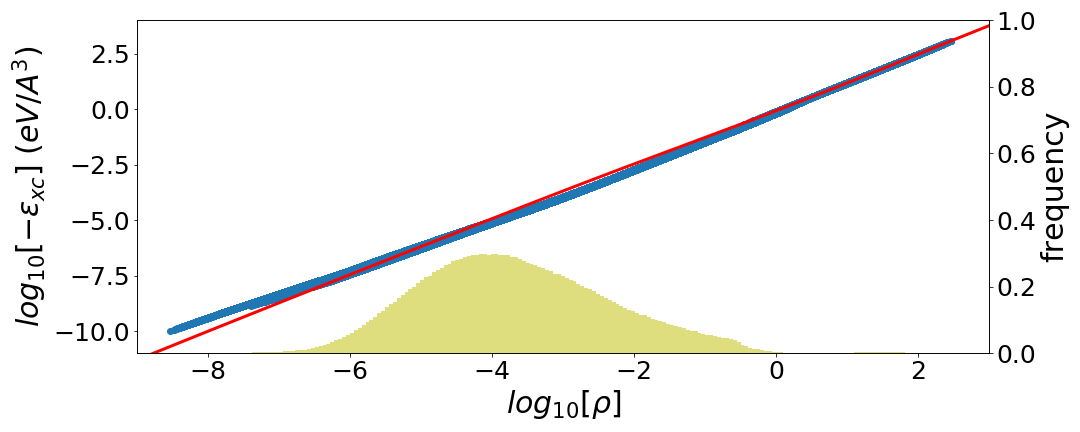}
        \caption{$\epsilon_{xc}$ vs. $\rho$ log scale}
        \label{fig:EngDensVsElecDensLogLog}
	\end{subfigure}%
    \begin{subfigure}{0.33\textwidth}
		\centering
		\includegraphics[width=\linewidth]{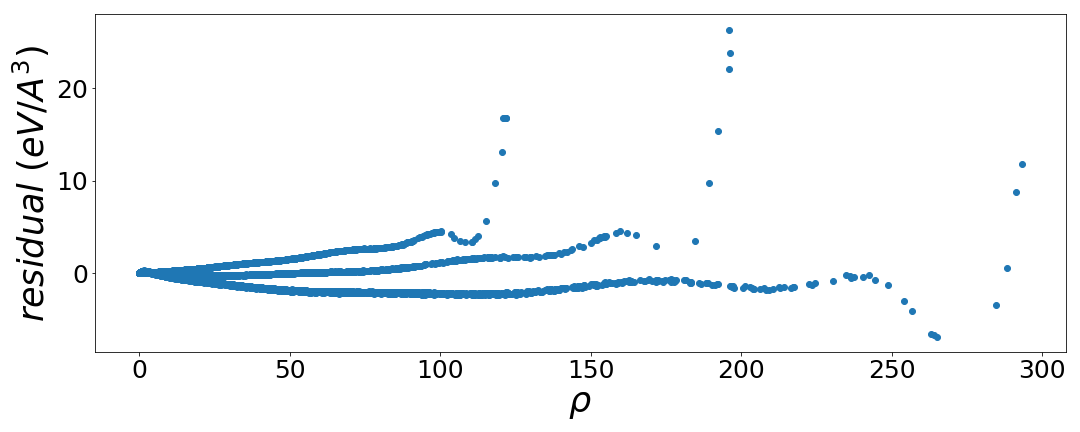}
        \caption{r-VWN residual vs. $\rho$}
        \label{fig:EngDensVsElecDensLogLog}
	\end{subfigure}
\caption{Plots of 3 million randomly sampled data points from training set to show the distribution, plus 10000 uniformly subsampled data points to highlight the high-density (core) regions. (a) Plot of $\epsilon_{xc}$ vs. $\rho$ in linear scale. The yellow bar plot shows the distribution of points. (b) Plot of $\epsilon_{xc}$ vs. $\rho$ in log scale. The yellow bar plot shows the distribution of points. (c) Plot of r-VWN model residual vs. $\rho$ in linear scale.}
\label{fig:DensityvsEnergyDensity}
\end{figure*}

The domain and range of the electron density and corresponding XC energy density span over 12 orders of magnitude, as seen in Fig. \ref{fig:DensityvsEnergyDensity}. This creates a substantial numerical challenge for machine-learning models since most implementations rely on double precision floats with a machine epsilon of $\sim 10^{-16}$. Practically, the situation is somewhat better, since the vast majority of the distribution (99.3\%) falls between $10^{-6.5} - 10^{0}$ (Fig. \ref{fig:DensityvsEnergyDensity}). However, even relatively small errors in the high-density region can have a substantial impact on the system-level energy, and training a machine-learning model that is accurate across this span is challenging. Nonetheless, physical models are known to approximate the energy density across this large domain/range. The VWN parameterization of the LDA model achieves this by using an analytical function that reproduces the behavior of the homogeneous electron gas (HEG)\cite{VoskoS.H1980Asel}:
\begin{equation}
\begin{split}
e_{XC,VWN} &= e_{X,VWN} + e_{C,VWN} \\
&=\rho \frac{C_1}{r_s} + \rho G(r_s,\gamma,\alpha_1,\beta_1,\beta_2,\beta_3,\beta_4)
\end{split}
\label{eqn:VWNFormalism}
\end{equation}
where $C_1, \gamma, \alpha_1, \beta_1, \beta_2, \beta_3, \beta_4$ are the parameters, $r_s$ is the Wigner–Seitz radius defined as:
\begin{equation}
r_s = (3/4 \pi \rho)^{1/3}
\label{eqn:rsDefinition}
\end{equation}
$G$ is defined as:
\begin{equation}
\begin{split}
G(r_s,\gamma,\alpha_1,\beta_1,\beta_2,\beta_3,\beta_4) = -2\gamma(1+\alpha_1r_s)\\
\times\ln\left\{1+\frac{1}{2\gamma r_s^{1/2}(\beta_1+r_s^{1/2}(\beta_2+r_s^{1/2}(\beta_3+r_s^{1/2}\beta_4)))}\right\}
\end{split}
\label{eqn:GDefinition}
\end{equation}

The behavior of the HEG is qualitatively similar to the B3LYP training data, but there are quantitative discrepancies. These residuals were minimized using non-linear optimization with the VWN parameters as initial guesses (see Sec. \ref{sec:methods}) providing an improved LDA approximation (r-VWN) to B3LYP for molecular systems of interest. The refitted model is illustrated in Fig. \ref{fig:DensityvsEnergyDensity}b, and the magnitude of the residuals is reduced by an order of magnitude as compared to the original energy density (Fig. \ref{fig:DensityvsEnergyDensity}a,c).

\begin{figure*}
	\centering
	\begin{subfigure}{0.5\textwidth}
		\centering
		\includegraphics[width=\linewidth]{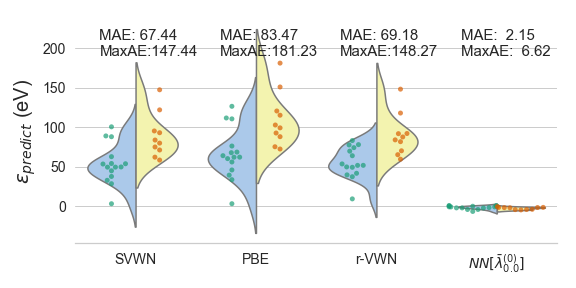}
        \caption{Energy prediction error distribution}
        \label{fig:LDAEnerPredErrorDist}
	\end{subfigure}%
	\begin{subfigure}{0.5\textwidth}
		\centering
		\includegraphics[width=\linewidth]{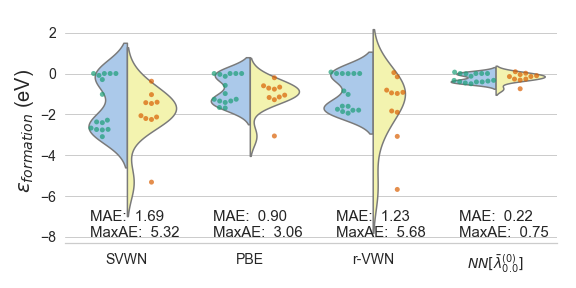}
        \caption{Formation energy prediction error distribution}
        \label{fig:LDAFormEnerPredErrorDist}
	\end{subfigure} 
    \caption{Results for local density based models. $a)$ $b)$ Error distributions for system-level energy prediction error and formation energy prediction error, respectively. Blue points/curves correspond to the training molecules, orange points/yellow curves correspond to the test and extrapolation molecules. $MAE$ denotes the mean absolute error of all systems, and $MaxAE$ denotes the maximum of absolute error }
	\label{fig:LDAResults1}
\end{figure*}

The r-VWN model is applied to the test and training sets, and the results are compared to the VWN and common PBE GGA functionals in Fig. \ref{fig:LDAResults1}. The results show that the energy prediction error stays approximately constant ($MAE_{VWN}$ = 67.44 eV, $MAE_{r-VWN}$ = 69.18 eV). Interestingly, both VWN and r-VWN have lower energy errors than PBE, and the ``test'' set has higher errors on average for both VWN and PBE despite the fact that they are not trained on the training set (i.e. the test set has inherently larger system-level errors). The trend between VWN and r-VWN is similar for the formation energy errors, where both systematically underestimate the B3LYP formation energy, and the r-VWN model has less systematic error ($MAE_{VWN}$ = 1.69 eV, $MAE_{r-VWN}$ = 1.23 eV). The magnitude of formation energy errors are also $\sim$2 orders of magnitude smaller than the system-level errors, due to cancellation of error. The formation energy errors for PBE are somewhat lower than even the r-VWN model, opposite of the trend for system-level energies, indicating that PBE relies more on cancellation of error between systems than the LDA models. 

The residual learning framework is also applied to the LDA model space ($NN[\bar{\lambda}_{(0.00)}^{(0)}]$) to provide a control for the convolutional descriptor models. Although the $r-VWN$ model doesn't show significant improvement in accuracy, it provides a good approximation to B3LYP XC energy density, and learning the residual is easier than learning the energy density directly due to a reduction in the range of the dependent variable. The results, also shown in Fig. \ref{fig:LDAResults1}, show a significant improvement in the system-level energy and a less dramatic but still significant improvement in the formation energy. Interestingly the formation energies from the NN model are more accurate than the GGA results, despite the fact that the GGA model space contains more information. The cancellation of error indicates that the NN model is not over-fit, and the improved performance indicates that there is room for improvement of LDA models by increasing flexibility, consistent with prior studies \cite{Tozer_1996}. 
However, as shown in Figure \ref{fig:EngDensVsElecDensLogLog}, the local energy residuals for the $r-VWN$ model shows three tails, which are attributed to the core regions of C, O, and N. In both cases the main error arises from the fact that the local energy is not a single-valued function of the local density, especially in the core regions. This suggests that the limiting factor to further improve accuracy is not the flexibility of the XC model, but the information contained in the model space. This motivates the inclusion of additional descriptors (Sec. \ref{sec:MCSHdesc}).

\begin{figure}
	\centering
	\includegraphics[width=\linewidth]{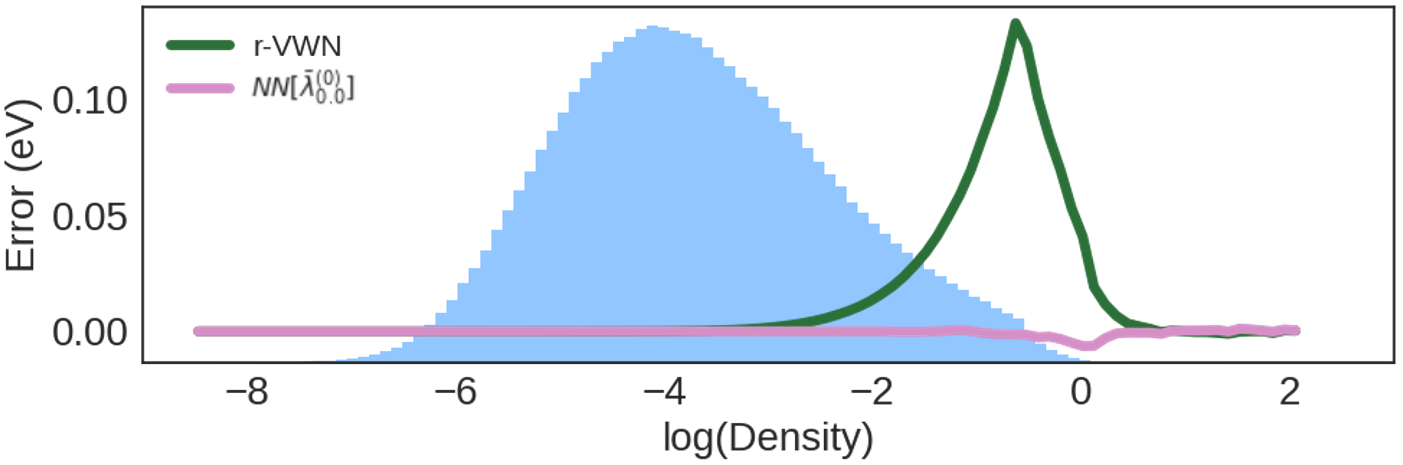}
    \caption{Plot of the sum of error for the models across different density scales for the same randomly sampled data points.}
	\label{fig:LDASumErrorDist}
\end{figure}

To get more insight into the origins of the errors for the $r-VWN$ and $NN[\bar{\lambda}_{(0.00)}^{(0)}]$ functionals we examine the contribution to the system-level error as a function of density. From Figure \ref{fig:DensityvsEnergyDensity} it is clear that the domain and range of both the electron and energy density span many orders of magnitude, and that there is a wide range in the number of points that occur at different electron densities, with the vast majority falling between $10^{-6.5} - 10^0$. The system-level energy is ultimately computed by integrating (approximated by summation) the local energy density, hence the system-level error will depend on a trade-off between the size of the error at a given density and the number of points with that density. This is illustrated in Fig. \ref{fig:LDASumErrorDist}, where the contribution to the system-level error is plotted as a function of the electron density. The results indicate that for the r-VWN model nearly all of the system-level error occurs in the density region of $10^{-3} - 10^{1}$ eV/\AA{}$^3$, corresponding to the valence/bonding regions of the molecular system. This is intuitive, since this is where chemical bonding occurs, and where there are an appreciable number of data points (29.5\%), the energy density is relatively large ($10^{-4} - 10^{1} eV/$\AA{}$^3$), and is multi-valued (see Fig. \ref{fig:EngDensVsElecDensLogLog}). In comparison, the error from the NN[LDA] model is much smaller, and concentrated in the region of $10^{-1} - 10^{1} eV/$\AA{}$^3$. This is attributed to the near-core regions where the energy density is relatively large and multi-valued, making it impossible for the neural network to capture the behavior.

This trade-off between the error and the number of sample points highlights the importance of the inclusion of randomly sampled data in the sub-sampling routine, and the selection of a proper choice of objective function during training. The randomly sampled data effectively weights the error at each density by the number of points similar to the weight that will be used to compute the system-level energy. Enough randomly sampled data must be included to ensure that regions with low electron/energy density contribute to the objective function, but if too much randomly-sampled data is included it will overwhelm the contribution from the tails of the distribution. The ratio of random to uniform sub-samples is chosen heuristically to minimize the system level error in this work. A multi-step training process is also employed, with multiple types of objective function. The details of the training procedure can be found in the Supplementary Information.
Ultimately, the results of the $r-VWN$ and $NN[\bar{\lambda}_{(0.00)}^{(0)}]$ models indicate that information beyond the local electron density must be included in the model space to significantly improve accuracy.

\subsubsection{NN models with MCSH descriptors}
\label{sec:MCSHdesc}

To capture more non-local information about the electronic environments MCSH descriptors are used, and NN models are applied to connect the descriptors to the energy density. For each set a NN model with 2 hidden layers of 100 nodes each and ReLU activation functions is used as the non-linear model for the XC energy density, denoted as NN[\MCSHDescr] where $R$ denotes the cutoff radius and $n$ denotes the order of the MCSHs used (see Eq. \ref{eq:MCSHDescriptorSetDefinition}). Each NN model is trained using a consistent training procedure, as described in the Supplementary Information. The hypothesis that including more semi-local information in the model space will systematically improve the model accuracy is tested by comparing system-level sum of local absolute error ($\varepsilon_{absolute}$), energy prediction error ($\varepsilon_{predict}$) and formation energy prediction error ($\varepsilon_{formation}$) as defined in Section \ref{sec:methods} with a consistent NN architecture. Systematic improvement is defined as improvement for each individual system without exception, while general improvement is defined as a decrease in the mean absolute error. 

\begin{figure*}
	\centering
	\begin{subfigure}{0.95\textwidth}
		\centering
		\includegraphics[width=\linewidth]{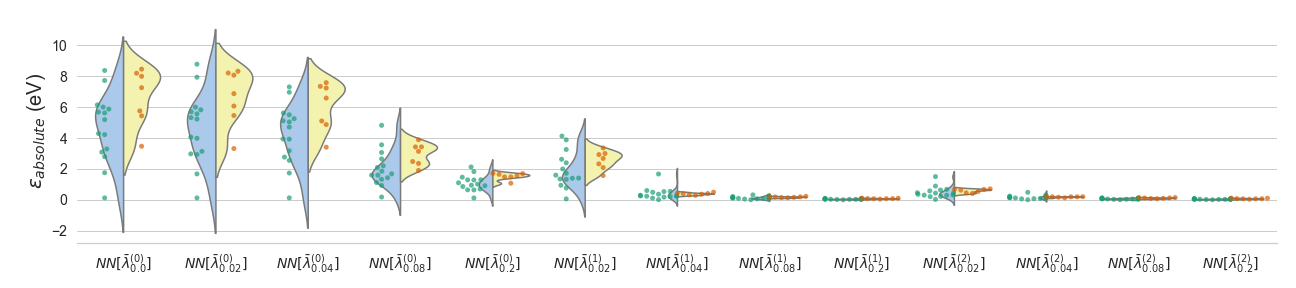}
        \caption{Sum of absolute energy prediction error distribution}
        \label{fig:MCSHEnerAbsErrorDist}
	\end{subfigure}
	\begin{subfigure}{0.95\textwidth}
		\centering
		\includegraphics[width=\linewidth]{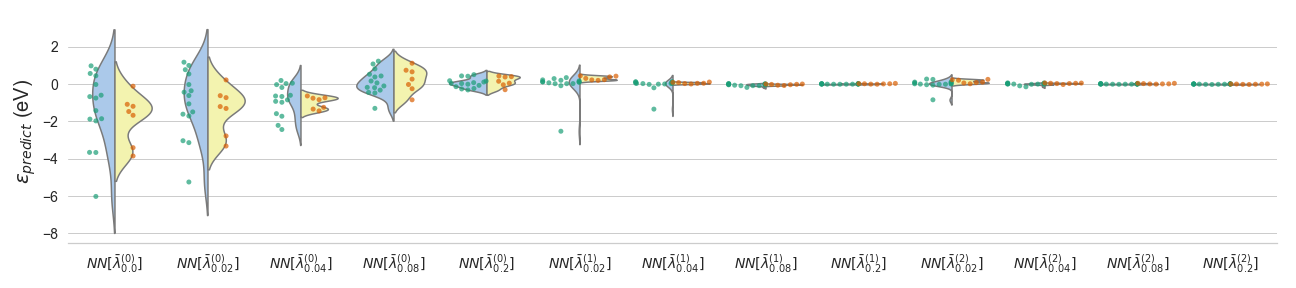}
        \caption{Energy prediction error distribution}
        \label{fig:MCSHEnerPredErrorDist}
	\end{subfigure}
    \begin{subfigure}{0.95\textwidth}
		\centering
		\includegraphics[width=\linewidth]{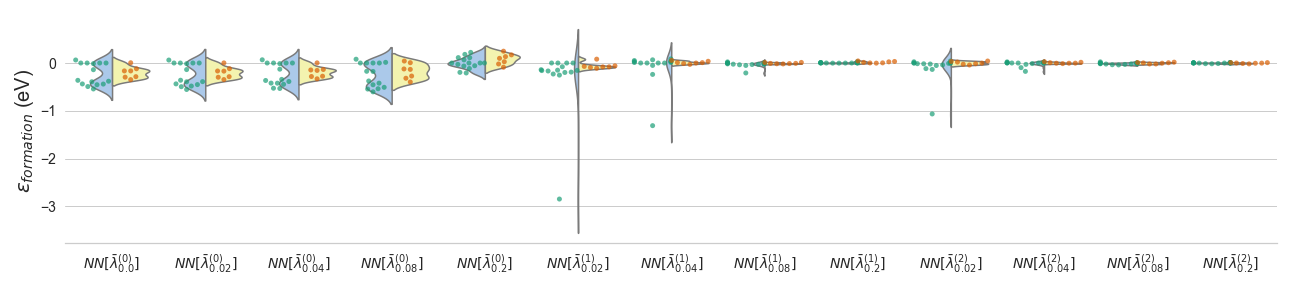}
        \caption{Formation energy prediction error distribution}
        \label{fig:MCSHFormEnerPredErrorDist}
	\end{subfigure}
	\begin{subfigure}{0.95\textwidth}
		\centering
		\includegraphics[width=\linewidth]{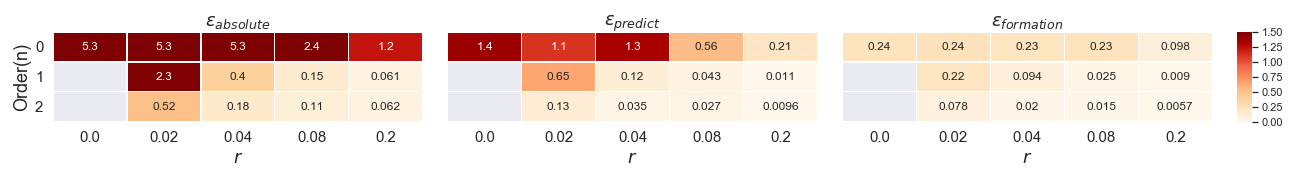}
        \caption{Mean absolute error (MAE) of the error metrics with all models}
        \label{fig:MCSHErrorMAE}
	\end{subfigure}
    \caption{Results for MCSH descriptor based models. $a)$ $b)$ $c)$ Error distributions for sum of absolute error, system-level energy prediction error and formation energy prediction error. Blue points/curves correspond to the 15 training molecular systems, orange points/yellow curves correspond to the 7 test molecular systems. $d)$  Statistical analysis of the errors of the 15 training and 7 test molecular systems. The 3 exreapolation systems are not considered here.}
	\label{fig:MCSHResults}
\end{figure*}

The results for general improvement are shown in Figure \ref{fig:MCSHResults}, indicating that the general accuracy always improves as more descriptors are added. The detailed result for each model and system are given in the Supporting Information. In this section we focus on the models' performances for the 15 training and 7 test systems; the 3 extrapolation systems are described in Sec. \ref{sec:outliersAnalysis}. Based on Figure \ref{fig:MCSHErrorMAE}, general improvement in the model accuracy is observed as angular features from zero-th order to first order are included and the range is increased from 0 to 0.2 \AA{}. The inclusion of the first-order angular features has a drastic impact, where the accuracy of the first-order model with a range of 0.02 \AA{} is comparable to the zeroth-order model with a range of 0.08 \AA{}. The first-order angular feature is needed to express the reduced gradient, and the grid spacing is 0.02 \AA{}, so the NN[$\bar{\lambda}^{(1)}_{0.02}$] model is analogous to the GGA model space. A further and substantial improvement is observed as the range is increased, with a minimum MAE of 0.061 eV achieved at a range of 0.2 \AA{}. The inclusion of descriptors of second-order angular features further improves the model, particularly at short ranges, but the improvement is less drastic. 

\begin{figure*}
	\centering
	\begin{subfigure}{0.4\textwidth}
		\centering
		\includegraphics[width=\linewidth]{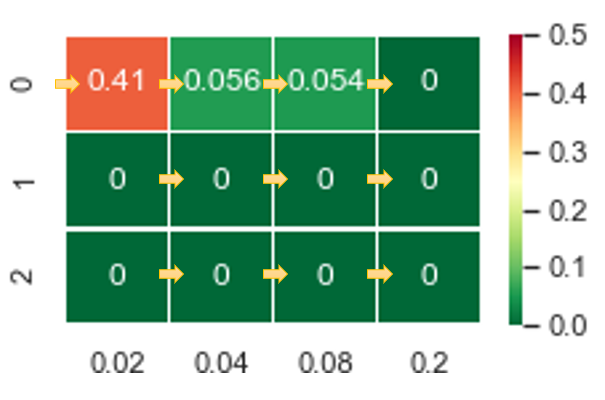}
        \caption{Range}
        \label{fig:MCSHEnerAbsErrorDist}
	\end{subfigure}
	\begin{subfigure}{0.4\textwidth}
		\centering
		\includegraphics[width=\linewidth]{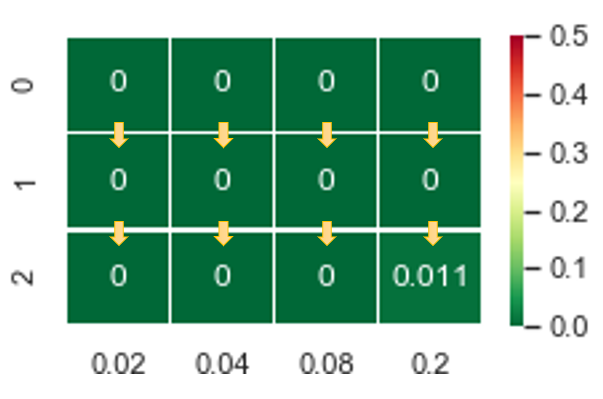}
        \caption{Angular feature}
        \label{fig:MCSHEnerPredErrorDist}
	\end{subfigure}
    \caption{Systematic improvability test. The numbers denote the maximum deviation from systematic improvement for sum of absolute error as compared to previous models. A value of 0 indicates that the model improves systematically since no system gets worse. a) each model is compared to all other models with same order of angular features and shorter range as indicated by the arrows  (e.g. $NN[\bar{\lambda}_{(0.04)}^{(0)}]$ model is compared with $NN[\bar{\lambda}_{(0.02)}^{(0)}]$ and $NN[\bar{\lambda}_{(0.00)}^{(0)}]$)  b) each model is compared to all other models with lower order of angular features and same range as indicated by the arrows (e.g. $NN[\bar{\lambda}_{(0.04)}^{(2)}]$ model is compared with $NN[\bar{\lambda}_{(0.04)}^{(1)}]$ and $NN[\bar{\lambda}_{(0.04)}^{(0)}]$).}
	\label{fig:ImprovabilityTest}
\end{figure*}

The results for the systematic improvability test for the sum of absolute error are shown in Figure \ref{fig:ImprovabilityTest}, where the number represents the maximum increase in error for any given system when the order of the angular feature or spatial range is increased; a value of 0 represents a systematic improvement since the error of every system is decreased without exception. The results show that systematic improvement is often, but not always, observed when additional descriptors are added. In particular, systematic improvability is not observed for zeroth-order angular features as range is increased from 0 to 0.08 \AA{}. This is hypothesized to occur because the added descriptors contain relatively little additional information, causing statistical noise to play a larger role in training. The randomly-selected training data represents only 0.55\% of the total data, and neural networks are initialized with random weights, causing the resulting models to favor some electronic environments over others due to randomness. This could possibly be overcome by using a static training set and a systematic strategy for initializing the neural networks, or by adding descriptors with more information. The latter strategy is shown to work here, as systematic improvability is achieved when higher-order angular descriptors and/or longer-range information are included. One exception to this trend is observed when moving from descriptors of first-order angular features to that of second-order angular features at a range of 0.2 \AA{}. This is attributed to the fact that the dimensionality of the descriptor space increases substantially from 21-41, but the flexibility of the NN model is not increased. The points will be more separated in a higher-dimensional space, as supported by the fact that the number of uniformly-sampled data from the $\bar{\lambda}_{(0.2)}^{(n)}$ sets are an order of magnitude higher than for the $\bar{\lambda}_{(\leq 0.08)}^{(n)}$ descriptors (see Supplementary Information). This forces the model to interpolate over larger distances, and will generally require a more complex NN model. These results provide evidence supporting the hypothesis that systematic XC model improvement can be achieved by systematically increasing the range and rotational order of convolutional descriptors, though optimization of the regression model architecture and training procedure is an important consideration.

Systematic and general improvement of the absolute error is promising, but the physical quantities of energy prediction error and formation energy prediction error are of practical interest. The absolute error provides an upper bound for these quantities, which can take advantage of cancellation of error within a single system (energy prediction error) and across systems (formation energy prediction error). Indeed, the general accuracy of these quantities is greatly improved as compared to the absolute error as shown in Figure \ref{fig:MCSHResults}. The MAE of the prediction error reaches ``chemical accuracy'' (0.043 eV) at a spatial range of 0.08 \AA{} for the first-order rotational descriptors, and 0.04 \AA{} with second-order rotational descriptors. A longer range is needed to reduce the maximum error to below chemical accuracy, but this can be achieved with both first- and second-order rotational descriptors at a range of 0.2 \AA{}. While this general decrease in error is promising, it should be noted that the systematic improvement is not observed at the prediction energy level. This arises due to the fact that cancellation of error plays a varying role in different chemical systems depending on the frequency with which different electronic environments occur. This variation in cancellation of error will also occur with other types of XC functionals, and explains the tremendous difficulty of achieving systematic improvement in the field of XC functional design. The counter-intuitive nature of cancellation of error is even more apparent when comparing prediction energy errors and formation energy errors. Formation energies generally benefit from cancellation of error across different systems, particularly in the core regions since the atomic composition of a molecule is utilized to compute the formation energy. This is apparent in the general improvement of between the prediction energy and the formation energy. However, when examining systematic improvement it is clear that some systems exhibit drastically larger formation-energy errors than prediction errors. This arises due to the anti-cancellation of error between a reference system and the system of interest, and highlights an additional consideration for the design of functionals with systematic improvements in formation energies. Nonetheless, several models ($\bar{\lambda}_{(0.2)}^{(1)}$, $\bar{\lambda}_{(0.08)}^{(2)}$, $\bar{\lambda}_{(0.2)}^{(2)}$) are capable of reducing even the maximum formation-energy error of the convolutional descriptor models to within chemical accuracy.

\subsubsection{Outliers and Extrapolation}
\label{sec:outliersAnalysis}

The results discussed in Sec. \ref{sec:MCSHdesc} are generally positive, although there is a noticeable outlier in the training set for some models, and the ``universality'' of the model is not clear since the test set contains similar chemistry to the training set. In this section we examine the outlier (\ce{C_2H_6}) to gain insight into where the approach fails, and attempt to extrapolate the model to three systems with different chemistry: $CH_3NO_2$, $glycine$ and $NCCN$. These compounds contain nitro groups, amine groups, and multiple cyano groups that are not present in the train or test data and hence probe the machine-learning model's ability to generalize to new chemistries.

\begin{figure}
	\centering
    \includegraphics[width=\linewidth]{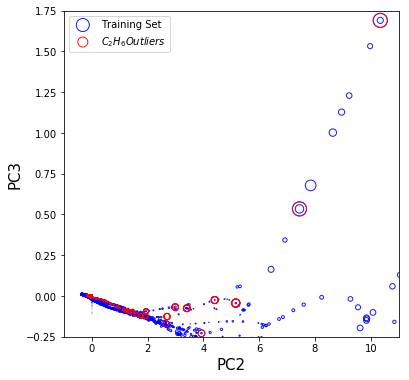}
    \caption{Comparison between the electronic environment of $C-C$ bonding region and that of the whole training set as characterized by the $\bar{\lambda}_{(0.04)}^{(1)}$ descriptor set. The training set is represented by uniformly sampled points plus 3,000,000 randomly sampled points. Principle component analysis (PCA) model is trained with $C-C$ bonding region data points and applied to both datasets. The plot of $2^{nd}$ and $3^{rd}$ principle components are shown here, where the red circles correspond to $C-C$ bonding environments and blue circles correspond to training data, the sizes of the circles correspond to the absolute prediction error of the $NN[\bar{\lambda}_{(0.04)}^{(1)}]$ model }
	\label{fig:C2H6PCAResult}
\end{figure}

First, we address the \ce{C_2H_6} system that appears as an outlier in the training set. This is generally surprising, since machine-learning models tend to perform well on data they are trained to. Notably, the $C_2H_6$ system is not a clear outlier for $\varepsilon_{absolute}$ (Figure \ref{fig:MCSHEnerAbsErrorDist}), which is directly related to the objective functions used in training (see SI), confirming that this is not a failure of the NN training procedure. However, \ce{C_2H_6} becomes an outlier in $\varepsilon_{predict}$ (Figure \ref{fig:MCSHEnerPredErrorDist}), and even more substantially in $\varepsilon_{formation}$ (Figure \ref{fig:MCSHFormEnerPredErrorDist}). This indicates that the issue arises due to a lack of cancellation of error in the prediction energy, and/or anti-cancellation of error in the formation energy. This is attributed to a combination of two factors: electronic environments that are not sufficiently distinguished by descriptors and under-representation of these electronic environments. These factors are illustrated using the $NN[\bar{\lambda}_{(0.04)}^{(1)}]$ model. Since domain decomposition method is used in model training (see Sec. \ref{sec:methods}), it is possible to determine that most of the prediction error (1.51 eV out of 1.67 eV for $\varepsilon_{absolute}$ and -1.32 eV out of -1.34 eV for $\varepsilon_{predict}$) can be attributed to the $C-C$ bonding region of the system. The electronic environments in this region were projected onto a low-dimensional space using principal component analysis (PCA) and compared to the environments in the entire training set. Figure \ref{fig:C2H6PCAResult} shows the model error as a function of two principal components, and illustrates that several points have substantially smaller errors for the training set as compared to C-C bonding region of \ce{C_2H_6}. This indicates that the objective function is multi-valued at these locations in descriptor space, forcing the model to make a tradeoff in accuracy between the two possible outcomes. This tradeoff will depend on the relative frequency of the two types of environments that are present in the training set. In this study \ce{C_2H_6} is the only molecule in the training set with a C-C single bond, causing these environments to be under-represented and not favored by the model. This could be remedied to some extent by including more examples of C-C bonds in the training set, though this would simply balance the error between systems rather than reducing it. Alternatively, the inclusion of more descriptors enables the model to distinguish between these environments and reduce the error for both; this is evident from the fact that both the prediction and formation energy errors for the \ce{C_2H_6} molecule reduce substantially as higher-order and longer-range descriptors are included (Fig. \ref{fig:MCSHResults}).

\begin{figure*}
	\centering
    \includegraphics[width=\linewidth]{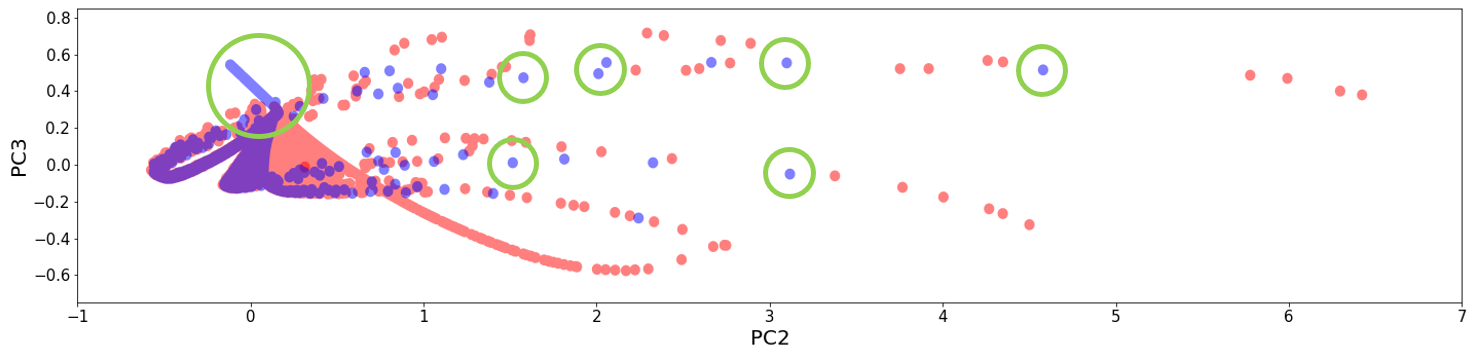}
    \caption{Comparison between the electronic environment of $CH_3NO_2$ system and that of the whole training set as characterized by the $\bar{\lambda}_{(0.04)}^{(1)}$ descriptor set. The $CH_3NO_2$ system is represented by the uniformly sampled points of the system, and the training set is represented by uniformly sampled points plus 3,000,000 randomly sampled points. Principle component analysis (PCA) model is trained with the training data points and applied to both datasets. The plot of $2^{nd}$ and $3^{rd}$ principle components are shown here, where the blue dots correspond to $CH_3NO_2$ system and red dots correspond to training data. Green circles highlight regions where the $CH_3NO_2$ system is outside the domain of the training data.}
	\label{fig:OutliersPCAResult}
\end{figure*}

The results for the extrapolation set ($CH_3NO_2$, $glycine$ and $NCCN$) are shown in Table \ref{tab:OutliersAnalysis}, and it is clear that the errors are generally larger by as much as an order of magnitude. This situation is common in machine learning, and generally arises when the training data is not representative of the test data. This occurs because the NN model can only interpolate between training examples, and will become unreliable if used for extrapolation \cite{HaleyP.J.1992Elom}. In this case the extrapolation system contains several chemical environments that are not observed in the test systems, so it is not surprising that unique electronic environments arise. This is quantitatively illustrated in Fig. \ref{fig:OutliersPCAResult} in the case of the $NN[\bar{\lambda}_{(0.04)}^{(1)}]$ model for \ce{CH_3NO_2}, where it is clear that there are a large number of electronic environments that fall outside the domain of the training data. This is found to be consistent across the other extrapolation systems and in higher dimensions, as provided in the Supplementary Information. One straightforward solution to this issue is to add additional training systems that capture the chemistries of interest. This highlights the general limitation of machine-learning models that they are only as good as the data they are trained on. However, the problem can also be mitigated by increasing the dimensionality of the descriptor space. This is evident from the model performance, where the prediction error for these outlier systems is reduced to 0.08 eV, 0.19 eV and 0.12 eV, respectively, for the $NN[\bar{\lambda}_{(0.2)}^{(2)}]$ model. This occurs because as more information about the rotational and radial variations of the electron density is added these outlier systems become more similar to environments that exist in the training data. In these higher-dimensional spaces the new systems appear more like interpolations between existing environments as opposed to extrapolations beyond the domain of all training data. This phenomenon suggests that sufficiently large convolutional descriptor spaces, combined with diverse training data sets and comprehensive testing, may enable the construction of universal machine-learning XC functionals.

\begin{table}[]
\begin{tabular}{C{1.7cm}C{1.1cm}C{1.4cm}C{1.2cm}C{1.15cm}C{0.9cm}}
\toprule
Model & Error & $CH_3NO_2$ & $glycine$ & $NCCN$ & MAE         \\ \midrule
$NN[\bar \lambda^{(1)}_{0.04}]$ & $\varepsilon_{abs.}$ & 2.01 & 0.95 & 0.94 & 0.4 \\
 & $\varepsilon_{pred.}$ & 1.38 & 0.28 & -0.21 & 0.12 \\
 & $\varepsilon_{form.}$ & 1.27 & 0.19 & -0.22 & 0.09 \\
 & & & & \\
$NN[\bar \lambda^{(1)}_{0.08}]$ & $\varepsilon_{abs.}$ & 1.24 & 0.76 & 0.49 & 0.15 \\
 & $\varepsilon_{pred.}$ & 0.18 & 0.11 & 0.24 & 0.04 \\
 & $\varepsilon_{form.}$ & 0.24 & 0.16 & 0.22 & 0.02 \\
 & & & & \\
$NN[\bar \lambda^{(1)}_{0.2}]$ & $\varepsilon_{abs.}$ & 0.24 & 0.55 & 0.13 & 0.06 \\
 & $\varepsilon_{pred.}$ & 0.08 & 0.37 & 0.03 & 0.01 \\
 & $\varepsilon_{form.}$ & 0.08 & 0.38 & 0.02 & 0.01 \\
 & & & & \\
$NN[\bar \lambda^{(2)}_{0.04}]$ & $\varepsilon_{abs.}$ & 3.66 & 0.56 & 0.49 & 0.18 \\
 & $\varepsilon_{pred.}$ & -3.14 & 0.12 & 0.12 & 0.03 \\
 & $\varepsilon_{form.}$ & -3.19 & 0.06 & 0.12 & 0.02 \\
 & & & & \\
$NN[\bar \lambda^{(2)}_{0.08}]$ & $\varepsilon_{abs.}$ & 1.42 & 0.61 & 0.14 & 0.09 \\
 & $\varepsilon_{pred.}$ & 1.08 & 0.3 & 0.02 & 0.01 \\
 & $\varepsilon_{form.}$ & 1.05 & 0.26 & -0.02 & 0.01 \\
 & & & & \\
$NN[\bar \lambda^{(2)}_{0.2}]$ & $\varepsilon_{abs.}$ & 0.35 & 0.7 & 0.12 & 0.06 \\
 & $\varepsilon_{pred.}$ & 0.08 & 0.19 & -0.01 & 0.01 \\
 & $\varepsilon_{form.}$ & 0.09 & 0.19 & -0.01 & 0.01 \\
 \bottomrule
\end{tabular}
\caption{Errors of the outlier systems in the test set. The unit is eV. MAE is the mean absolute error of the corresponding error metric for the 15 training molecular systems plus 7 test systems}
\label{tab:OutliersAnalysis}
\end{table}

\section{Conclusions}

This work introduces convolutional descriptors as a promising new paradigm for the construction of model spaces for XC functionals. Convolutional descriptors provide a systematically expandable and theoretically complete feature space for constructing XC functionals in a finite difference representation. They are orbital-free and can be computed with $N\log(N)$ computational complexity. Furthermore, convolutional descriptors can be combined with non-linear regression models to construct machine-learning functionals. Using neural networks is particularly promising, since the universal approximation theorem ensures that NNs can represent an arbitrarily complex function. The resulting models are conceptually similar to convNets, suggesting that deep learning approaches are a promising route for functional development. A sub-class of convolutional descriptors, Maxwell-Cartesian spherical harmonics (MCSH) descriptors, were employed to construct and test a range of machine-learned orbital-free approximations to the hybrid B3LYP functional based on data from a total of 25 small-molecule systems containing C, H, O, and N. These descriptors provide a numerically stable and rotationally invariant route to capturing rotational and radial variations in the electron density. The machine-learning models are constructed from model spaces based on the descriptors with increasing range from 0.02 \AA{} - 0.2 \AA{} and progressively finer angular features from zero-order to second-order. The results show that the average accuracy of the models improves as either the range or rotation symmetry is increased. A systematic improvement in the absolute error is typically observed for both training and test sets, but the improvement in system-level energy and formation energy are not systematic due to cancellation of error.

In addition to these promising initial results, this work also identifies several challenges that must be addressed in the construction of XC functionals based on convolutional descriptors and/or machine learning. One challenge that is general to any approach that utilizes localized XC energy density is the ability to generate training data. Machine-learning approaches are most powerful when they are based on data from high-level methods for which no analytical form exists; however, these approaches are typically based on non-local integrals, so projecting the XC energy density to a finite difference grid is challenging. Approaches for this have been reported \cite{Zhao1994Fedt, Burke_1998, CLB98}, but implementations are not openly available. Another related challenge is the fact that these high-level methods are typically all-electron, resulting in rapidly varying electron/energy density near the core regions. Accurately representing this with an finite difference grid requires very fine grid spacings (0.02 \AA{} in this work). In this case although the theoretical scaling of convolutions is $N\log(N)$, the size of $N$ is so large that the approach is much slower than the underlying B3LPY calculation. Similar concerns will be faced for other models seeking to reproduce the results of wavefunction-based theories, and routes to extract the XC contribution of valence electrons will be critical to training models that are consistent with pseudopotentials commonly used in practical DFT calculations. Finally, the optimization and numerical performance of machine-learning models must be considered. In this work NN models were used, leading to challenges in deconvoluting the error due to an insufficient model space and error due to sub-optimal hyperparameters or training procedures. The machine-learning models must also achieve a high accuracy over a large numerical range due to the large number of points with relatively low energy/electron density, and the quantities of interest (e.g. formation energy) rely on cancellation of error and may require specialized objective functions. This may present challenges in the application of out-of-the-box machine-learning models to the problem of XC functionals.

This work indicates that the combination of convolutional descriptors and machine learning models is a theoretically appealing framework for XC functional design. Despite practical challenges, the framework provides a route to empirically investigate fundamental questions about the nature of the XC energy. For example, this work provides empirical evidence that the exact exchange contribution of the B3LYP functional can be represented to within chemical accuracy (of system-level energies) for an orbital-free functional with a spatial range of $<$0.2 \AA{} for small C, H, O, N molecules. Further examination of these numerical approximations may provide inspiration for new physical or empirical XC models with improved accuracy and practicality. In addition, there are many possible routes to improvement of the accuracy of these machine learning models. The choice of convolutional descriptors could be improved by inclusion of higher-order spherical harmonics, longer radial distances, decreasing grid-point space, integration with pseudopotentials, or the use of deep-learning convNets to automatically extract the optimal convolutional descriptors from the data. Training data can be extracted from high-level wavefunction theories, and convolutional models can be easily implemented in solid-state codes, providing a route to data-driven wavefunction embedding \cite{ManbyFrederickR2012ASED, LibischFlorian2014Ecws, HuangChen2011Qmet, GovindN.1998Aaie}. Moreover, the MCSH descriptor sets introduced in this study are not specific to electron density, but could be applied generally to any 3D functions that are inherently rotationally invariant. This includes many problems in physics since rotational and translational invariance are common. These exciting possibilities suggest that further research into convolutional-based machine-learning functionals is a worthwhile addition to the already numerous strategies for density functional design.

\section{Acknowledgements}

We acknowledge Daniel G. A. Smith and David Sherrill for assistance in extracting grid-resolved B3LYP XC densities, and Polo Chau and Fred Hohman for assistance in visualizing electron densities and descriptors. This material is based upon work supported by the U.S. Department of Energy, Office of Science, Office of Basic Energy Sciences Computational Chemical Sciences program under Award Number DE-SC0019410. The authors are also grateful for a GPU generously provided by the NVIDIA GPU Grant program that was used to train the neural networks. 
\bibliographystyle{unsrtnat}
\bibliography{main}

\begin{thebibliography}{96}
\providecommand{\natexlab}[1]{#1}
\providecommand{\url}[1]{\texttt{#1}}
\expandafter\ifx\csname urlstyle\endcsname\relax
  \providecommand{\doi}[1]{doi: #1}\else
  \providecommand{\doi}{doi: \begingroup \urlstyle{rm}\Url}\fi

\bibitem[Hohenberg and Kohn(1964)]{PhysRev.136.B864}
P.~Hohenberg and W.~Kohn.
\newblock Inhomogeneous electron gas.
\newblock \emph{Phys. Rev.}, 136:\penalty0 B864--B871, Nov 1964.
\newblock \doi{10.1103/PhysRev.136.B864}.

\bibitem[Kohn and Sham(1965)]{PhysRev.140.A1133}
W.~Kohn and L.~J. Sham.
\newblock Self-consistent equations including exchange and correlation effects.
\newblock \emph{Phys. Rev.}, 140:\penalty0 A1133--A1138, Nov 1965.
\newblock \doi{10.1103/PhysRev.140.A1133}.

\bibitem[Perdew(2001)]{Perdew_2001}
John~P. Perdew.
\newblock Jacob's ladder of density functional approximations for the
  exchange-correlation energy.
\newblock In \emph{{AIP} Conference Proceedings}. {AIP}, 2001.
\newblock \doi{10.1063/1.1390175}.

\bibitem[Becke(1988{\natexlab{a}})]{BeckeA.D.1988Ceoa}
A.~D. Becke.
\newblock Correlation energy of an inhomogeneous electron gas: A coordinate
  space model.
\newblock \emph{The Journal of Chemical Physics}, 88\penalty0 (2):\penalty0
  1053--1062, January 1988{\natexlab{a}}.
\newblock ISSN 0021-9606.

\bibitem[Perdew and Yue(1986)]{PerdewJohnP.1986Aasd}
John~P. Perdew and Wang Yue.
\newblock Accurate and simple density functional for the electronic exchange
  energy: Generalized gradient approximation.
\newblock \emph{Physical Review B}, 33\penalty0 (12):\penalty0 8800--8802, June
  1986.
\newblock ISSN 0163-1829.

\bibitem[Perdew and Yue(1989)]{Perdew1989EAas}
Perdew and Yue.
\newblock Erratum: Accurate and simple density functional for the electronic
  exchange energy: Generalized gradient approximation.
\newblock \emph{Physical review. B, Condensed matter}, 40\penalty0 (5), August
  1989.
\newblock ISSN 0163-1829.

\bibitem[Medvedev et~al.(2017)Medvedev, Bushmarinov, Sun, Perdew, and
  Lyssenko]{MedvedevMichaelG2017RtCo}
Michael~G. Medvedev, Ivan~S. Bushmarinov, Jianwei Sun, John~P. Perdew, and
  Konstantin~A. Lyssenko.
\newblock Density functional theory is straying from the path toward the exact
  functional.
\newblock \emph{Science}, 355\penalty0 (6320):\penalty0 49--52, jan 2017.
\newblock \doi{10.1126/science.aah5975}.

\bibitem[Becke(1988{\natexlab{b}})]{BeckeA.D.1988Deaw}
A.D. Becke.
\newblock Density-functional exchange-energy approximation with correct
  asymptotic behavior.
\newblock \emph{Physical Review A}, 38\penalty0 (6):\penalty0 3098--3100,
  1988{\natexlab{b}}.
\newblock ISSN 10502947.

\bibitem[Becke(1993)]{BeckeA.D.1993DtIT}
A.D. Becke.
\newblock Density-functional thermochemistry. iii. the role of exact exchange.
\newblock \emph{The Journal of Chemical Physics}, 98\penalty0 (7):\penalty0
  5648--5652, 1993.
\newblock ISSN 00219606.

\bibitem[Lee et~al.(1988)Lee, Yang, and Parr]{LeeC.1988DotC}
C.~Lee, W.~Yang, and R.G. Parr.
\newblock Development of the colle-salvetti correlation-energy formula into a
  functional of the electron density.
\newblock \emph{Physical Review B}, 37\penalty0 (2):\penalty0 785--789, 1988.
\newblock ISSN 01631829.

\bibitem[Heyd et~al.(2003)Heyd, Scuseria, and Ernzerhof]{HeydJochen2003Hfbo}
Jochen Heyd, Gustavo~E. Scuseria, and Matthias Ernzerhof.
\newblock Hybrid functionals based on a screened coulomb potential.
\newblock \emph{The Journal of Chemical Physics}, 118\penalty0 (18):\penalty0
  8207--8215, May 2003.
\newblock ISSN 0021-9606.

\bibitem[Krukau et~al.(2006)Krukau, Vydrov, Izmaylov, and
  Scuseria]{KrukauAliaksandrV.2006Iote}
Aliaksandr~V. Krukau, Oleg~A. Vydrov, Artur~F. Izmaylov, and Gustavo~E.
  Scuseria.
\newblock Influence of the exchange screening parameter on the performance of
  screened hybrid functionals.
\newblock \emph{The Journal of Chemical Physics}, 125\penalty0 (22), December
  2006.
\newblock ISSN 0021-9606.

\bibitem[Zhao and Truhlar(2006)]{ZhaoY2006Dffs}
Y~Zhao and DG~Truhlar.
\newblock Density functional for spectroscopy: No long-range self-interaction
  error, good performance for rydberg and charge-transfer states, and better
  performance on average than b3lyp for ground states.
\newblock \emph{Journal Of Physical Chemistry A}, 110\penalty0 (49):\penalty0
  13126--13130, December 2006.
\newblock ISSN 1089-5639.

\bibitem[Becke(2003)]{BeckeAxelD.2003Armo}
Axel~D. Becke.
\newblock A real-space model of nondynamical correlation.
\newblock \emph{The Journal of Chemical Physics}, 119\penalty0 (6):\penalty0
  2972--2977, August 2003.
\newblock ISSN 0021-9606.

\bibitem[Armiento and Mattsson(2005)]{Armiento_2005}
R.~Armiento and A.~E. Mattsson.
\newblock Functional designed to include surface effects in self-consistent
  density functional theory.
\newblock \emph{Physical Review B}, 72\penalty0 (8), aug 2005.
\newblock \doi{10.1103/physrevb.72.085108}.

\bibitem[de~Silva and Corminboeuf(2015)]{de_Silva_2015}
Piotr de~Silva and Cl{\'{e}}mence Corminboeuf.
\newblock Communication: A new class of non-empirical explicit density
  functionals on the third rung of jacob's ladder.
\newblock \emph{The Journal of Chemical Physics}, 143\penalty0 (11):\penalty0
  111105, sep 2015.
\newblock \doi{10.1063/1.4931628}.

\bibitem[Zhao et~al.(2005)Zhao, Lynch, and Truhlar]{ZhaoYan2005Medf}
Yan Zhao, Benjamin~J. Lynch, and Donald~G. Truhlar.
\newblock Multi-coefficient extrapolated density functional theory for
  thermochemistry and thermochemical kinetics.
\newblock \emph{Physical Chemistry Chemical Physics}, 7\penalty0 (1):\penalty0
  43--52, December 2005.
\newblock ISSN 1463-9076.

\bibitem[Angyan et~al.(2005)Angyan, Gerber, Savin, and
  Toulouse]{AngyanJanosG.2005vdWf}
Janos~G. Angyan, Iann~C. Gerber, Andreas Savin, and Julien Toulouse.
\newblock van der waals forces in density functional theory: Perturbational
  long-range electron-interaction corrections.
\newblock \emph{Physical Review. A}, 72\penalty0 (1), 2005.
\newblock ISSN 1050-2947.

\bibitem[Grimme(2006{\natexlab{a}})]{GrimmeStefan2006Shdf}
Stefan Grimme.
\newblock Semiempirical hybrid density functional with perturbative
  second-order correlation.
\newblock \emph{The Journal of Chemical Physics}, 124\penalty0 (3), January
  2006{\natexlab{a}}.
\newblock ISSN 0021-9606.

\bibitem[Wu and Yang(2002)]{WuQin2002Ectd}
Qin Wu and Weitao Yang.
\newblock Empirical correction to density functional theory for van der waals
  interactions.
\newblock \emph{The Journal of Chemical Physics}, 116\penalty0 (2):\penalty0
  515--524, January 2002.
\newblock ISSN 0021-9606.

\bibitem[Grimme(2004)]{GrimmeStefan2004Adov}
Stefan Grimme.
\newblock Accurate description of van der waals complexes by density functional
  theory including empirical corrections.
\newblock \emph{Journal of Computational Chemistry}, 25\penalty0 (12):\penalty0
  1463--1473, September 2004.
\newblock ISSN 0192-8651.

\bibitem[Grimme(2006{\natexlab{b}})]{GrimmeStefan2006SGdf}
Stefan Grimme.
\newblock Semiempirical gga type density functional constructed with a long
  range dispersion correction.
\newblock \emph{Journal of Computational Chemistry}, 27\penalty0 (15):\penalty0
  1787--1799, November 2006{\natexlab{b}}.
\newblock ISSN 0192-8651.

\bibitem[Andersson et~al.(1996)Andersson, Andersson, and
  Lundqvist]{AnderssonY.1996VdWi}
Y.~Andersson, D.C. Andersson, and B.I. Lundqvist.
\newblock Van der waals interactions in density-functional theory.
\newblock \emph{Physical Review Letters}, 76\penalty0 (1):\penalty0 102--105,
  1996.
\newblock ISSN 00319007.

\bibitem[Dobson and Dinte(1996)]{DobsonJ.F.1996Csil}
J.F. Dobson and B.P. Dinte.
\newblock Constraint satisfaction in local and gradient susceptibility
  approximations: Application to a van der waals density functional.
\newblock \emph{Physical Review Letters}, 76\penalty0 (11):\penalty0
  1780--1783, 1996.
\newblock ISSN 0031-9007.

\bibitem[Dion et~al.(2004)Dion, Rydberg, Schroder, Langreth, and
  Lundqvist]{DionM2004VdWd}
M~Dion, H~Rydberg, E~Schroder, DC~Langreth, and Bi~Lundqvist.
\newblock Van der waals density functional for general geometries.
\newblock \emph{Physical Review Letters}, 92\penalty0 (24), June 2004.
\newblock ISSN 0031-9007.

\bibitem[Thonhauser et~al.(2007)Thonhauser, Cooper, Li, Puzder, Hyldgaard, and
  Langreth]{ThonhauserT.2007VdWd}
T.~Thonhauser, V.R. Cooper, S.~Li, A.~Puzder, P.~Hyldgaard, and D.C. Langreth.
\newblock Van der waals density functional: Self-consistent potential and the
  nature of the van der waals bond.
\newblock \emph{Physical Review B - Condensed Matter and Materials Physics},
  76\penalty0 (12), September 2007.
\newblock ISSN 10980121.

\bibitem[Witkin(1987)]{Witkin_1987}
Andrew~P. Witkin.
\newblock Scale-space filtering.
\newblock In \emph{Readings in Computer Vision}, pages 329--332. Elsevier,
  1987.
\newblock \doi{10.1016/b978-0-08-051581-6.50036-2}.

\bibitem[Lindeberg(1993)]{Lindeberg_1993}
Tony Lindeberg.
\newblock \emph{Scale-Space Theory in Computer Vision}.
\newblock Springer, 1993.
\newblock ISBN 9780792394181.

\bibitem[LeCun et~al.(1990)LeCun, Boser, Denker, Henderson, Howard, Hubbard,
  and Jackel]{NIPS1989_293}
Yann LeCun, Bernhard~E. Boser, John~S. Denker, Donnie Henderson, R.~E. Howard,
  Wayne~E. Hubbard, and Lawrence~D. Jackel.
\newblock Handwritten digit recognition with a back-propagation network.
\newblock In D.~S. Touretzky, editor, \emph{Advances in Neural Information
  Processing Systems 2}, pages 396--404. Morgan-Kaufmann, 1990.

\bibitem[Lecun et~al.(1998)Lecun, Bottou, Bengio, and
  Haffner]{Lecun98gradient-basedlearning}
Yann Lecun, Léon Bottou, Yoshua Bengio, and Patrick Haffner.
\newblock Gradient-based learning applied to document recognition.
\newblock In \emph{Proceedings of the IEEE}, pages 2278--2324, 1998.

\bibitem[Krizhevsky et~al.(2017)Krizhevsky, Sutskever, and
  Hinton]{KrizhevskySutskeverHinton17cacm}
Alex Krizhevsky, Ilya Sutskever, and Geoffrey~E. Hinton.
\newblock Imagenet classification with deep convolutional neural networks.
\newblock \emph{Communications of the ACM}, 60\penalty0 (6):\penalty0 84--90,
  2017.
\newblock ISSN 0001-0782.
\newblock \doi{10.1145/3065386}.

\bibitem[Simard et~al.(2003)Simard, Steinkraus, and Platt]{SimardP.Y.2003Bpfc}
P.Y. Simard, D.~Steinkraus, and J.C. Platt.
\newblock Best practices for convolutional neural networks applied to visual
  document analysis.
\newblock volume 2003-, pages 958--963, USA, 2003. IEEE.
\newblock ISBN 0769519601.

\bibitem[Vaillant et~al.(1994)Vaillant, Monrocq, and Cun]{Vaillant94anoriginal}
Régis Vaillant, Christophe Monrocq, and Yann~Le Cun.
\newblock An original approach for the localization of objects in images, 1994.

\bibitem[Nowlan and Platt(1995)]{Nowlan95aconvolutional}
Steven~J. Nowlan and John~C. Platt.
\newblock A convolutional neural network hand tracker.
\newblock In \emph{Advances in Neural Information Processing Systems 7}, pages
  901--908. Morgan Kaufmann, 1995.

\bibitem[Lawrence et~al.(1997)Lawrence, Giles, Tsoi, and
  Back]{LawrenceS.1997Frac}
S.~Lawrence, C.L. Giles, Ah~Chung Tsoi, and A.D. Back.
\newblock Face recognition: a convolutional neural-network approach.
\newblock \emph{Neural Networks, IEEE Transactions on}, 8\penalty0
  (1):\penalty0 98--113, January 1997.
\newblock ISSN 1045-9227.

\bibitem[Perdew et~al.(1996)Perdew, Burke, and Ernzerhof]{PerdewJ.P.1996GGAM}
J.P. Perdew, K.~Burke, and M.~Ernzerhof.
\newblock Generalized gradient approximation made simple.
\newblock \emph{Physical Review Letters}, 77\penalty0 (18):\penalty0
  3865--3868, 1996.
\newblock ISSN 0031-9007.

\bibitem[Perdew et~al.(1997)Perdew, Burke, and Ernzerhof]{PerdewJohnP.1997GGAM}
John~P. Perdew, Kieron Burke, and Matthias Ernzerhof.
\newblock Generalized gradient approximation made simple [phys. rev. lett. 77,
  3865 (1996)].
\newblock \emph{Physical Review Letters}, 78\penalty0 (7):\penalty0 1396--1396,
  February 1997.
\newblock ISSN 0031-9007.

\bibitem[Wellendorff et~al.(2012)Wellendorff, Lundgaard, M\o{}gelh\o{}j,
  Petzold, Landis, N\o{}rskov, Bligaard, and Jacobsen]{PhysRevB.85.235149}
Jess Wellendorff, Keld~T. Lundgaard, Andreas M\o{}gelh\o{}j, Vivien Petzold,
  David~D. Landis, Jens~K. N\o{}rskov, Thomas Bligaard, and Karsten~W.
  Jacobsen.
\newblock Density functionals for surface science: Exchange-correlation model
  development with bayesian error estimation.
\newblock \emph{Phys. Rev. B}, 85:\penalty0 235149, Jun 2012.
\newblock \doi{10.1103/PhysRevB.85.235149}.

\bibitem[Zhao and Truhlar(2008)]{ZhaoYan2008TMso}
Yan Zhao and Donald Truhlar.
\newblock The m06 suite of density functionals for main group thermochemistry,
  thermochemical kinetics, noncovalent interactions, excited states, and
  transition elements: two new functionals and systematic testing of four
  m06-class functionals and 12 other functionals.
\newblock \emph{Theoretical Chemistry Accounts}, 120\penalty0 (1):\penalty0
  215--241, May 2008.
\newblock ISSN 1432-881X.

\bibitem[Peverati and Truhlar(2011)]{PeveratiRoberto2011ItAo}
Roberto Peverati and Donald~G. Truhlar.
\newblock Improving the accuracy of hybrid meta-gga density functionals by
  range separation.
\newblock \emph{The Journal of Physical Chemistry Letters}, 2\penalty0
  (21):\penalty0 2810--2817, November 2011.
\newblock ISSN 1948-7185.

\bibitem[Peverati and Truhlar(2012)]{PeveratiRoberto2012EFwG}
Roberto Peverati and Donald~G Truhlar.
\newblock Exchange-correlation functional with good accuracy for both
  structural and energetic properties while depending only on the density and
  its gradient.
\newblock \emph{Journal of chemical theory and computation}, 8\penalty0 (7),
  July 2012.
\newblock ISSN 1549-9618.

\bibitem[Yu et~al.(2016)Yu, He, Li, and Truhlar]{YuHaoyuS.2016MAKg}
Haoyu~S. Yu, Xiao He, Shaohong~L. Li, and Donald~G. Truhlar.
\newblock Mn15: A kohnsham global-hybrid exchangecorrelation density functional
  with broad accuracy for multi-reference and single-reference systems and
  noncovalent interactions.
\newblock \emph{Chemical Science}, 7\penalty0 (8):\penalty0 5032--5051, July
  2016.
\newblock ISSN 2041-6520.

\bibitem[Snyder et~al.(2012)Snyder, Rupp, Hansen, Müller, and
  Burke]{SnyderJohnC2012Fdfw}
John~C Snyder, Matthias Rupp, Katja Hansen, Klaus-Robert Müller, and Kieron
  Burke.
\newblock Finding density functionals with machine learning.
\newblock \emph{Physical review letters}, 108\penalty0 (25), June 2012.
\newblock ISSN 1079-7114.

\bibitem[Huan et~al.(2017)Huan, Batra, Chapman, Krishnan, Chen, and
  Ramprasad]{HuanTranDoan2017Ausf}
Tran~Doan Huan, Rohit Batra, James Chapman, Sridevi Krishnan, Lihua Chen, and
  Rampi Ramprasad.
\newblock A universal strategy for the creation of machine learning-based
  atomistic force fields.
\newblock \emph{npj Computational Materials}, 3\penalty0 (1), 2017.
\newblock ISSN 2057-3960.

\bibitem[Yao et~al.(2018)Yao, Herr, Toth, Mckintyre, and
  Parkhill]{YaoKun2018TTmc}
Kun Yao, John~E Herr, DavidW Toth, Ryker Mckintyre, and John Parkhill.
\newblock The tensormol-0.1 model chemistry: a neural network augmented with
  long-range physics.
\newblock \emph{Chemical science.}, 9\penalty0 (8):\penalty0 2261--2269, 2018.
\newblock ISSN 2041-6520.

\bibitem[Rowe et~al.(2018)Rowe, Csányi, Alfè, and
  Michaelides]{RowePatrick2018Doam}
Patrick Rowe, Gábor Csányi, Dario Alfè, and Angelos Michaelides.
\newblock Development of a machine learning potential for graphene.
\newblock \emph{Physical review.}, 97\penalty0 (5), 2018.
\newblock ISSN 2469-9950.

\bibitem[Brockherde et~al.(2017)Brockherde, Vogt, Li, Tuckerman, Burke, and
  Müller]{BrockherdeFelix2017BtKe}
Felix Brockherde, Leslie Vogt, Li~Li, Mark~E Tuckerman, Kieron Burke, and
  Klaus-Robert Müller.
\newblock Bypassing the kohn-sham equations with machine learning.
\newblock \emph{Nature communications.}, 8\penalty0 (1), 2017.
\newblock ISSN 2041-1723.

\bibitem[Hegde and Bowen(2017)]{HegdeGanesh2017MatD}
Ganesh Hegde and R~Chris Bowen.
\newblock Machine-learned approximations to density functional theory
  hamiltonians.
\newblock \emph{Scientific reports.}, 7, 2017.
\newblock ISSN 2045-2322.

\bibitem[Himmetoglu(2016)]{HimmetogluBurak2016Tbml}
Burak Himmetoglu.
\newblock Tree based machine learning framework for predicting ground state
  energies of molecules.
\newblock \emph{Journal of chemical physics.}, 145\penalty0 (13), 2016.
\newblock ISSN 0021-9606.

\bibitem[Rupp(2015)]{RuppMatthias2015MLfQ}
Matthias Rupp.
\newblock Machine learning for quantum mechanics in a nutshell.
\newblock \emph{International journal of quantum chemistry.}, 115\penalty0
  (16):\penalty0 1058--1073, 2015.
\newblock ISSN 0020-7608.

\bibitem[Hansen et~al.(2015)Hansen, Biegler, Ramakrishnan, Pronobis, von
  Lilienfeld, Müller, and Tkatchenko]{HansenKatja2015MLPo}
Katja Hansen, Franziska Biegler, Raghunathan Ramakrishnan, Wiktor Pronobis,
  O~Anatole von Lilienfeld, Klaus-Robert Müller, and Alexandre Tkatchenko.
\newblock Machine learning predictions of molecular properties: Accurate
  many-body potentials and nonlocality in chemical space.
\newblock \emph{The journal of physical chemistry letters.}, 6\penalty0
  (12):\penalty0 2326--2331, 2015.
\newblock ISSN 1948-7185.

\bibitem[Yao and Parkhill(2016)]{YaoKun2016KEoH}
Kun Yao and John Parkhill.
\newblock Kinetic energy of hydrocarbons as a function of electron density and
  convolutional neural networks.
\newblock \emph{Journal of chemical theory and computation : JCTC.},
  12\penalty0 (3):\penalty0 1139--1147, 2016.
\newblock ISSN 1549-9618.

\bibitem[Mills et~al.(2017)Mills, Spanner, and Tamblyn]{MillsKyle2017Dlat}
Kyle Mills, Michael Spanner, and Isaac Tamblyn.
\newblock Deep learning and the schrödinger equation.
\newblock \emph{Physical review.}, 96\penalty0 (4), 2017.
\newblock ISSN 2469-9926.

\bibitem[Pereira et~al.(2017)Pereira, Xiao, Latino, Wu, Zhang, and Aires-de
  Sousa]{PereiraFlorbela2017MLMt}
Florbela Pereira, Kaixia Xiao, Diogo A R~S Latino, Chengcheng Wu, Qingyou
  Zhang, and Joao Aires-de Sousa.
\newblock Machine learning methods to predict density functional theory b3lyp
  energies of homo and lumo orbitals.
\newblock \emph{Journal of chemical information and modeling.}, 57\penalty0
  (1):\penalty0 11--21, 2017.
\newblock ISSN 1549-9596.

\bibitem[Bartók et~al.(2013)Bartók, Gillan, Manby, and
  Csányi]{BartokAlbertP2013Mafo}
Albert~P Bartók, Michael~J Gillan, Frederick~R Manby, and Gábor Csányi.
\newblock Machine-learning approach for one- and two-body corrections to
  density functional theory: Applications to molecular and condensed water.
\newblock \emph{Physical review.}, 88\penalty0 (5), 2013.
\newblock ISSN 1098-0121.

\bibitem[Seino et~al.(2018)Seino, Kageyama, Fujinami, Ikabata, and
  Nakai]{SeinoJunji2018Smke}
Junji Seino, Ryo Kageyama, Mikito Fujinami, Yasuhiro Ikabata, and Hiromi Nakai.
\newblock Semi-local machine-learned kinetic energy density functional with
  third-order gradients of electron density.
\newblock \emph{Journal of chemical physics.}, 148\penalty0 (24), 2018.
\newblock ISSN 0021-9606.

\bibitem[Gao et~al.(2016)Gao, Li, Li, Li, Fang, Li, Hu, Lu, and
  Su]{GaoTing2016Amlc}
Ting Gao, Hongzhi Li, Wenze Li, Lin Li, Chao Fang, Hui Li, LiHong Hu, Yinghua
  Lu, and Zhong-Min Su.
\newblock A machine learning correction for dft non-covalent interactions based
  on the s22, s66 and x40 benchmark databases.
\newblock \emph{Journal of cheminformatics.}, 8\penalty0 (1), 2016.
\newblock ISSN 1758-2946.

\bibitem[Liu et~al.(2017)Liu, Wang, Du, Hu, Zheng, and Chen]{LiuQin2017ItPo}
Qin Liu, JingChun Wang, PengLi Du, LiHong Hu, Xiao Zheng, and GuanHua Chen.
\newblock Improving the performance of long-range-corrected
  exchange-correlation functional with an embedded neural network.
\newblock \emph{The journal of physical chemistry.}, 121\penalty0
  (38):\penalty0 7273--7281, 2017.
\newblock ISSN 1089-5639.

\bibitem[Tozer et~al.(1996)Tozer, Ingamells, and Handy]{Tozer_1996}
David~J. Tozer, Victoria~E. Ingamells, and Nicholas~C. Handy.
\newblock Exchange-correlation potentials.
\newblock \emph{The Journal of Chemical Physics}, 105\penalty0 (20):\penalty0
  9200--9213, nov 1996.
\newblock \doi{10.1063/1.472753}.

\bibitem[Wellendorff et~al.(2014)Wellendorff, Lundgaard, Jacobsen, and
  Bligaard]{WellendorffJess2014mAas}
Jess Wellendorff, Keld~T Lundgaard, Karsten~W Jacobsen, and Thomas Bligaard.
\newblock mbeef: An accurate semi-local bayesian error estimation density
  functional.
\newblock \emph{Journal of chemical physics.}, 140\penalty0 (14), 2014.
\newblock ISSN 0021-9606.

\bibitem[Lundgaard et~al.(2016)Lundgaard, Wellendorff, Voss, Jacobsen, and
  Bligaard]{PhysRevB.93.235162}
Keld~T. Lundgaard, Jess Wellendorff, Johannes Voss, Karsten~W. Jacobsen, and
  Thomas Bligaard.
\newblock mbeef-vdw: Robust fitting of error estimation density functionals.
\newblock \emph{Phys. Rev. B}, 93:\penalty0 235162, Jun 2016.
\newblock \doi{10.1103/PhysRevB.93.235162}.

\bibitem[Aldegunde et~al.(2016)Aldegunde, Kermode, and
  Zabaras]{AldegundeManuel2016Doae}
Manuel Aldegunde, James~R Kermode, and Nicholas Zabaras.
\newblock Development of an exchange–correlation functional with uncertainty
  quantification capabilities for density functional theory.
\newblock \emph{Journal of computational physics}, 311:\penalty0 173--195,
  2016.
\newblock ISSN 0021-9991.

\bibitem[Vapnik(1995)]{VapnikVladimirNaumovich1995Tnos}
Vladimir~Naumovich Vapnik.
\newblock \emph{The nature of statistical learning theory}.
\newblock Springer, New York, 1995.
\newblock ISBN 0387945598.

\bibitem[Murphy(2012)]{MurphyKevinP.2012Mlap}
Kevin~P. Murphy.
\newblock \emph{Machine learning a probabilistic perspective}.
\newblock Adaptive computation and machine learning. MIT Press, Cambridge,
  Mass., 2012.
\newblock ISBN 9780262305242.

\bibitem[Rasmussen(2006)]{RasmussenCarlEdward2006Gpfm}
Carl~Edward Rasmussen.
\newblock \emph{Gaussian processes for machine learning}.
\newblock Adaptive computation and machine learning. MIT Press, Cambridge,
  Mass., 2006.
\newblock ISBN 9786612097966.

\bibitem[McCulloch and Pitts(1990)]{McCullochWarren1990Alco}
Warren McCulloch and Walter Pitts.
\newblock A logical calculus of the ideas immanent in nervous activity.
\newblock \emph{Bulletin of Mathematical Biology}, 52\penalty0 (1):\penalty0
  99--115, January 1990.
\newblock ISSN 0092-8240.

\bibitem[Hornik et~al.(1989{\natexlab{a}})Hornik, Stinchcombe, and
  White]{HornikKurt1989Mfna}
Kurt Hornik, Maxwell Stinchcombe, and Halbert White.
\newblock Multilayer feedforward networks are universal approximators.
\newblock \emph{Neural Networks}, 2\penalty0 (5):\penalty0 359--366,
  1989{\natexlab{a}}.
\newblock ISSN 0893-6080.

\bibitem[Hornik et~al.(1989{\natexlab{b}})Hornik, Stinchcombe, and
  White]{HornikEtAl89}
K.~Hornik, M.~Stinchcombe, and H.~White.
\newblock Multilayer feedforward networks are universal approximators.
\newblock \emph{Neural Networks}, 2\penalty0 (5):\penalty0 359--366,
  1989{\natexlab{b}}.

\bibitem[Cybenko(1989)]{citeulike:3561150}
G.~Cybenko.
\newblock {Approximation by superpositions of a sigmoidal function}.
\newblock \emph{Mathematics of Control, Signals, and Systems (MCSS)},
  2\penalty0 (4):\penalty0 303--314, 1989.
\newblock ISSN 0932-4194.
\newblock \doi{10.1007/BF02551274}.

\bibitem[Worrall et~al.(2016)Worrall, Garbin, Turmukhambetov, and
  Brostow]{WorrallDanielE.2016HNDT}
Daniel~E. Worrall, Stephan~J. Garbin, Daniyar Turmukhambetov, and Gabriel~J.
  Brostow.
\newblock Harmonic networks: Deep translation and rotation equivariance.
\newblock 2016.

\bibitem[Applequist(2002)]{ApplequistJon2002Mshi}
Jon Applequist.
\newblock Maxwell–cartesian spherical harmonics in multipole potentials and
  atomic orbitals.
\newblock \emph{Theoretical Chemistry Accounts}, 107\penalty0 (2):\penalty0
  103--115, 2002.
\newblock ISSN 1432-881X.

\bibitem[Parrish et~al.(2017)Parrish, Burns, Smith, Simmonett, DePrince,
  Hohenstein, Bozkaya, Sokolov, Di~Remigio, Richard, Gonthier, James,
  McAlexander, Kumar, Saitow, Wang, Pritchard, Verma, Schaefer, Patkowski,
  King, Valeev, Evangelista, Turney, Crawford, and
  Sherrill]{doi:10.1021/acs.jctc.7b00174}
Robert~M. Parrish, Lori~A. Burns, Daniel G.~A. Smith, Andrew~C. Simmonett,
  A.~Eugene DePrince, Edward~G. Hohenstein, Uğur Bozkaya, Alexander~Yu.
  Sokolov, Roberto Di~Remigio, Ryan~M. Richard, Jérôme~F. Gonthier, Andrew~M.
  James, Harley~R. McAlexander, Ashutosh Kumar, Masaaki Saitow, Xiao Wang,
  Benjamin~P. Pritchard, Prakash Verma, Henry~F. Schaefer, Konrad Patkowski,
  Rollin~A. King, Edward~F. Valeev, Francesco~A. Evangelista, Justin~M. Turney,
  T.~Daniel Crawford, and C.~David Sherrill.
\newblock Psi4 1.1: An open-source electronic structure program emphasizing
  automation, advanced libraries, and interoperability.
\newblock \emph{Journal of Chemical Theory and Computation}, 13\penalty0
  (7):\penalty0 3185--3197, 2017.
\newblock \doi{10.1021/acs.jctc.7b00174}.
\newblock PMID: 28489372.

\bibitem[Johnson(2011)]{CCCBDB}
Russell~D. Johnson.
\newblock {NIST} computational chemistry comparison and benchmark database,
  August 2011.

\bibitem[Jones et~al.(2001--)Jones, Oliphant, Peterson, et~al.]{Scipy}
Eric Jones, Travis Oliphant, Pearu Peterson, et~al.
\newblock {SciPy}: Open source scientific tools for {Python}, 2001--.
\newblock URL \url{http://www.scipy.org/}.
\newblock [Online; accessed Aug 03 2018].

\bibitem[Ramakrishnan et~al.(2015)Ramakrishnan, Dral, Rupp, and von
  Lilienfeld]{RamakrishnanRaghunathan2015BDMQ}
Raghunathan Ramakrishnan, Pavlo~O Dral, Matthias Rupp, and O~Anatole von
  Lilienfeld.
\newblock Big data meets quantum chemistry approximations: The $\delta$-machine
  learning approach.
\newblock \emph{Journal of chemical theory and computation : JCTC.},
  11\penalty0 (5):\penalty0 2087--2096, 2015.
\newblock ISSN 1549-9618.

\bibitem[Vosko et~al.(1980)Vosko, Wilk, and Nusair]{VoskoS.H1980Asel}
S.~H Vosko, L.~Wilk, and M.~Nusair.
\newblock Accurate spin dependent electron liquid correlation energies for
  local spin density calculations: a critical analysis.
\newblock August 1980.

\bibitem[Gao and Han(2012)]{GaoFuchang2012ItNs}
Fuchang Gao and Lixing Han.
\newblock Implementing the nelder-mead simplex algorithm withadaptive
  parameters.
\newblock \emph{Computational Optimization and Applications}, 51\penalty0
  (1):\penalty0 259--277, January 2012.
\newblock ISSN 0926-6003.

\bibitem[Kingma and Ba(2014)]{KingmaDiederikP.2014AAMf}
Diederik~P. Kingma and Jimmy Ba.
\newblock Adam: A method for stochastic optimization.
\newblock December 2014.

\bibitem[Chollet et~al.(2015)]{chollet2015keras}
Fran\c{c}ois Chollet et~al.
\newblock Keras.
\newblock \url{https://keras.io}, 2015.

\bibitem[Nair and Hinton(2010)]{Nair:2010:RLU:3104322.3104425}
Vinod Nair and Geoffrey~E. Hinton.
\newblock Rectified linear units improve restricted boltzmann machines.
\newblock In \emph{Proceedings of the 27th International Conference on
  International Conference on Machine Learning}, ICML'10, pages 807--814, USA,
  2010. Omnipress.
\newblock ISBN 978-1-60558-907-7.

\bibitem[Kolb et~al.(2017)Kolb, Lentz, and Kolpak]{Kolb_2017}
Brian Kolb, Levi~C. Lentz, and Alexie~M. Kolpak.
\newblock Discovering charge density functionals and structure-property
  relationships with {PROPhet}: A general framework for coupling machine
  learning and first-principles methods.
\newblock \emph{Scientific Reports}, 7\penalty0 (1), apr 2017.
\newblock \doi{10.1038/s41598-017-01251-z}.

\bibitem[Burke et~al.(1998)Burke, Cruz, and Lam]{Burke_1998}
Kieron Burke, Federico~G. Cruz, and Kin-Chung Lam.
\newblock Unambiguous exchange-correlation energy density.
\newblock \emph{The Journal of Chemical Physics}, 109\penalty0 (19):\penalty0
  8161--8167, nov 1998.
\newblock \doi{10.1063/1.477479}.

\bibitem[Mortensen et~al.(2005)Mortensen, Hansen, and
  Jacobsen]{ISI:000226735900040}
J.~J. Mortensen, L.~B. Hansen, and K.~W. Jacobsen.
\newblock Real-space grid implementation of the projector augmented wave
  method.
\newblock \emph{Phys. Rev. B}, 71\penalty0 (3):\penalty0 035109, JAN 2005.
\newblock ISSN 1098-0121.
\newblock \doi{10.1103/PhysRevB.71.035109}.

\bibitem[Ghosh and Suryanarayana(2017{\natexlab{a}})]{GhoshSwarnava2017SAae1}
Swarnava Ghosh and Phanish Suryanarayana.
\newblock Sparc: Accurate and efficient finite-difference formulation and
  parallel implementation of density functional theory: Extended systems.
\newblock \emph{Computer Physics Communications}, 216:\penalty0 109--125,
  2017{\natexlab{a}}.
\newblock ISSN 0010-4655.

\bibitem[Ghosh and Suryanarayana(2017{\natexlab{b}})]{GhoshSwarnava2017SAae2}
Swarnava Ghosh and Phanish Suryanarayana.
\newblock Sparc: Accurate and efficient finite-difference formulation and
  parallel implementation of density functional theory: Isolated clusters.
\newblock \emph{Computer Physics Communications}, 212\penalty0 (C):\penalty0
  189--204, 2017{\natexlab{b}}.
\newblock ISSN 0010-4655.

\bibitem[Perdew and Constantin(2007)]{Perdew_2007}
John~P. Perdew and Lucian~A. Constantin.
\newblock Laplacian-level density functionals for the kinetic energy density
  and exchange-correlation energy.
\newblock \emph{Physical Review B}, 75\penalty0 (15), apr 2007.
\newblock \doi{10.1103/physrevb.75.155109}.

\bibitem[Cramer(2004)]{CramerChristopherJ.2004Eocc}
Christopher~J. Cramer.
\newblock \emph{Essentials of computational chemistry : theories and models}.
\newblock Wiley, Chichester, West Sussex, England ; Hoboken, NJ, 2nd ed..
  edition, 2004.
\newblock ISBN 0470091827.

\bibitem[Thomas et~al.(2018)Thomas, Smidt, Kearnes, Yang, Li, Kohlhoff, and
  Riley]{ThomasNathaniel2018TfnR}
Nathaniel Thomas, Tess Smidt, Steven Kearnes, Lusann Yang, Li~Li, Kai Kohlhoff,
  and Patrick Riley.
\newblock Tensor field networks: Rotation- and translation-equivariant neural
  networks for 3d point clouds.
\newblock 2018.

\bibitem[Maxwell(1873)]{maxwell1873treatise}
J.C. Maxwell.
\newblock \emph{A Treatise on Electricity and Magnetism}.
\newblock Number v. 1 in A Treatise on Electricity and Magnetism. Clarendon
  Press, 1873.

\bibitem[Haley and Soloway(1992)]{HaleyP.J.1992Elom}
P.J. Haley and D.~Soloway.
\newblock Extrapolation limitations of multilayer feedforward neural networks.
\newblock volume~4, pages 25--30. IEEE Publishing, 1992.
\newblock ISBN 0780305590.

\bibitem[Zhao et~al.(1994)Zhao, Morrison, and Parr]{Zhao1994Fedt}
Zhao, Morrison, and Parr.
\newblock From electron densities to kohn-sham kinetic energies, orbital
  energies, exchange-correlation potentials, and exchange-correlation energies.
\newblock \emph{Physical review. A, Atomic, molecular, and optical physics},
  50\penalty0 (3):\penalty0 2138--2142, 1994.
\newblock ISSN 1050-2947.

\bibitem[Cruz et~al.(1998)Cruz, Lam, and Burke]{CLB98}
Federico~G. Cruz, Kin-Chung Lam, and Kieron Burke.
\newblock Exchange-correlation energy density from virial theorem.
\newblock \emph{The Journal of Physical Chemistry A}, 102\penalty0
  (25):\penalty0 4911--4917, 1998.
\newblock \doi{10.1021/jp980950v}.

\bibitem[Manby et~al.(2012)Manby, Stella, Goodpaster, and
  Miller]{ManbyFrederickR2012ASED}
Frederick~R Manby, Martina Stella, Jason~D Goodpaster, and Thomas~F Miller.
\newblock A simple, exact density-functional-theory embedding scheme.
\newblock \emph{Journal of chemical theory and computation}, 8\penalty0 (8),
  2012.
\newblock ISSN 1549-9626.

\bibitem[Libisch et~al.(2014)Libisch, Huang, and
  Carter]{LibischFlorian2014Ecws}
Florian Libisch, Chen Huang, and Emily~A. Carter.
\newblock Embedded correlated wavefunction schemes: theory and
  applications.(report).
\newblock \emph{Accounts of Chemical Research}, 47\penalty0 (9):\penalty0
  2768--2775, 2014.
\newblock ISSN 0001-4842.

\bibitem[Huang et~al.(2011)Huang, Pavone, and Carter]{HuangChen2011Qmet}
Chen Huang, Michele Pavone, and Emily~A. Carter.
\newblock Quantum mechanical embedding theory based on a unique embedding
  potential.
\newblock \emph{The Journal of Chemical Physics}, 134\penalty0 (15), 2011.
\newblock ISSN 0021-9606.

\bibitem[Govind et~al.(1998)Govind, Wang, Da~Silva, and
  Carter]{GovindN.1998Aaie}
N.~Govind, Y.A. Wang, A.J.R. Da~Silva, and E.A. Carter.
\newblock Accurate ab initio energetics of extended systems via explicit
  correlation embedded in a density functional environment.
\newblock \emph{Chemical Physics Letters}, 295\penalty0 (1):\penalty0 129--134,
  1998.
\newblock ISSN 0009-2614.

\end{thebibliography}

\end{document}


\maketitle

\section{Subsampling Procedure}
\begin{figure}[!h]
\centering
\includegraphics[width=0.75\textwidth]{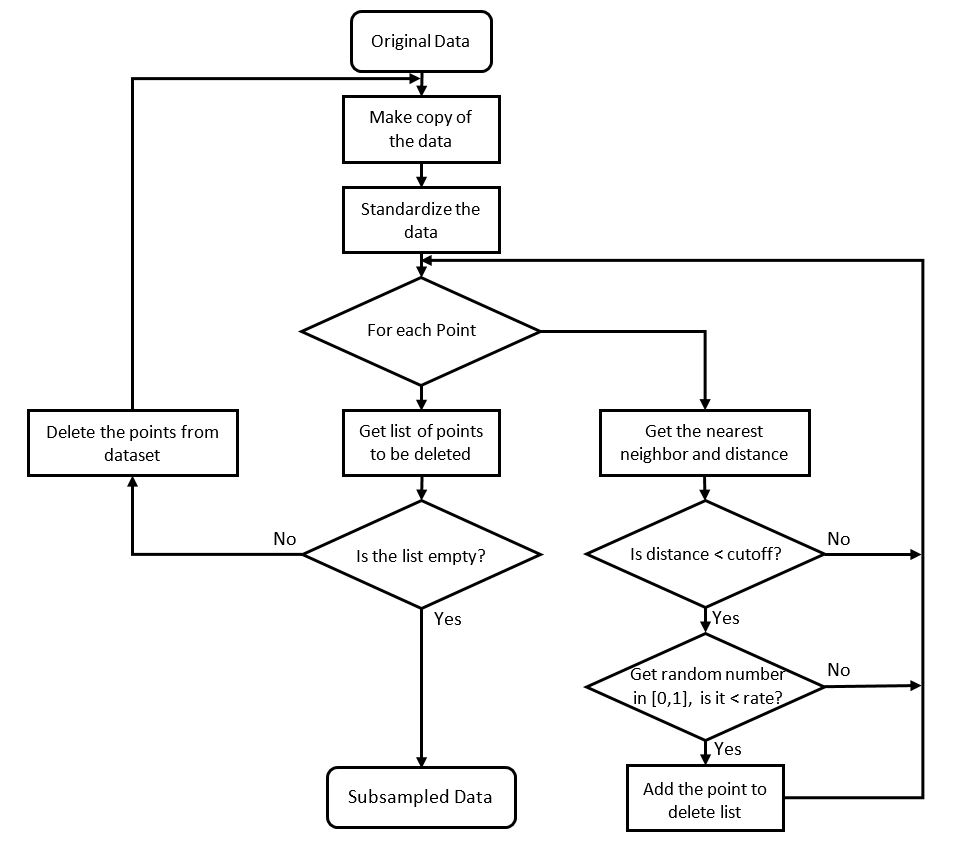}
\caption{Schematic of subsampling procedure}
\label{fig:SubsamplingProcedure}
\end{figure}

The sampling algorithm is an iterative process based on a kD-tree data representation. In each iteration, the dataset is normalized with the standard scaler (mean = 0, standard deviation = 1) and a kD-tree is constructed for the dataset using pykdtree library. High-dimensional datasets ($d>10$) are also pre-processed by using principal component analysis (PCA) to reduce the dimensionality by finding the minimum number of principal components that capture 99.9999\% of the variance prior to kD-tree construction.  The resulting kD-tree is queried to find the nearest neighbor for each data point and the distance between them. If the distance is below a certain cutoff distance, the neighbor is removed with some probability. The process is iterated until there are no more points to be removed. The algorithm has two hyper-parameters: cutoff distance and deletion probability. 
The cutoff distance controls the sparsity of the resulting representative dataset with higher cutoff distances resulting in fewer sub-sampled points, and the deletion probability controls robustness with lower deletion rates being more robust but resulting in slower execution.
For this work a cutoff distance of 0.01 standard deviation and a deletion probability of 0.2 were found to provide a good balance of sub-sampling efficiency and speed. This results in a sub-sample size of roughly 30,000 points per molecular dataset. All preprocessing and dimensionality reduction are used only to select the sub-sample; the resulting sub-sampled dataset retains the original dimensionality and numerical values.
The illustration of the subsampling procedure is shown in Figure \ref{fig:SubsamplingProcedure}

\clearpage
\section{MCSH Descriptor Illustrations}

\begin{figure*}[!h]
	\centering
    \begin{subfigure}{0.105\textwidth}
		\centering
        \includegraphics[width=\linewidth]{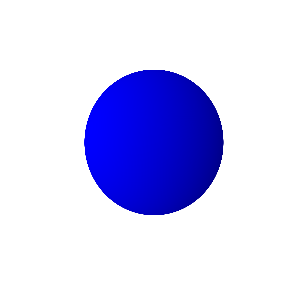}
        \caption{$S^{(0)}_{P(000)}$}
        \label{fig:LocalDensityStencil}
	\end{subfigure}%
    \begin{subfigure}{0.30\textwidth}
		\centering
        \includegraphics[width=\linewidth]{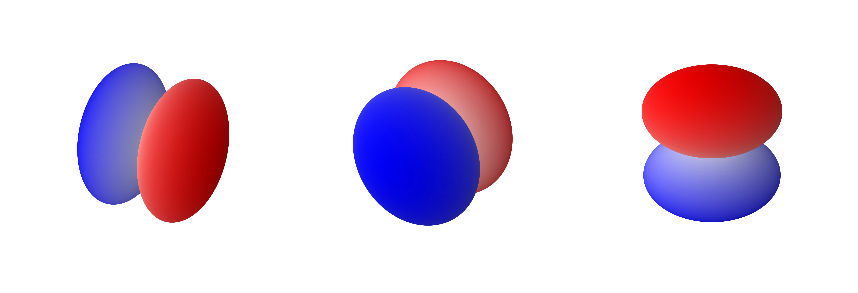}
        \caption{$S^{(1)}_{P(100)}$}
        \label{fig:AveDens006Stencil}
	\end{subfigure}%
    \begin{subfigure}{0.30\textwidth}
		\centering
        \includegraphics[width=\linewidth]{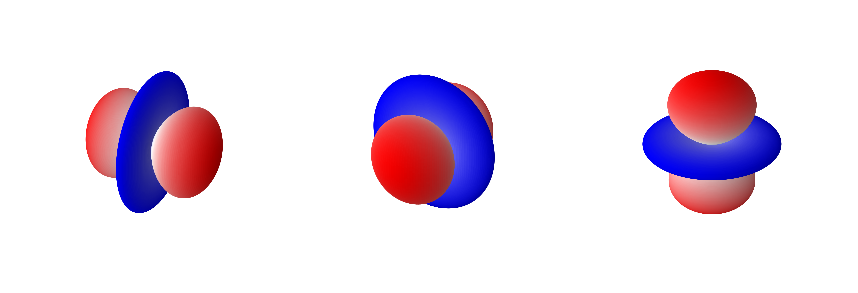}
        \caption{$S^{(2)}_{P(200)}$}
        \label{fig:LocalDensityStencil}
	\end{subfigure}%
    \begin{subfigure}{0.30\textwidth}
		\centering
        \includegraphics[width=\linewidth]{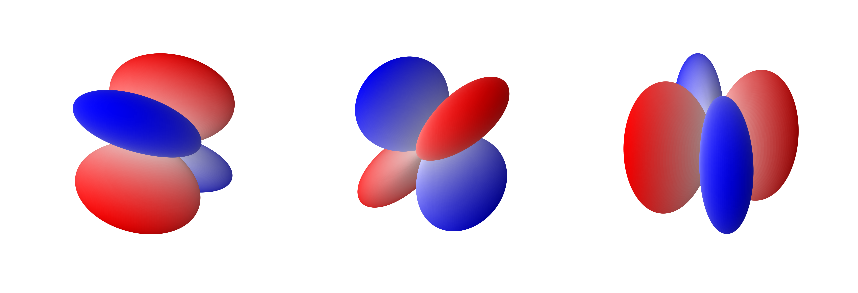}
        \caption{$S^{(2)}_{P(110)}$}
        \label{fig:AveDens006Stencil}
	\end{subfigure}
    \begin{subfigure}{0.30\textwidth}
		\centering
        \includegraphics[width=\linewidth]{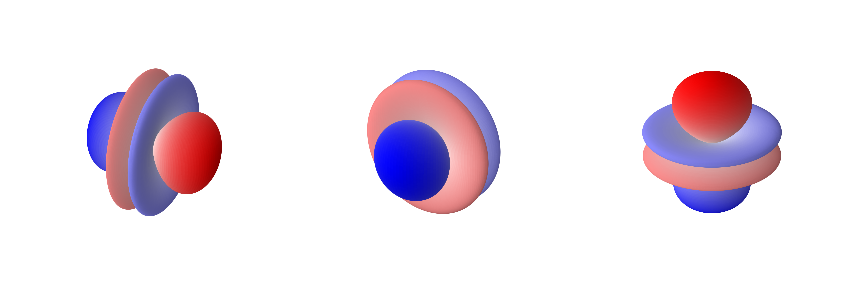}
        \caption{$S^{(3)}_{P(300)}$}
        \label{fig:AveDens006Stencil}
	\end{subfigure}%
    \begin{subfigure}{0.60\textwidth}
		\centering
        \includegraphics[width=\linewidth]{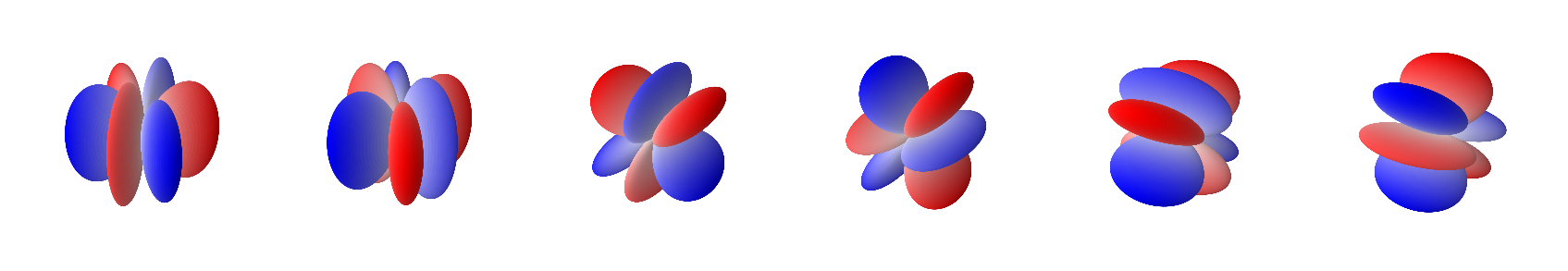}
        \caption{$S^{(3)}_{P(210)}$}
        \label{fig:LocalDensityStencil}
	\end{subfigure}%
    \begin{subfigure}{0.107\textwidth}
		\centering
        \includegraphics[width=\linewidth]{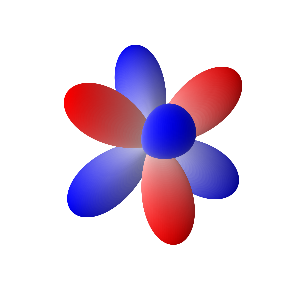}
        \caption{$S^{(3)}_{P(111)}$}
        \label{fig:AveDens006Stencil}
	\end{subfigure}
    \begin{subfigure}{0.30\textwidth}
		\centering
        \includegraphics[width=\linewidth]{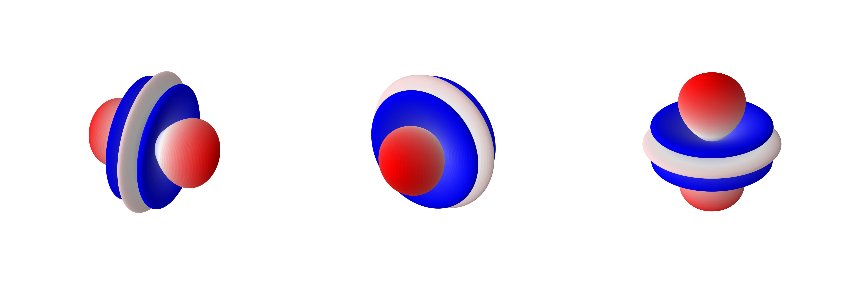}
        \caption{$S^{(4)}_{P(400)}$}
        \label{fig:AveDens006Stencil}
	\end{subfigure}%
    \begin{subfigure}{0.60\textwidth}
		\centering
        \includegraphics[width=\linewidth]{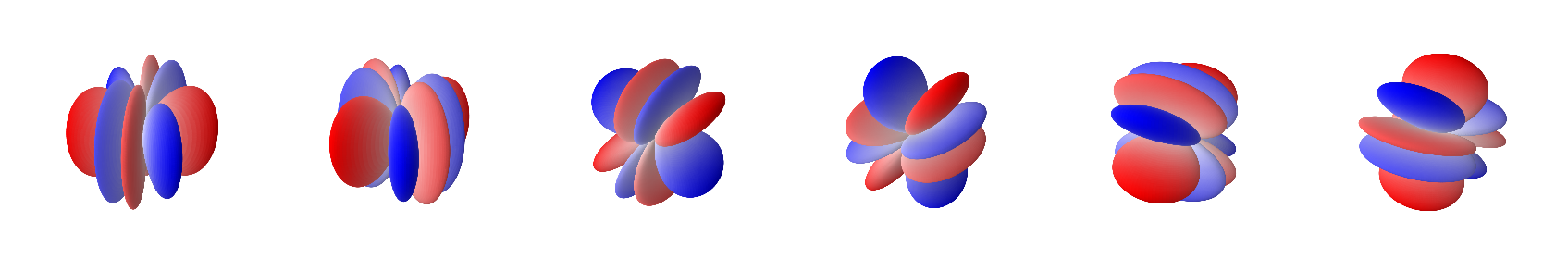}
        \caption{$S^{(4)}_{P(310)}$}
        \label{fig:AveDens006Stencil}
	\end{subfigure}
    \begin{subfigure}{0.30\textwidth}
		\centering
        \includegraphics[width=\linewidth]{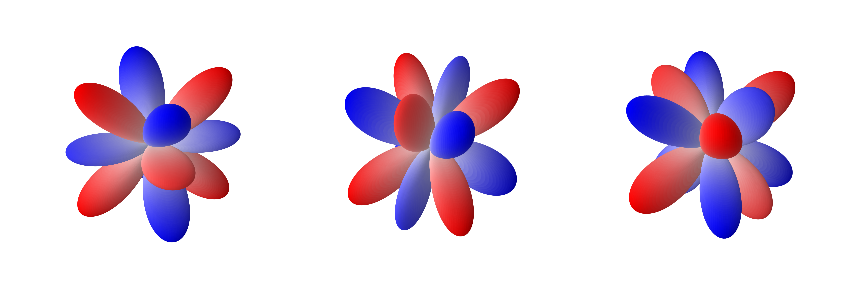}
        \caption{$S^{(4)}_{P(211)}$}
        \label{fig:LocalDensityStencil}
	\end{subfigure}%
    \begin{subfigure}{0.30\textwidth}
		\centering
        \includegraphics[width=\linewidth]{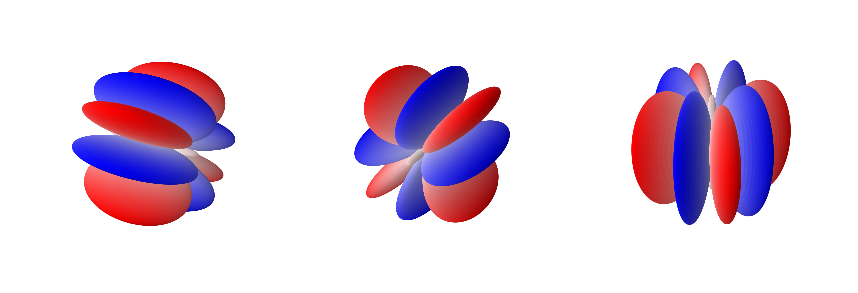}
        \caption{$S^{(4)}_{P(220)}$}
        \label{fig:AveDens006Stencil}
	\end{subfigure}
    \caption{Alternate graphical illustrations of the first 4 orders of Maxwell-Cartesian spherical harmonics (MCSH) descriptors denoted by $S^{(n)}_{P(ijk)}$. $n$ is the order, $P(ijk)$ denotes the permutation group of the index $ijk$.}
    \label{fig:MCSH2}
\end{figure*}

\clearpage
\section{Descriptor Rotation Invariance Test}
The descriptors are tested by calculating the descriptors of \ce{H_2} molecular system rotated about various points and comparing the distributions of descriptor values. \ce{H2} electron densities are computed with single point calculations at the B3LYP/cc-pvdz level. 

MCSH descriptors with angular orders of 0, 1, 2, 3 and spatial ranges of 0.06 \AA{}, 0.1 \AA{}, 0.14 \AA{} are tested. The descriptors are calculated at 21 equally-spaced points from -1.0 \AA{} to 0.0 \AA{} along the molecular bond coordinate of the of the system (where 0 is the mid-point of the two hydrogen atoms). At each point, the system was rotated in 343 different ways around the point (30 degree increment for all three dimension up to 180 degree, hence $7^3 = 343$ possible rotations), and the descriptors at this point were calculated for every rotation and compared to the average value of all rotations at that point. The descriptor values are plotted against the average values as parity plots below. Minor deviations are expected due to numerical noise, which will increase as the angular order increases or as the spatial extent decreases.

\begin{figure*}[!h]
\hspace*{-0.97in}
\centering
\includegraphics[width=1.25\textwidth]{Invariance_test/H2_MCSH0,1_parity_plot.png}
\caption{Parity plot of descriptor values for the $M^{(0)}_{r,000}$ descriptor for rotated systems. The x-axis represents the average value across all rotations, and the y-axis represents the value at specific rotations. Parity indicates rotational invariance. Colors denote different points along the molecular bond coordinate with higher numbers being closer to the center of the bond.}
\label{fig:H2MCSHDescriptors0-1}
\end{figure*}

\begin{figure*}[!h]
\hspace*{-0.97in}
\centering
\includegraphics[width=1.25\textwidth]{Invariance_test/H2_MCSH1,1_parity_plot.png}
\caption{Parity plot of descriptor values for the $M^{(1)}_{r,100}$ descriptor for rotated systems. See Fig. \ref{fig:H2MCSHDescriptors0-1} caption for details.}
\label{fig:H2MCSHDescriptors1-1}
\end{figure*}

\begin{figure*}[!h]
\hspace*{-0.97in}
\centering
\includegraphics[width=1.25\textwidth]{Invariance_test/H2_MCSH2,1_parity_plot.png}
\caption{Parity plot of descriptor values for the $M^{(2)}_{r,200}$ descriptor for rotated systems. See Fig. \ref{fig:H2MCSHDescriptors0-1} caption for details.}
\label{fig:H2MCSHDescriptors2-1}
\end{figure*}

\begin{figure*}[!h]
\hspace*{-0.97in}
\centering
\includegraphics[width=1.25\textwidth]{Invariance_test/H2_MCSH2,2_parity_plot.png}
\caption{Parity plot of descriptor values for the $M^{(2)}_{r,110}$ descriptor for rotated systems. See Fig. \ref{fig:H2MCSHDescriptors0-1} caption for details.}
\label{fig:H2MCSHDescriptors2-2}
\end{figure*}

\begin{figure*}[!h]
\hspace*{-0.97in}
\centering
\includegraphics[width=1.25\textwidth]{Invariance_test/H2_MCSH3,1_parity_plot.png}
\caption{Parity plot of descriptor values for the $M^{(3)}_{r,300}$ descriptor for rotated systems. See Fig. \ref{fig:H2MCSHDescriptors0-1} caption for details.}
\label{fig:H2MCSHDescriptors3-1}
\end{figure*}

\begin{figure*}[!h]
\hspace*{-0.97in}
\centering
\includegraphics[width=1.25\textwidth]{Invariance_test/H2_MCSH3,2_parity_plot.png}
\caption{Parity plot of descriptor values for the $M^{(3)}_{r,210}$ descriptor for rotated systems. See Fig. \ref{fig:H2MCSHDescriptors0-1} caption for details.}
\label{fig:H2MCSHDescriptors3-2}
\end{figure*}

\begin{figure*}[!h]
\hspace*{-0.97in}
\centering
\includegraphics[width=1.25\textwidth]{Invariance_test/H2_MCSH3,3_parity_plot.png}
\caption{Parity plot of descriptor values for the $M^{(3)}_{r,111}$ descriptor for rotated systems. See Fig. \ref{fig:H2MCSHDescriptors0-1} caption for details.}
\label{fig:H2MCSHDescriptors3-3}
\end{figure*}

\clearpage
\section{Training Procedure}
\subsection{Overview}
The neural network model training was done using the Python \texttt{Keras} package with a \texttt{Tensorflow} backend. The training process was broken down into two consecutive phases, corresponding to two loss functions and different numbers of training data points.

\subsection{Data Transformation}
The absolute value of the descriptors and energy densities goes down to $10^{-10}$, and most data points are in the range of $10^{-5} - 10^{-1}$ (for density and energy density), leading to numerically small values of the error. Re-scaling the data by a factor of $1,000,000$ (equivalent to changing volume units) was found to improve the training performance.

\subsection{Loss Function}
Two types of loss functions are used in training: mean squared error (MSE) and sum of absolute error (SAE). MSE is biased towards high energy density regions that occur near the atomic cores. The difference in values for the multivalued $\rho \to \epsilon_{XC}$ function is most extreme towards the high energy density regions, leading to large errors that can drown out the error in the valence region. SAE is less biased toward high density regions, and improves performance in the valence regions.

\subsection{Training Process}
Phase 1

\begin{itemize}
\item Goal: Initial training the model to get the overall performance right with a focus on high-density regions
\item Training data: uniformly subsampled data + 3,000,000 randomly sampled data
\item Loss function: mean squared error
\item Optimizer: Adam
\item Leaning rate: 0.001 (max 600 steps), 0.0001 (max 600 steps), 0.00001 (max 600 steps)
\end{itemize}

Phase 2
\begin{itemize}
\item Goal: Fine tune the model so it performs well across all regions including the valence regions
\item Training data: uniformly subsampled data + 10,000,000 randomly sampled data
\item Loss function: sum of absolute error
\item Optimizer: Adam
\item Leaning rate: 0.0001 (max 1600 steps), 0.00001 (max 1600 steps), 0.000001 (max 800 steps)
\end{itemize}

Each step is roughly 30 seconds of training, thus corresponds to about 10 epochs for phase 1 and 3 epochs for phase 2.  The training loss of the model is checked after each step, and the model is updated if the loss is better than the best prior model. If the model is not updated after certain number steps (the max step as shown above), the learning rate is lowered.

\pagebreak
\section{Training Data}

\begin{table}[!h]
\centering
\begin{tabular}{C{2.5cm}C{4cm}C{4cm}C{4cm}}
\toprule
Descriptor Set & Uniformly subsampled data points & Total training data points (MSE phase) & Total training data points (SAE phase) \\ \midrule
$\bar{\lambda}_{(0.00)}^{(0)}$ & 10831 & 3010831 & 10010831  \\
& & & \\
$\bar{\lambda}_{(0.02)}^{(0)}$ & 10869 & 3010869 & 10010869 \\
& & & \\
$\bar{\lambda}_{(0.04)}^{(0)}$ & 10893 & 3010893 & 10010893 \\
& & & \\
$\bar{\lambda}_{(0.08)}^{(0)}$ & 11151 & 3011151 & 10011151 \\
& & & \\
$\bar{\lambda}_{(0.2)}^{(0)}$ & 249688 & 3249688 & 10249688 \\
& & & \\
$\bar{\lambda}_{(0.02)}^{(1)}$ & 28856 & 3028856 & 10028856 \\
& & & \\
$\bar{\lambda}_{(0.04)}^{(1)}$ & 30702 & 3030702 & 10030702\\
& & & \\
$\bar{\lambda}_{(0.08)}^{(1)}$ & 33138 & 3033138 & 10033138\\
& & & \\
$\bar{\lambda}_{(0.2)}^{(1)}$ & 704379 & 3704379 & 10704379 \\
& & & \\
$\bar{\lambda}_{(0.02)}^{(2)}$ & 141728 & 3141728 & 10141728\\
& & & \\
$\bar{\lambda}_{(0.04)}^{(2)}$ & 152572 & 3152572 & 10152572\\
& & & \\
$\bar{\lambda}_{(0.08)}^{(2)}$ & 692243 & 3692243 & 10692243\\
& & & \\
$\bar{\lambda}_{(0.2)}^{(2)}$ & 1067195 & 4067195 & 11067195\\                  
 \bottomrule
\end{tabular}
\caption{Number of training data points used in each phase of training for different models.}
\label{tab:TrainingData}
\end{table}

\pagebreak
\section{Detailed Result of the Models}

\begin{table}[!h]
\centering
\resizebox{\textwidth}{!}{%
\begin{tabular}{C{1.7cm}C{1.1cm}C{1.3cm}C{1.3cm}C{1.3cm}C{1.35cm}C{1.2cm}C{1.1cm}C{1.1cm}C{1.1cm}C{1.1cm}C{1.1cm}C{1.1cm}C{0.9cm}}
\toprule
Model & Error & $C_2H_2$ & $C_2H_4$ & $C_2H_6$ & $CH_3OH$ & $CH_4$ & $CO$ & $CO_2$ & $H_2$ & $H_2O$ & $HCN$ & $HNC$ & MAE        \\ \midrule
$NN[\bar \lambda^{(0)}_{0.0}]$ & $\varepsilon_{abs.}$ & 5.63 & 5.88 & 6.14 & 5.68 & 3.09 & 5.19 & 7.73 & 0.13 & 2.8 & 4.29 & 4.22 & 5.31 \\
 & $\varepsilon_{pred.}$ & 0.57 & 0.8 & 0.98 & -1.42 & 0.44 & -1.86 & -3.66 & -0.02 & -1.88 & -0.67 & -0.6 & 1.73 \\
 & $\varepsilon_{form.}$ & -0.39 & -0.14 & 0.06 & -0.01 &  -  & -0.49 & -0.44 &  -  &  -  & -0.44 & -0.36 & 0.23 \\
 & & & & & & & & & & & & &\\
$NN[\bar \lambda^{(0)}_{0.02}]$ & $\varepsilon_{abs.}$ & 5.32 & 5.57 & 5.83 & 5.7 & 2.94 & 5.23 & 7.94 & 0.13 & 2.98 & 4.06 & 3.99 & 5.26 \\
 & $\varepsilon_{pred.}$ & 0.77 & 1.0 & 1.17 & -1.06 & 0.54 & -1.49 & -3.04 & -0.03 & -1.61 & -0.44 & -0.36 & 1.47 \\
 & $\varepsilon_{form.}$ & -0.39 & -0.14 & 0.06 & -0.01 &  -  & -0.5 & -0.45 &  -  &  -  & -0.44 & -0.36 & 0.23 \\
 & & & & & & & & & & & & &\\
$NN[\bar \lambda^{(0)}_{0.04}]$ & $\varepsilon_{abs.}$ & 5.05 & 5.25 & 5.51 & 5.12 & 2.77 & 4.71 & 6.97 & 0.13 & 2.55 & 3.94 & 3.93 & 4.82 \\
 & $\varepsilon_{pred.}$ & -0.18 & 0.02 & 0.19 & -0.6 & 0.05 & -0.98 & -1.59 & -0.03 & -0.67 & -0.93 & -0.85 & 0.91 \\
 & $\varepsilon_{form.}$ & -0.37 & -0.13 & 0.07 & -0.01 &  -  & -0.45 & -0.42 &  -  &  -  & -0.42 & -0.34 & 0.23 \\
 & & & & & & & & & & & & &\\
$NN[\bar \lambda^{(0)}_{0.08}]$ & $\varepsilon_{abs.}$ & 1.6 & 1.86 & 2.21 & 2.65 & 1.12 & 2.09 & 3.56 & 0.18 & 1.69 & 1.43 & 1.33 & 2.31 \\
 & $\varepsilon_{pred.}$ & 0.81 & 1.08 & 1.23 & 0.16 & 0.52 & -0.19 & -0.45 & -0.1 & -0.48 & 0.38 & 0.43 & 0.53 \\
 & $\varepsilon_{form.}$ & -0.54 & -0.17 & 0.08 & 0.01 &  -  & -0.55 & -0.42 &  -  &  -  & -0.51 & -0.46 & 0.24 \\
 & & & & & & & & & & & & &\\
$NN[\bar \lambda^{(0)}_{0.2}]$ & $\varepsilon_{abs.}$ & 0.95 & 1.11 & 1.3 & 1.3 & 0.64 & 0.86 & 1.47 & 0.12 & 0.65 & 1.02 & 0.92 & 1.23 \\
 & $\varepsilon_{pred.}$ & 0.42 & 0.44 & 0.5 & 0.11 & 0.16 & 0.08 & 0.04 & -0.07 & -0.24 & 0.03 & 0.18 & 0.21 \\
 & $\varepsilon_{form.}$ & -0.12 & -0.03 & 0.11 & 0.11 &  -  & -0.06 & 0.07 &  -  &  -  & -0.21 & -0.06 & 0.1 \\
 & & & & & & & & & & & & &\\
$NN[\bar \lambda^{(1)}_{0.02}]$ & $\varepsilon_{abs.}$ & 1.6 & 1.72 & 4.13 & 2.4 & 0.94 & 2.0 & 3.26 & 0.06 & 1.35 & 1.41 & 1.4 & 2.13 \\
 & $\varepsilon_{pred.}$ & 0.09 & 0.22 & -2.53 & 0.12 & 0.17 & 0.11 & 0.34 & 0.02 & 0.2 & 0.02 & 0.02 & 0.3 \\
 & $\varepsilon_{form.}$ & -0.19 & -0.08 & -2.85 & -0.23 &  -  & -0.2 & -0.15 &  -  &  -  & -0.16 & -0.15 & 0.23 \\
 & & & & & & & & & & & & &\\
$NN[\bar \lambda^{(1)}_{0.04}]$ & $\varepsilon_{abs.}$ & 0.24 & 0.24 & 1.67 & 0.53 & 0.11 & 0.3 & 0.49 & 0.01 & 0.21 & 0.25 & 0.26 & 0.4 \\
 & $\varepsilon_{pred.}$ & 0.05 & 0.02 & -1.34 & -0.2 & -0.02 & 0.06 & 0.12 & -0.01 & 0.05 & 0.03 & 0.03 & 0.12 \\
 & $\varepsilon_{form.}$ & 0.07 & 0.05 & -1.31 & -0.24 &  -  & 0.01 & 0.01 &  -  &  -  & 0.03 & 0.03 & 0.09 \\
 & & & & & & & & & & & & &\\
$NN[\bar \lambda^{(1)}_{0.08}]$ & $\varepsilon_{abs.}$ & 0.1 & 0.1 & 0.32 & 0.17 & 0.05 & 0.14 & 0.22 & 0.01 & 0.1 & 0.11 & 0.11 & 0.15 \\
 & $\varepsilon_{pred.}$ & 0.01 & 0.03 & -0.18 & -0.04 & 0.02 & -0.06 & -0.09 & 0.0 & -0.04 & -0.02 & -0.02 & 0.04 \\
 & $\varepsilon_{form.}$ & -0.01 & 0.0 & -0.21 & -0.02 &  -  & -0.03 & -0.02 &  -  &  -  & -0.03 & -0.02 & 0.02 \\
 & & & & & & & & & & & & &\\
$NN[\bar \lambda^{(1)}_{0.2}]$ & $\varepsilon_{abs.}$ & 0.05 & 0.04 & 0.05 & 0.06 & 0.02 & 0.06 & 0.07 & 0.01 & 0.03 & 0.05 & 0.05 & 0.06 \\
 & $\varepsilon_{pred.}$ & 0.0 & -0.0 & -0.01 & -0.01 & -0.01 & -0.01 & -0.0 & -0.0 & -0.01 & 0.01 & 0.01 & 0.01 \\
 & $\varepsilon_{form.}$ & 0.01 & 0.0 & -0.0 & -0.0 &  -  & -0.0 & 0.0 &  -  &  -  & 0.01 & 0.0 & 0.01 \\
 & & & & & & & & & & & & &\\
$NN[\bar \lambda^{(2)}_{0.02}]$ & $\varepsilon_{abs.}$ & 0.44 & 0.44 & 1.51 & 0.64 & 0.23 & 0.46 & 0.68 & 0.03 & 0.3 & 0.32 & 0.36 & 0.52 \\
 & $\varepsilon_{pred.}$ & 0.27 & 0.25 & -0.84 & -0.02 & 0.11 & 0.11 & 0.01 & -0.0 & 0.01 & 0.07 & 0.09 & 0.13 \\
 & $\varepsilon_{form.}$ & 0.04 & 0.03 & -1.07 & -0.13 &  -  & -0.02 & -0.12 &  -  &  -  & -0.04 & -0.01 & 0.08 \\
 & & & & & & & & & & & & &\\
$NN[\bar \lambda^{(2)}_{0.04}]$ & $\varepsilon_{abs.}$ & 0.14 & 0.12 & 0.49 & 0.32 & 0.06 & 0.14 & 0.19 & 0.01 & 0.08 & 0.13 & 0.15 & 0.18 \\
 & $\varepsilon_{pred.}$ & 0.02 & 0.01 & -0.1 & -0.15 & -0.0 & 0.02 & 0.06 & -0.0 & 0.03 & 0.0 & 0.0 & 0.03 \\
 & $\varepsilon_{form.}$ & 0.01 & 0.0 & -0.1 & -0.18 &  -  & -0.01 & -0.0 &  -  &  -  & 0.0 & 0.0 & 0.02 \\
 & & & & & & & & & & & & &\\
$NN[\bar \lambda^{(2)}_{0.08}]$ & $\varepsilon_{abs.}$ & 0.07 & 0.06 & 0.12 & 0.09 & 0.04 & 0.08 & 0.11 & 0.01 & 0.05 & 0.06 & 0.07 & 0.09 \\
 & $\varepsilon_{pred.}$ & 0.0 & 0.01 & 0.03 & 0.01 & 0.0 & -0.01 & -0.01 & -0.01 & 0.0 & -0.0 & -0.01 & 0.01 \\
 & $\varepsilon_{form.}$ & -0.02 & -0.0 & 0.02 & -0.0 &  -  & -0.04 & -0.04 &  -  &  -  & -0.02 & -0.02 & 0.01 \\
 & & & & & & & & & & & & &\\
$NN[\bar \lambda^{(2)}_{0.2}]$ & $\varepsilon_{abs.}$ & 0.04 & 0.04 & 0.06 & 0.05 & 0.02 & 0.05 & 0.06 & 0.01 & 0.03 & 0.04 & 0.05 & 0.06 \\
 & $\varepsilon_{pred.}$ & 0.01 & -0.0 & -0.02 & -0.01 & -0.0 & -0.02 & -0.01 & -0.0 & -0.0 & 0.01 & 0.0 & 0.01 \\
 & $\varepsilon_{form.}$ & 0.01 & 0.0 & -0.02 & -0.0 &  -  & -0.01 & -0.01 &  -  &  -  & 0.0 & 0.0 & 0.01 \\
 \bottomrule
\end{tabular}}
\caption{All error metrics for each training and testing system with each different descriptor set. Units are in eV. MAE is the mean absolute error of the corresponding error metric for the 15 training molecular systems plus 7 test systems}
\label{tab:ModelResult1}
\end{table}

 .\\
 \\
 \\
 \\
\begin{table}[!th]
\centering
\resizebox{\textwidth}{!}{%
\begin{tabular}{C{1.7cm}C{1.1cm}C{1.1cm}C{1.2cm}C{1.2cm}C{1.1cm}|C{1.2cm}C{1.5cm}C{1.2cm}C{1.3cm}C{1.3cm}C{1.65cm}C{1.4cm}C{0.9cm}}
\toprule
Model & Error & $N_2$ & $N_2O$ & $NH_3$ & $O_3$ & $N_2H_4$ & $HCOOH$ & $H_2O_2$ & $H_2CO$ & $H_2CCO$ & $CH_3CHO$ & $CH_3CN$ & MAE         \\ \midrule

$NN[\bar \lambda^{(0)}_{0.0}]$ & $\varepsilon_{abs.}$ & 3.3 & 6.01 & 1.76 & 8.37 & 3.47 & 8.0 & 5.76 & 5.44 & 8.2 & 8.46 & 7.26 & 5.31 \\
 & $\varepsilon_{pred.}$ & -1.97 & -3.67 & -0.75 & -6.02 & -1.47 & -3.41 & -3.86 & -1.67 & -1.19 & -1.09 & -0.12 & 1.73 \\
 & $\varepsilon_{form.}$ & -0.54 & -0.38 &  -  & -0.45 & 0.01 & -0.17 & -0.12 & -0.29 & -0.3 & -0.17 & -0.35 & 0.23 \\
 & & & & & & & & & & & & &\\
$NN[\bar \lambda^{(0)}_{0.02}]$ & $\varepsilon_{abs.}$ & 3.14 & 6.0 & 1.68 & 8.78 & 3.32 & 8.21 & 6.07 & 5.46 & 8.07 & 8.33 & 6.88 & 5.26 \\
 & $\varepsilon_{pred.}$ & -1.71 & -3.14 & -0.62 & -5.25 & -1.21 & -2.78 & -3.32 & -1.31 & -0.73 & -0.62 & 0.22 & 1.47 \\
 & $\varepsilon_{form.}$ & -0.55 & -0.39 &  -  & -0.48 & 0.0 & -0.17 & -0.12 & -0.29 & -0.3 & -0.17 & -0.35 & 0.23 \\
 & & & & & & & & & & & & &\\
$NN[\bar \lambda^{(0)}_{0.04}]$ & $\varepsilon_{abs.}$ & 3.18 & 5.62 & 1.74 & 7.31 & 3.41 & 7.24 & 5.11 & 4.87 & 7.35 & 7.59 & 6.59 & 4.82 \\
 & $\varepsilon_{pred.}$ & -1.72 & -2.21 & -0.64 & -2.44 & -1.25 & -1.34 & -1.44 & -0.84 & -0.74 & -0.64 & -0.76 & 0.91 \\
 & $\varepsilon_{form.}$ & -0.53 & -0.39 &  -  & -0.53 & 0.0 & -0.15 & -0.14 & -0.28 & -0.28 & -0.16 & -0.33 & 0.23 \\
 & & & & & & & & & & & & &\\
$NN[\bar \lambda^{(0)}_{0.08}]$ & $\varepsilon_{abs.}$ & 1.6 & 3.07 & 0.93 & 4.83 & 1.91 & 3.89 & 3.43 & 2.36 & 3.14 & 3.43 & 2.48 & 2.31 \\
 & $\varepsilon_{pred.}$ & -0.18 & -0.32 & 0.06 & -1.3 & 0.27 & -0.25 & -0.84 & -0.02 & 0.66 & 0.74 & 1.12 & 0.53 \\
 & $\varepsilon_{form.}$ & -0.6 & -0.38 &  -  & -0.18 & 0.04 & -0.12 & 0.01 & -0.27 & -0.32 & -0.13 & -0.4 & 0.24 \\
 & & & & & & & & & & & & &\\
$NN[\bar \lambda^{(0)}_{0.2}]$ & $\varepsilon_{abs.}$ & 1.29 & 1.83 & 0.7 & 2.13 & 1.49 & 1.7 & 1.49 & 1.08 & 1.56 & 1.72 & 1.65 & 1.23 \\
 & $\varepsilon_{pred.}$ & -0.26 & -0.01 & -0.14 & -0.31 & -0.04 & 0.15 & -0.3 & 0.05 & 0.4 & 0.43 & 0.38 & 0.21 \\
 & $\varepsilon_{form.}$ & -0.2 & 0.22 &  -  & 0.18 & 0.17 & 0.25 & 0.1 & -0.02 & 0.03 & 0.13 & -0.09 & 0.1 \\
 & & & & & & & & & & & & &\\
$NN[\bar \lambda^{(1)}_{0.02}]$ & $\varepsilon_{abs.}$ & 1.33 & 2.63 & 0.77 & 3.89 & 1.57 & 3.36 & 2.68 & 2.09 & 2.93 & 3.0 & 2.33 & 2.13 \\
 & $\varepsilon_{pred.}$ & -0.09 & 0.12 & 0.07 & 0.29 & 0.2 & 0.44 & 0.3 & 0.22 & 0.38 & 0.42 & 0.24 & 0.3 \\
 & $\varepsilon_{form.}$ & -0.17 & -0.14 &  -  & -0.25 & 0.08 & -0.07 & -0.08 & -0.11 & -0.08 & -0.06 & -0.09 & 0.23 \\
 & & & & & & & & & & & & &\\
$NN[\bar \lambda^{(1)}_{0.04}]$ & $\varepsilon_{abs.}$ & 0.37 & 0.55 & 0.18 & 0.6 & 0.32 & 0.5 & 0.41 & 0.3 & 0.43 & 0.41 & 0.37 & 0.4 \\
 & $\varepsilon_{pred.}$ & 0.11 & 0.12 & 0.0 & 0.12 & 0.0 & 0.11 & 0.08 & 0.04 & 0.09 & 0.04 & 0.03 & 0.12 \\
 & $\varepsilon_{form.}$ & 0.09 & 0.04 &  -  & -0.05 & -0.01 & 0.01 & -0.03 & -0.0 & 0.05 & 0.01 & 0.04 & 0.09 \\
 & & & & & & & & & & & & &\\
$NN[\bar \lambda^{(1)}_{0.08}]$ & $\varepsilon_{abs.}$ & 0.14 & 0.21 & 0.07 & 0.26 & 0.14 & 0.23 & 0.19 & 0.14 & 0.21 & 0.2 & 0.16 & 0.15 \\
 & $\varepsilon_{pred.}$ & -0.06 & -0.08 & -0.0 & -0.09 & -0.01 & -0.08 & -0.06 & -0.04 & -0.0 & -0.02 & 0.0 & 0.04 \\
 & $\varepsilon_{form.}$ & -0.05 & -0.04 &  -  & 0.02 & -0.01 & -0.02 & 0.01 & -0.01 & 0.01 & -0.01 & -0.02 & 0.02 \\
 & & & & & & & & & & & & &\\
$NN[\bar \lambda^{(1)}_{0.2}]$ & $\varepsilon_{abs.}$ & 0.06 & 0.08 & 0.04 & 0.1 & 0.07 & 0.08 & 0.08 & 0.05 & 0.11 & 0.1 & 0.08 & 0.06 \\
 & $\varepsilon_{pred.}$ & 0.02 & 0.02 & 0.0 & 0.02 & 0.01 & -0.01 & 0.02 & -0.0 & 0.04 & 0.02 & 0.01 & 0.01 \\
 & $\varepsilon_{form.}$ & 0.01 & 0.01 &  -  & 0.03 & -0.0 & 0.0 & 0.02 & 0.0 & 0.04 & 0.03 & 0.01 & 0.01 \\
 & & & & & & & & & & & & &\\
$NN[\bar \lambda^{(2)}_{0.02}]$ & $\varepsilon_{abs.}$ & 0.3 & 0.58 & 0.21 & 0.89 & 0.41 & 0.71 & 0.63 & 0.45 & 0.67 & 0.68 & 0.54 & 0.52 \\
 & $\varepsilon_{pred.}$ & -0.04 & -0.06 & -0.01 & 0.02 & 0.02 & 0.1 & 0.06 & 0.08 & 0.23 & 0.25 & 0.2 & 0.13 \\
 & $\varepsilon_{form.}$ & -0.02 & -0.05 &  -  & -0.01 & 0.03 & -0.03 & 0.04 & -0.04 & -0.0 & 0.02 & -0.01 & 0.08 \\
 & & & & & & & & & & & & &\\
$NN[\bar \lambda^{(2)}_{0.04}]$ & $\varepsilon_{abs.}$ & 0.18 & 0.24 & 0.09 & 0.24 & 0.2 & 0.2 & 0.17 & 0.13 & 0.2 & 0.21 & 0.21 & 0.18 \\
 & $\varepsilon_{pred.}$ & 0.01 & 0.03 & -0.01 & 0.06 & -0.02 & 0.06 & 0.04 & 0.03 & 0.05 & 0.04 & 0.02 & 0.03 \\
 & $\varepsilon_{form.}$ & 0.02 & 0.02 &  -  & -0.03 & -0.0 & -0.0 & -0.01 & -0.0 & 0.02 & 0.01 & 0.03 & 0.02 \\
 & & & & & & & & & & & & &\\
$NN[\bar \lambda^{(2)}_{0.08}]$ & $\varepsilon_{abs.}$ & 0.07 & 0.11 & 0.04 & 0.13 & 0.09 & 0.13 & 0.1 & 0.07 & 0.15 & 0.15 & 0.12 & 0.09 \\
 & $\varepsilon_{pred.}$ & 0.0 & 0.01 & -0.0 & -0.0 & 0.01 & 0.01 & 0.01 & -0.0 & 0.04 & 0.04 & 0.02 & 0.01 \\
 & $\varepsilon_{form.}$ & -0.01 & -0.02 &  -  & -0.03 & 0.01 & -0.02 & -0.0 & -0.02 & 0.01 & 0.02 & 0.0 & 0.01 \\
 & & & & & & & & & & & & &\\
$NN[\bar \lambda^{(2)}_{0.2}]$ & $\varepsilon_{abs.}$ & 0.05 & 0.07 & 0.03 & 0.08 & 0.07 & 0.07 & 0.07 & 0.04 & 0.11 & 0.11 & 0.08 & 0.06 \\
 & $\varepsilon_{pred.}$ & 0.01 & 0.01 & 0.0 & 0.0 & 0.0 & -0.02 & -0.0 & -0.01 & 0.0 & -0.02 & 0.01 & 0.01 \\
 & $\varepsilon_{form.}$ & 0.01 & -0.0 &  -  & 0.0 & -0.0 & -0.01 & -0.0 & -0.0 & 0.01 & -0.02 & 0.01 & 0.01 \\

 \bottomrule
\end{tabular}}
\caption{Errors of all the systems with all models. The unit is eV. MAE is the mean absolute error of the corresponding error metric for the 15 training molecular systems plus 7 test systems}
\label{tab:ModelResult2}
\end{table}

 .\\
 \\
 \\
 \\
\begin{table}[!th]
\centering
\resizebox{0.7\textwidth}{!}{%
\begin{tabular}{C{1.7cm}C{1.1cm}C{1.8cm}C{1.8cm}C{1.8cm}C{1.2cm}}
\toprule
Model & Error & $CH_3NO2$ & $glycine$ & $NCCN$ & MAE         \\ \midrule
$NN[\bar \lambda^{(0)}_{0.0}]$ & $\varepsilon_{abs.}$ & 10.08 & 12.68 & 8.48 & 5.31 \\
 & $\varepsilon_{pred.}$ & -4.06 & -3.44 & -1.26 & 1.73 \\
 & $\varepsilon_{form.}$ & -0.09 & 0.06 & -0.81 & 0.23  \\
 & & & \\
$NN[\bar \lambda^{(0)}_{0.02}]$ & $\varepsilon_{abs.}$ & 10.17 & 12.67 & 8.02 & 5.26 \\
 & $\varepsilon_{pred.}$ & -3.3 & -2.58 & -0.78 & 1.47 \\
 & $\varepsilon_{form.}$ & -0.1 & 0.06 & -0.81 & 0.23  \\
 & & & \\
$NN[\bar \lambda^{(0)}_{0.04}]$ & $\varepsilon_{abs.}$ & 10.64 & 11.66 & 7.76 & 4.82 \\
 & $\varepsilon_{pred.}$ & -0.46 & -1.69 & -1.76 & 0.91 \\
 & $\varepsilon_{form.}$ & 1.34 & 0.03 & -0.78 & 0.23  \\
 & & & \\
$NN[\bar \lambda^{(0)}_{0.08}]$ & $\varepsilon_{abs.}$ & 5.39 & 6.05 & 2.81 & 2.31 \\
 & $\varepsilon_{pred.}$ & -0.06 & 0.87 & 0.97 & 0.53 \\
 & $\varepsilon_{form.}$ & -0.1 & 0.21 & -0.9 & 0.24  \\
 & & & \\
$NN[\bar \lambda^{(0)}_{0.2}]$ & $\varepsilon_{abs.}$ & 2.95 & 3.93 & 2.04 & 1.23 \\
 & $\varepsilon_{pred.}$ & 0.67 & 0.55 & 0.27 & 0.21 \\
 & $\varepsilon_{form.}$ & 0.83 & 0.48 & -0.27 & 0.1  \\
 & & & \\
$NN[\bar \lambda^{(1)}_{0.02}]$ & $\varepsilon_{abs.}$ & 21.74 & 5.18 & 7.3 & 2.13 \\
 & $\varepsilon_{pred.}$ & -16.8 & 0.9 & -4.34 & 0.3 \\
 & $\varepsilon_{form.}$ & -17.36 & 0.18 & -4.68 & 0.23  \\
 & & & \\
$NN[\bar \lambda^{(1)}_{0.04}]$ & $\varepsilon_{abs.}$ & 2.01 & 0.95 & 0.94 & 0.4 \\
 & $\varepsilon_{pred.}$ & 1.38 & 0.28 & -0.21 & 0.12 \\
 & $\varepsilon_{form.}$ & 1.27 & 0.19 & -0.22 & 0.09  \\
 & & & \\
$NN[\bar \lambda^{(1)}_{0.08}]$ & $\varepsilon_{abs.}$ & 1.24 & 0.76 & 0.49 & 0.15 \\
 & $\varepsilon_{pred.}$ & 0.18 & 0.11 & 0.24 & 0.04 \\
 & $\varepsilon_{form.}$ & 0.24 & 0.16 & 0.22 & 0.02  \\
 & & & \\
$NN[\bar \lambda^{(1)}_{0.2}]$ & $\varepsilon_{abs.}$ & 0.24 & 0.55 & 0.13 & 0.06 \\
 & $\varepsilon_{pred.}$ & 0.08 & 0.37 & 0.03 & 0.01 \\
 & $\varepsilon_{form.}$ & 0.08 & 0.38 & 0.02 & 0.01  \\
 & & & \\
$NN[\bar \lambda^{(2)}_{0.02}]$ & $\varepsilon_{abs.}$ & 6.86 & 1.53 & 3.04 & 0.52 \\
 & $\varepsilon_{pred.}$ & -5.69 & 0.38 & -2.27 & 0.13 \\
 & $\varepsilon_{form.}$ & -5.81 & 0.15 & -2.48 & 0.08  \\
 & & & \\
$NN[\bar \lambda^{(2)}_{0.04}]$ & $\varepsilon_{abs.}$ & 3.66 & 0.56 & 0.49 & 0.18 \\
 & $\varepsilon_{pred.}$ & -3.14 & 0.12 & 0.12 & 0.03 \\
 & $\varepsilon_{form.}$ & -3.19 & 0.06 & 0.12 & 0.02  \\
 & & & \\
$NN[\bar \lambda^{(2)}_{0.08}]$ & $\varepsilon_{abs.}$ & 1.42 & 0.61 & 0.14 & 0.09 \\
 & $\varepsilon_{pred.}$ & 1.08 & 0.3 & 0.02 & 0.01 \\
 & $\varepsilon_{form.}$ & 1.05 & 0.26 & -0.02 & 0.01  \\
 & & & \\
$NN[\bar \lambda^{(2)}_{0.2}]$ & $\varepsilon_{abs.}$ & 0.35 & 0.7 & 0.12 & 0.06 \\
 & $\varepsilon_{pred.}$ & 0.08 & 0.19 & -0.01 & 0.01 \\
 & $\varepsilon_{form.}$ & 0.09 & 0.19 & -0.01 & 0.01  \\

 \bottomrule
\end{tabular}}
\caption{Errors of all the systems with all models. The unit is eV. MAE is the mean absolute error of the corresponding error metric for the 15 training molecular systems plus 7 test systems}
\label{tab:ModelResult3}
\end{table}

\begin{figure}[!hb]
	\centering
	\begin{subfigure}{0.5\textwidth}
		\centering
		\includegraphics[width=\linewidth]{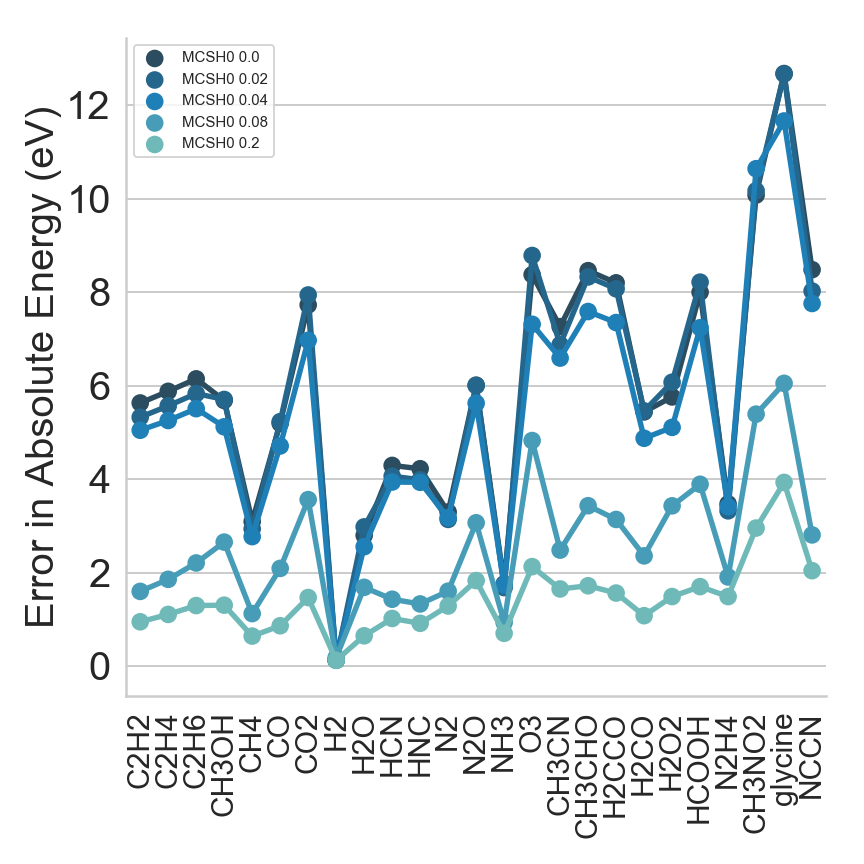}
        \label{fig:FactorResultAbs1}
	\end{subfigure}%
	\begin{subfigure}{0.5\textwidth}
		\centering
		\includegraphics[width=\linewidth]{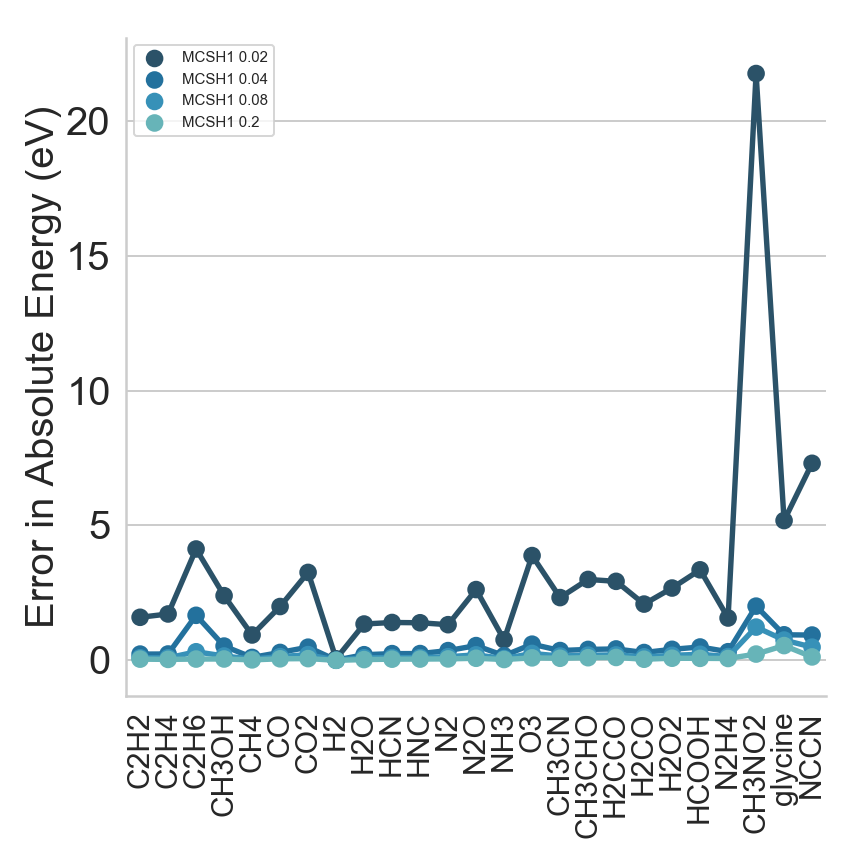}
        \label{fig:FactorResultAbs2}
	\end{subfigure}
	\begin{subfigure}{0.5\textwidth}
		\centering
		\includegraphics[width=\linewidth]{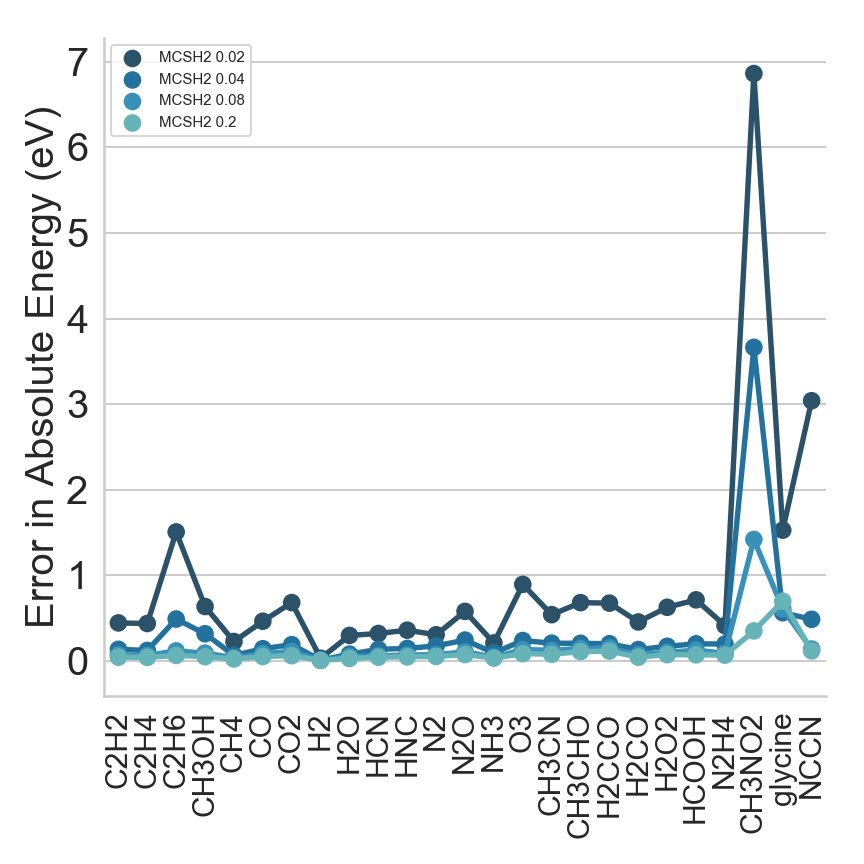}
        \label{fig:FactorResultAbs3}
	\end{subfigure}
    \caption{Factor plots of the results - sum of absolute error. The models are named as: ``MCSH\{order of angular feature\} \{range cutoff\}''}
	\label{fig:FactorResultAbs}
\end{figure}

\begin{figure}[!hb]
	\centering
	\begin{subfigure}{0.5\textwidth}
		\centering
		\includegraphics[width=\linewidth]{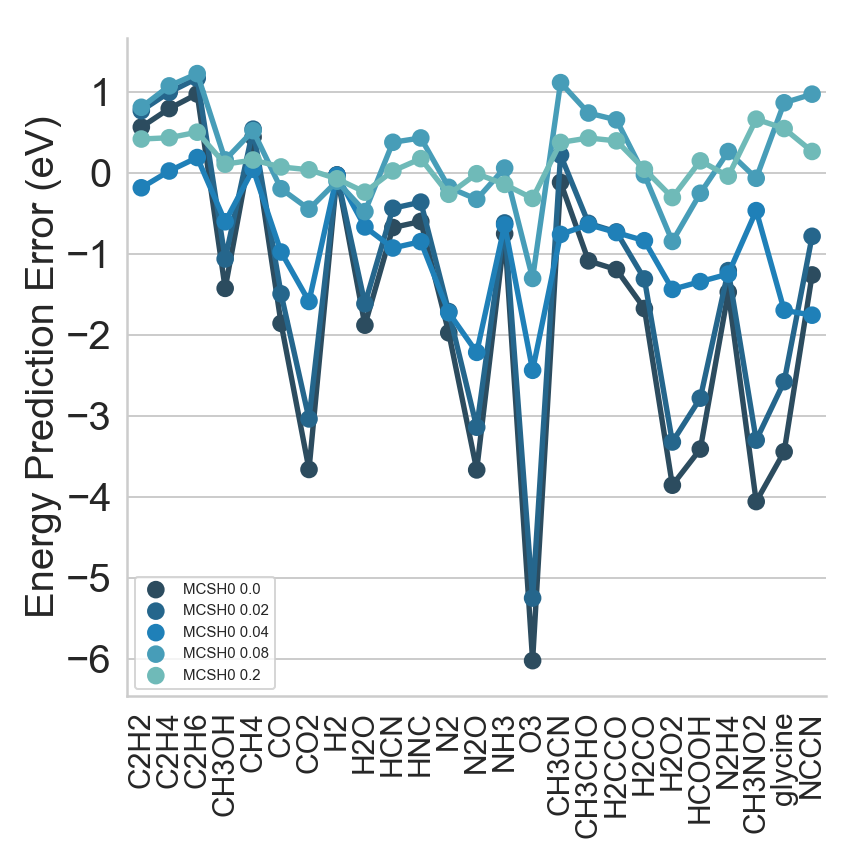}
        \label{fig:FactorResultPred1}
	\end{subfigure}%
	\begin{subfigure}{0.5\textwidth}
		\centering
		\includegraphics[width=\linewidth]{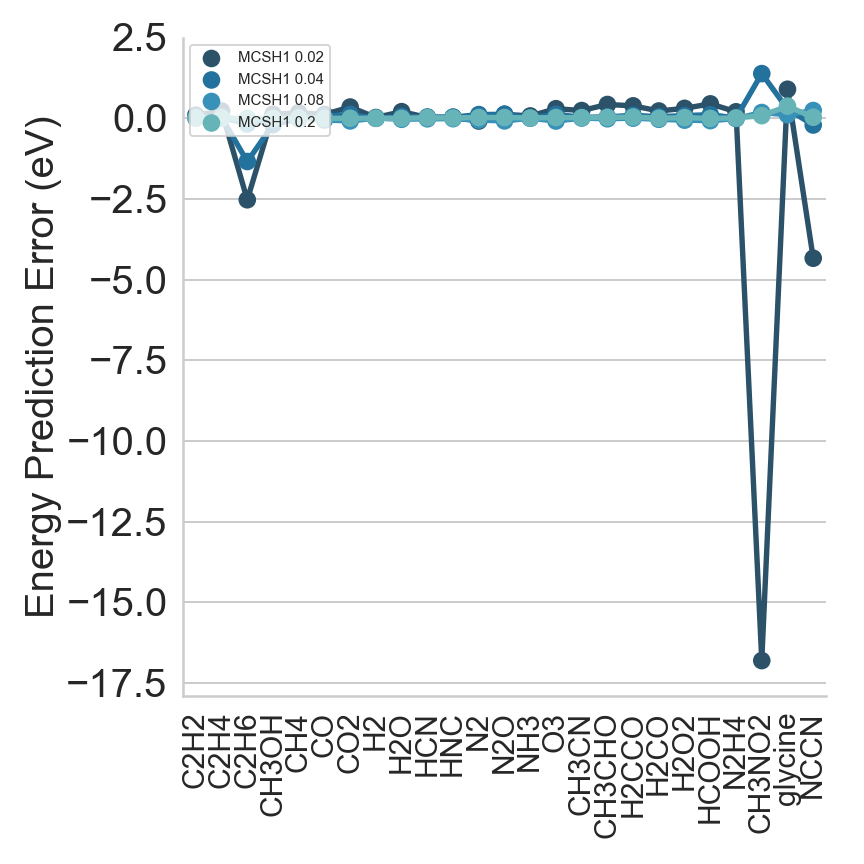}
        \label{fig:FactorResultPred2}
	\end{subfigure}
	\begin{subfigure}{0.5\textwidth}
		\centering
		\includegraphics[width=\linewidth]{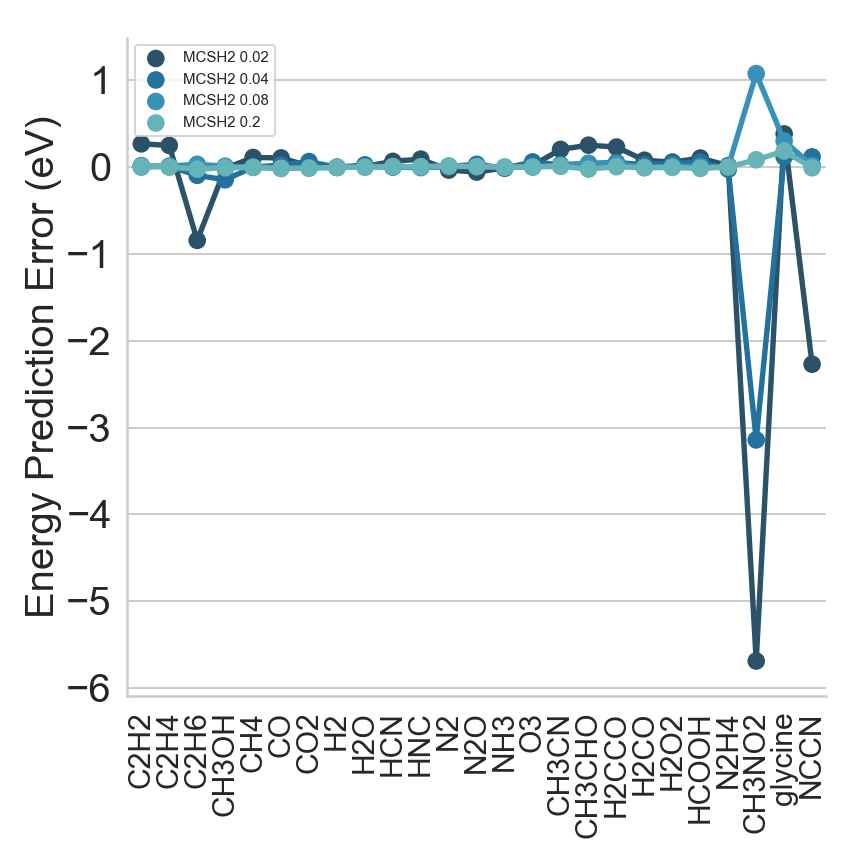}
        \label{fig:FactorResultPred3}
	\end{subfigure}
    \caption{Factor plots of the results - energy prediction error. The models are named as: ``MCSH\{order of angular feature\} \{range cutoff\}''}
	\label{fig:FactorResultPred}
\end{figure}

\begin{figure}[!hb]
	\centering
	\begin{subfigure}{0.5\textwidth}
		\centering
		\includegraphics[width=\linewidth]{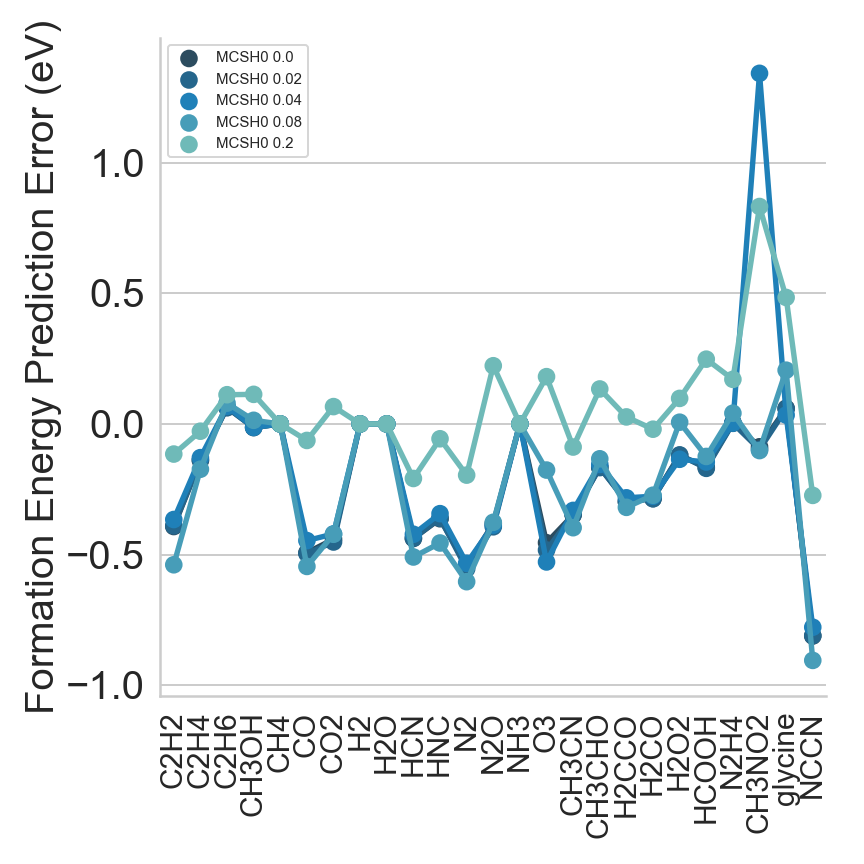}
        \label{fig:FactorResultForm1}
	\end{subfigure}%
	\begin{subfigure}{0.5\textwidth}
		\centering
		\includegraphics[width=\linewidth]{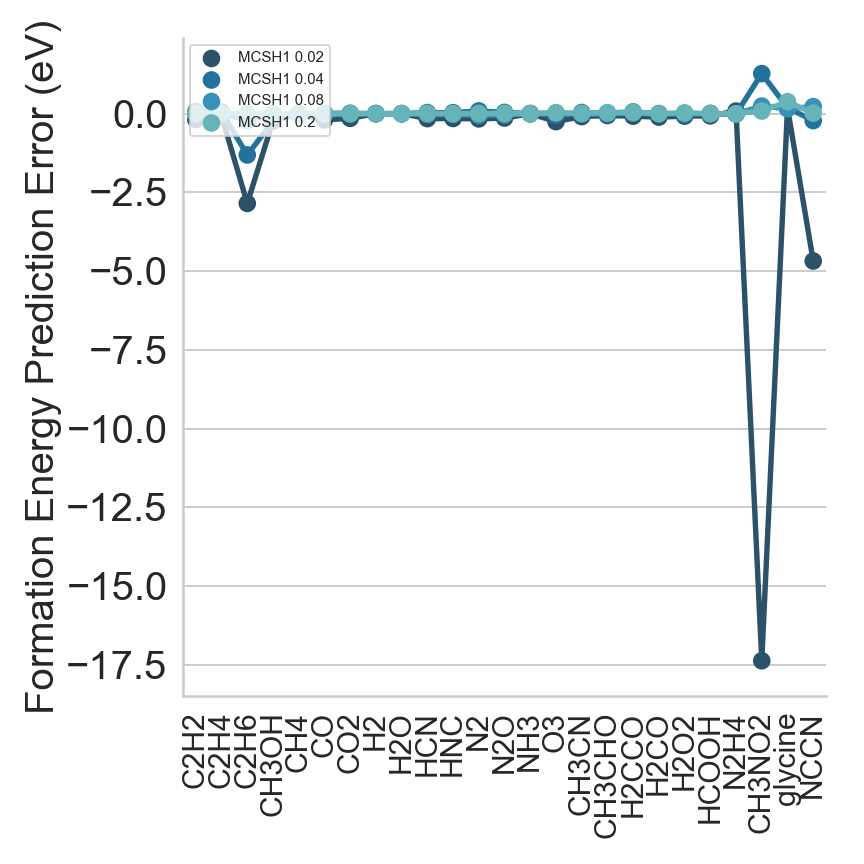}
        \label{fig:FactorResultForm2}
	\end{subfigure}
	\begin{subfigure}{0.5\textwidth}
		\centering
		\includegraphics[width=\linewidth]{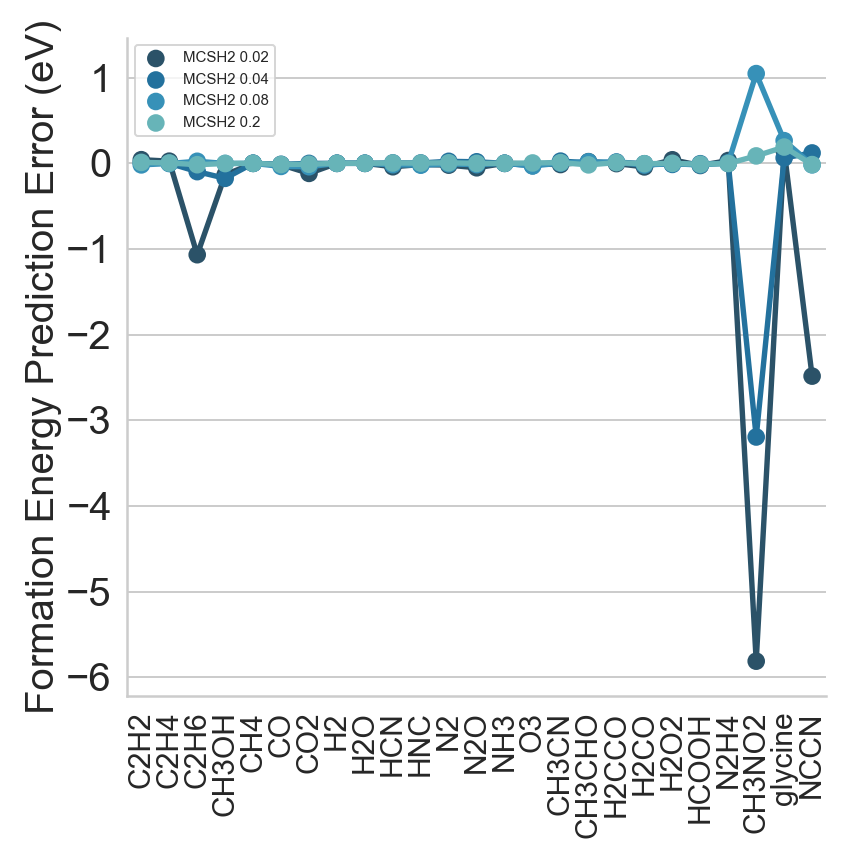}
        \label{fig:FactorResultForm3}
	\end{subfigure}
    \caption{Factor plots of the results - formation energy prediction error. The models are named as: ``MCSH\{order of angular feature\} \{range cutoff\}''}
	\label{fig:FactorResultForm}
\end{figure}

\clearpage
\section{Improvablity Test}

\begin{figure}[!hb]
	\centering
	\begin{subfigure}{0.4\textwidth}
		\centering
		\includegraphics[width=\linewidth]{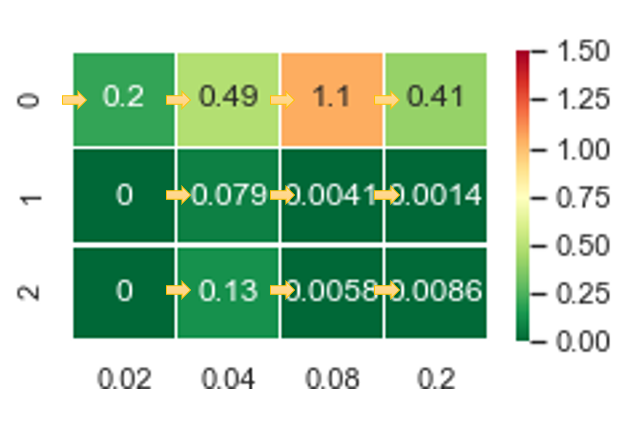}
        \caption{Range}
        \label{fig:MCSHEnerAbsErrorDist}
	\end{subfigure}
	\begin{subfigure}{0.4\textwidth}
		\centering
		\includegraphics[width=\linewidth]{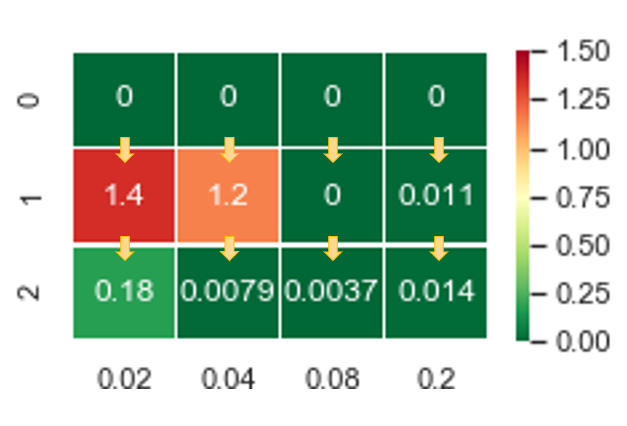}
        \caption{Angular feature}
        \label{fig:MCSHEnerPredErrorDist}
	\end{subfigure}
    \caption{Systematic improvability test for energy prediction error. The numbers denote the maximum deviation from systematic improvement for energy prediction error as compared to previous models.}
	\label{fig:ImprovabilityTestPred}
\end{figure}

\begin{figure}[!hb]
	\centering
	\begin{subfigure}{0.4\textwidth}
		\centering
		\includegraphics[width=\linewidth]{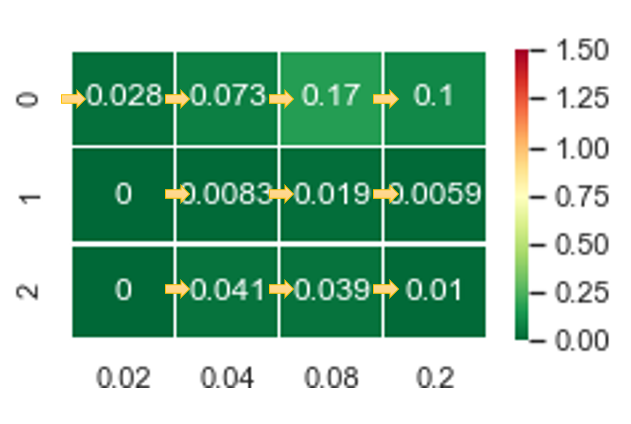}
        \caption{Range}
        \label{fig:MCSHEnerAbsErrorDist}
	\end{subfigure}
	\begin{subfigure}{0.4\textwidth}
		\centering
		\includegraphics[width=\linewidth]{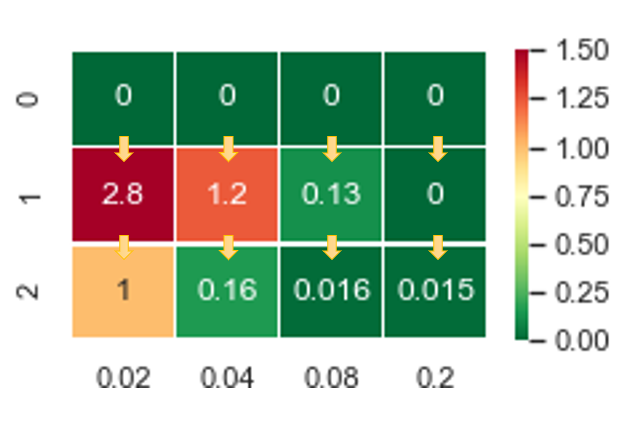}
        \caption{Angular feature}
        \label{fig:MCSHEnerPredErrorDist}
	\end{subfigure}
    \caption{Systematic improvability test for formation energy prediction error. The numbers denote the maximum deviation from systematic improvement for formation energy prediction error as compared to previous models.}
	\label{fig:ImprovabilityTestForm}
\end{figure}

\clearpage
\section{Result Validation Suite}
The models and tests are available via Github:
https://github.com/ray38/surrogate\_functional\_test

\clearpage
\section{Outlier and Extrapolation Analysis}

\subsection{Outlier error comparison}
\begin{figure}[!h]
	\centering
	\begin{subfigure}{0.48\textwidth}
		\centering
		\includegraphics[width=\linewidth]{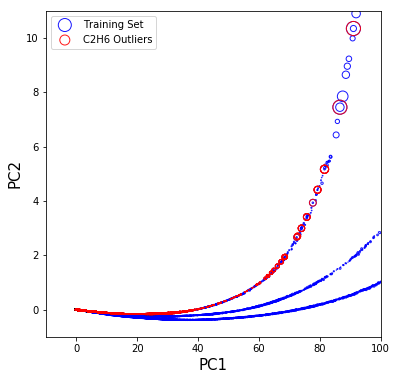}
        \caption{Energy prediction error distribution}
        \label{fig:LDAEnerPredErrorDist}
	\end{subfigure}%
	\begin{subfigure}{0.5\textwidth}
		\centering
		\includegraphics[width=\linewidth]{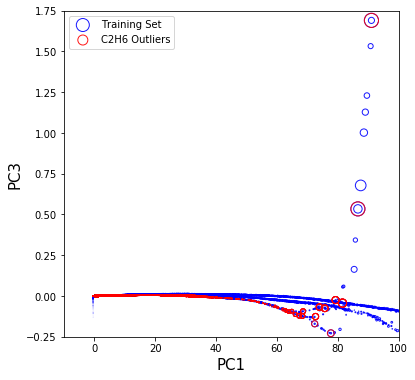}
        \caption{Energy prediction error distribution}
        \label{fig:LDAEnerPredErrorDist}
	\end{subfigure}
    \begin{subfigure}{0.5\textwidth}
		\centering
		\includegraphics[width=\linewidth]{Domain_detection/C2H6_232_PC2_PC3.png}
        \caption{Energy prediction error distribution}
        \label{fig:LDAEnerPredErrorDist}
	\end{subfigure}
\caption{Comparison between the electronic environment of $C-C$ bonding region and that of the whole training set as characterized by the $\bar{\lambda}_{(0.04)}^{(1)}$ descriptor set. The training set is represented by uniformly sampled points plus 3,000,000 randomly sampled points. Principle component analysis (PCA) model is trained with $C-C$ bonding region data points and applied to both datasets. The plot of first 3 principle components are shown here, where the red circles correspond to $C-C$ bonding environments and blue circles correspond to training data, the sizes of the circles correspond to the absolute prediction error of the $NN[\bar{\lambda}_{(0.04)}^{(1)}]$ model}
\end{figure}

\pagebreak
\subsection{Extrapolation comparison}

\begin{figure}[!h]
\centering
\includegraphics[width=\textwidth]{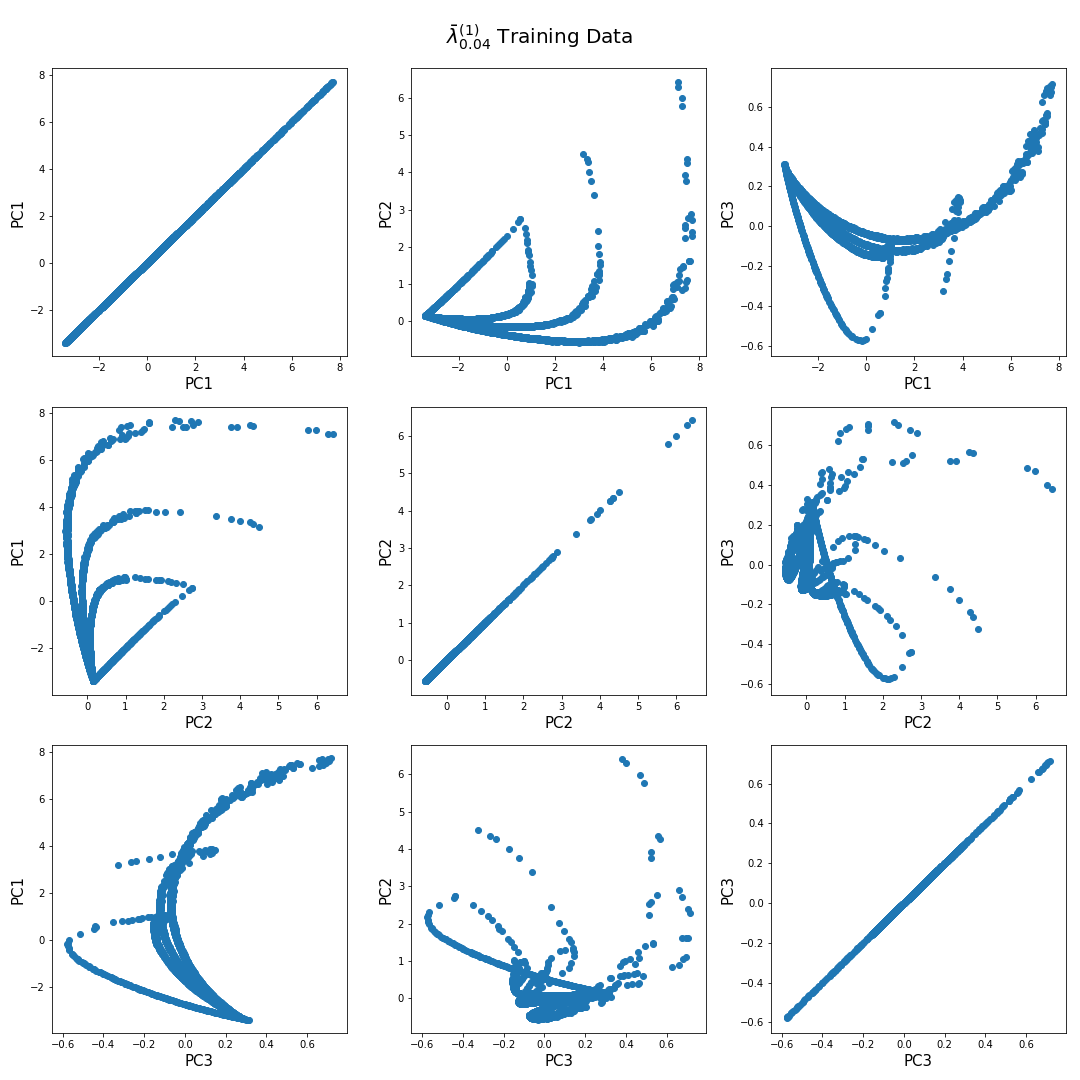}
\caption{Electronic environment captured in the whole training set as characterized by the $\bar{\lambda}_{(0.04)}^{(1)}$ descriptor set. The training set is represented by uniformly sampled points plus 3,000,000 randomly sampled points. Principle component analysis (PCA) model is trained with the training data points and applied to both datasets. The plot of first three principle components are shown here}
\label{fig:TestOutlierTraining}
\end{figure}

\begin{figure}[!h]
\centering
\includegraphics[width=\textwidth]{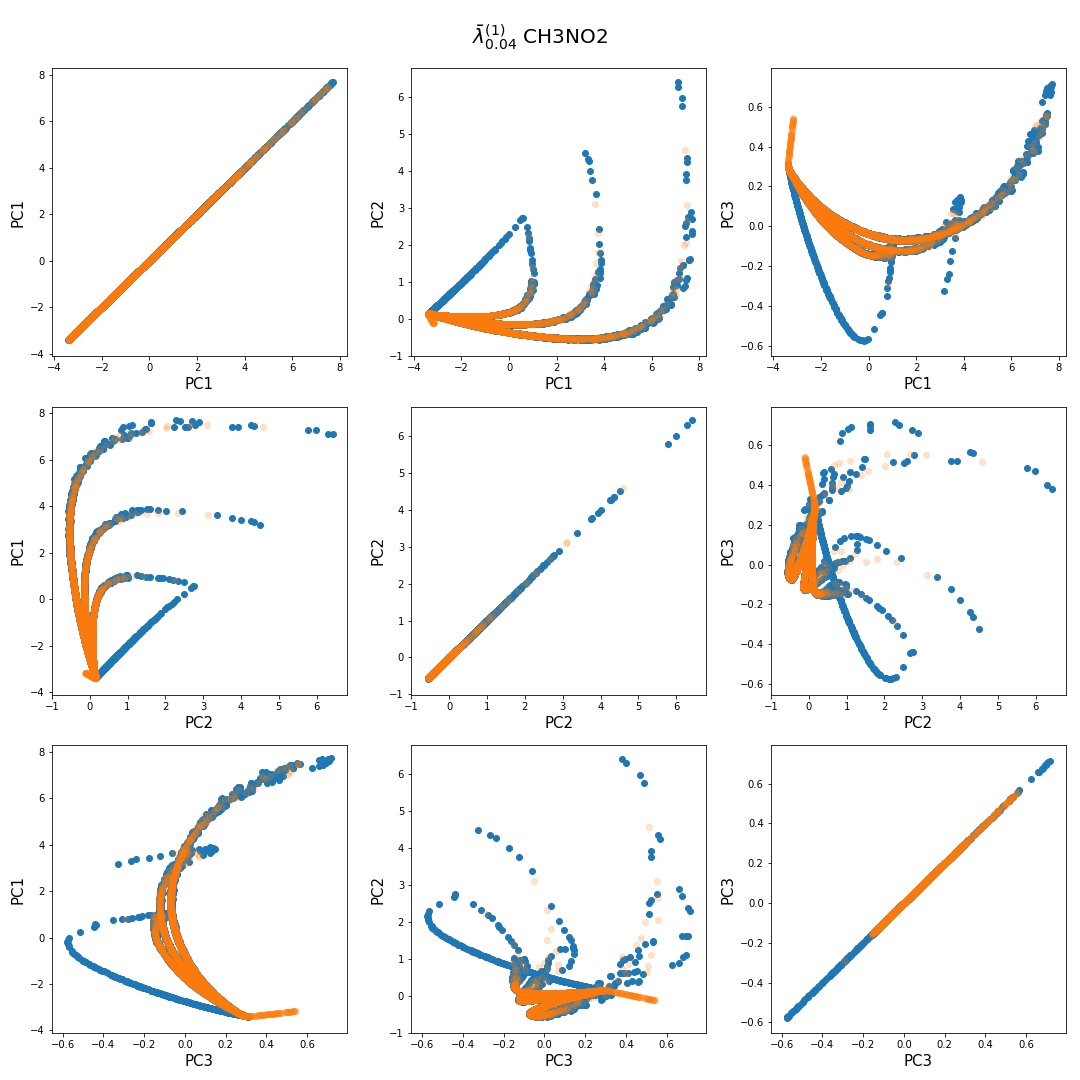}
\caption{Comparison between the electronic environment of $CH_3NO_2$ system and that of the whole training set as characterized by the $\bar{\lambda}_{(0.04)}^{(1)}$ descriptor set. The $CH_3NO_2$ system is represented by the uniformly sampled points of the system, and the training set is represented by uniformly sampled points plus 3,000,000 randomly sampled points. Principle component analysis (PCA) model is trained with the training data points and applied to both datasets. The plot of first three principle components are shown here, where the orange dots correspond to $CH_3NO_2$ system and blue dots correspond to training data.}
\label{fig:TestOutlierCH3NO2}
\end{figure}

\begin{figure}[!h]
\centering
\includegraphics[width=\textwidth]{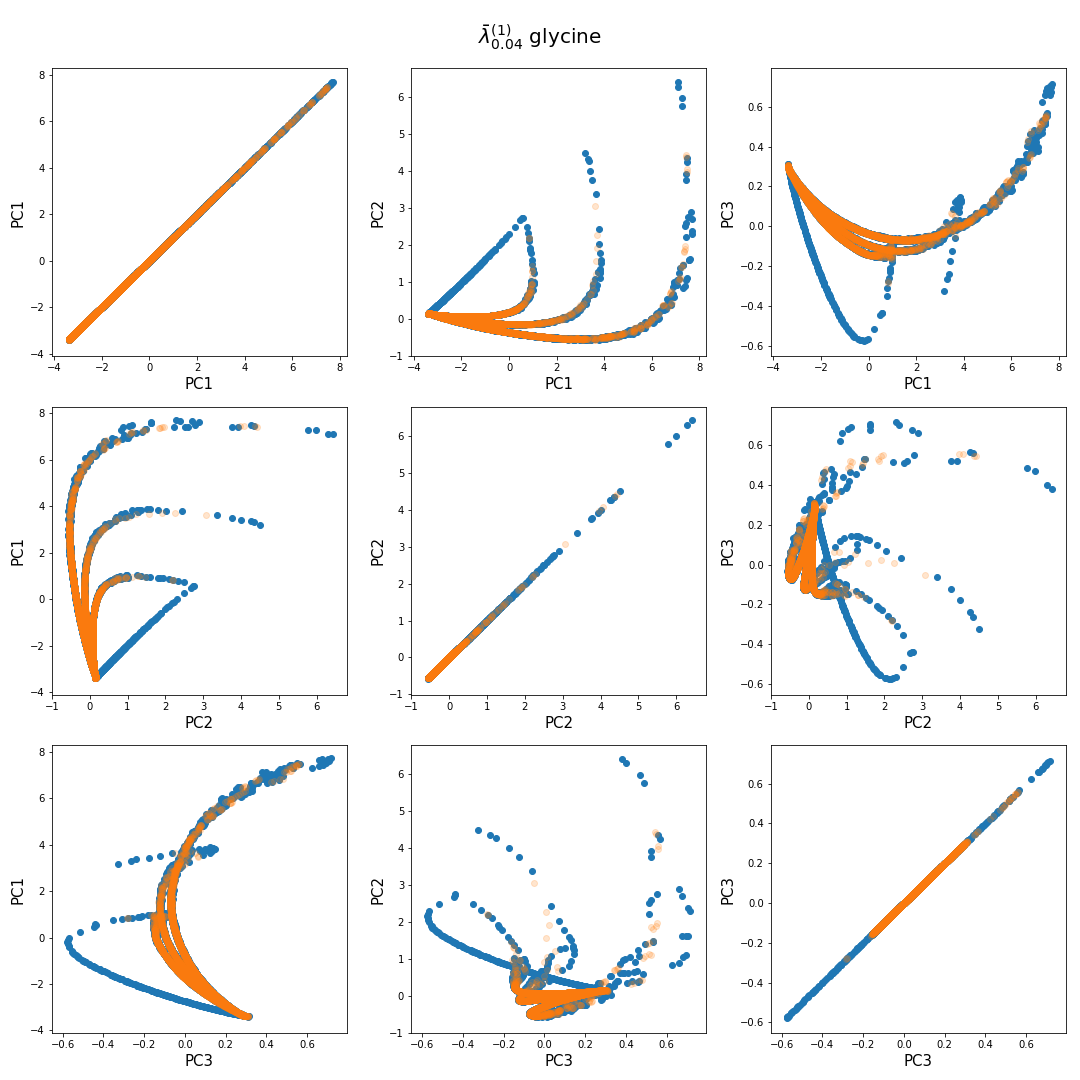}
\caption{Comparison between the electronic environment of $glycine$ system and that of the whole training set as characterized by the $\bar{\lambda}_{(0.04)}^{(1)}$ descriptor set. The $glycine$ system is represented by the uniformly sampled points of the system, and the training set is represented by uniformly sampled points plus 3,000,000 randomly sampled points. Principle component analysis (PCA) model is trained with the training data points and applied to both datasets. The plot of first three principle components are shown here, where the orange dots correspond to $glycine$ system and blue dots correspond to training data.}
\label{fig:TestOutlierGlycine}
\end{figure}

\begin{figure}[!h]
\centering
\includegraphics[width=\textwidth]{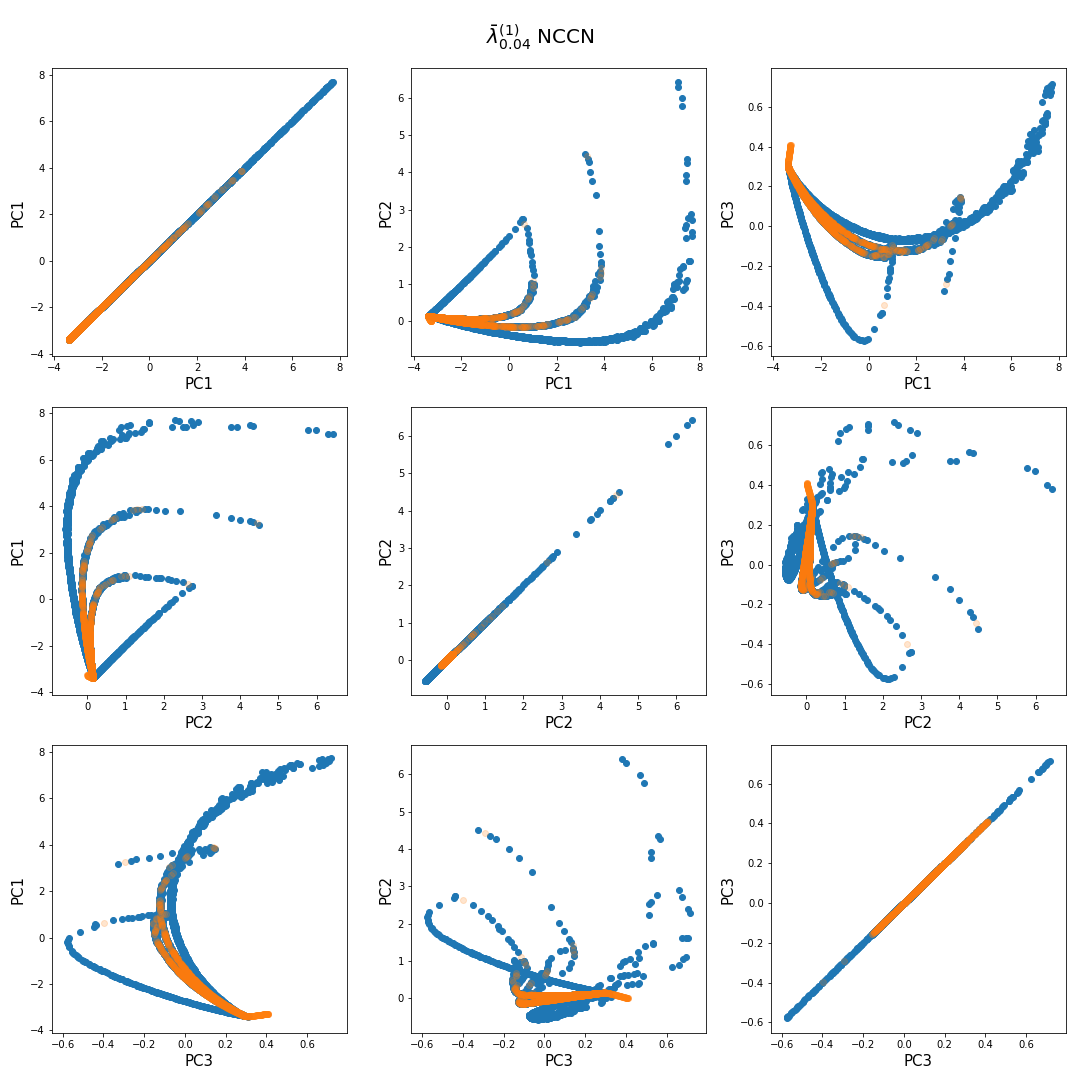}
\caption{Comparison between the electronic environment of $NCCN$ system and that of the whole training set as characterized by the $\bar{\lambda}_{(0.04)}^{(1)}$ descriptor set. The $NCCN$ system is represented by the uniformly sampled points of the system, and the training set is represented by uniformly sampled points plus 3,000,000 randomly sampled points. Principle component analysis (PCA) model is trained with the training data points and applied to both datasets. The plot of first three principle components are shown here, where the orange dots correspond to $NCCN$ system and blue dots correspond to training data.}
\label{fig:TestOutlierNCCN}
\end{figure}